\documentclass[sigconf,balance=false]{acmart}
\usepackage{popets}
\setcopyright{popets}
\copyrightyear{2024}

\acmYear{2024}
\acmVolume{2024}
\acmNumber{4}

\acmDOI{10.56553/popets-2024-XXXX}

\acmISBN{}
\acmConference{Proceedings on Privacy Enhancing Technologies}
\usepackage{tikz}
\usetikzlibrary{arrows,
                chains,
                positioning,
                shapes.geometric
                }

\usepackage{diagbox}
\usepackage{booktabs}
\usepackage{multirow}
\usepackage{graphicx}
\usepackage{xcolor,colortbl}
\usepackage{longtable}

\usepackage{array}
\newcolumntype{C}[1]{>{\centering\arraybackslash}p{#1}}

\usepackage{xurl}
\usepackage[inline]{enumitem} 

\begin{document}

\DeclareRobustCommand*\circledchar[2]{
  \tikz[baseline=(c.base)]
    \node[circle, white, font=\sffamily\bfseries, fill={#1},
    draw={#1}, inner sep=0pt, minimum size=1em](c){#2};
}
\newcommand*\circlea[1]{\circledchar{black}{#1}}

\date{}
\title[Honesty is the Best Policy]{Honesty is the Best Policy: On the Accuracy of Apple Privacy Labels Compared to Apps' Privacy Policies}%Privacy Labels vs. Privacy Policies} % TODO: replace with your title
%don't want date printed

%for single author (just remove % characters)
 % end author

\author{Mir Masood Ali}
\affiliation{
\institution{University of Illinois Chicago}\country{}}
\email{mali92@uic.edu}

 \author{David G. Balash}
\affiliation{
\institution{University of Richmond}
\country{}
}
\email{david.balash@richmond.edu}

\author{Monica Kodwani}
\affiliation{\institution{The George Washington University}\country{}}
\email{monicakodwani@gwu.edu}

\author{Chris Kanich}
\affiliation{\institution{University of Illinois Chicago}\country{}}
\email{ckanich@uic.edu}

\author{Adam J. Aviv}
\affiliation{\institution{The George Washington University}\country{}}
\email{aaviv@gwu.edu}

%-------------------------------------------------------------------------------
\begin{abstract}
%-------------------------------------------------------------------------------
Apple introduced \textit{privacy labels} in Dec. 2020 as a way for developers to report the privacy behaviors of their apps. While Apple does not validate labels, they also require developers to provide a privacy policy, which offers an important comparison point. In this paper, we fine-tuned BERT-based language models to extract privacy policy features for 474,669 apps on the iOS App Store, comparing the output to the privacy labels. We identify discrepancies between the policies and the labels, particularly as they relate to data collected linked to users. We find that 228K apps' privacy policies may indicate data collection linked to users than what is reported in the privacy labels. More alarming, a large number (97\%) of the apps with a {\em Data Not Collected} privacy label have a privacy policy indicating otherwise. We provide insights into potential sources for discrepancies, including the use of templates and confusion around Apple's definitions and requirements. These results suggest that significant work is still needed to help developers more accurately label their apps. Our system can be incorporated as a first-order check to inform developers when privacy labels are possibly misapplied. 

%linked according to the policy, but are not 
%
% \reminder{metdata sentence}
%
% Our approach correctly identified 97.1\% of the apps that had a privacy labels indicating that they collect data that is linked to the user. We also identified 56.81\% apps that had failed to declare that they collect data that is linked to the user. 
%Additionally, while 37.15\% of the apps that we analyzed declared on the App Store that they did not collect any data, we found that 95.58\% of these apps participated in some form of data collection, and 93.19\% of them collected data that is linked to the user.
%

% immaturity of privacy labels and their current application as evidenced by the divergence from reported behaving the the privacy policies, which often vetted for legal purposes. Privacy labels appear to receive no such oversight, suggesting that there may be broader disparities that we are not capturing and potentially misinforming users of the privacy functionality of apps. 
\end{abstract}

\maketitle

% Square showing the overlap between privacy types on the App Store vs. Polisis 
\newcommand{\privacyLabelPolicyOverlap}[0]{
    \begin{table*}[t]
    \centering
    \caption{The number of apps with three of the privacy types associated with their data collection practices, as stated in privacy labels, against practices found in privacy policies. Please note that three of the \textit{Privacy Types} shown here, \textit{Data Used to Track You}, \textit{Data Linked to You} and \textit{Data Not Linked to You}, are not mutually exclusive. \textcolor{red}{(values)} indicate the number of apps that did \textit{not} also declare the corresponding privacy type found by Polisis. } 
    \label{tab:label-policy-overlap}
    
    \resizebox{\linewidth}{!}{
    \begin{tabular}{|c|c|c|c|c|}
    \hline
    \diagbox[width=8em]{\textbf{Policy}}{\textbf{Label}} & \textbf{Data Used to Track You} & \textbf{Data Linked to You} & \textbf{Data Not Linked to You} & \textbf{Data Not Collected} \\ \hline
    \textbf{Data Used to Track You} & \textbf{53,359}                   & 83,160 \textcolor{red}{(48,039)}     &   91,665 \textcolor{red}{(53,912)} &  45,074  \textcolor{red}{(45,074)} \\ \hline
    \textbf{Data Linked to You}     & 97,333 \textcolor{red}{(34,294)} & \textbf{168,121} &     188,041 \textcolor{red}{(97,029)}     & 131,310 \textcolor{red}{(131,310)}       \\ \hline
    \textbf{Data Not Linked to You} & 44,479 \textcolor{red}{(13,636)}  & 77,171  \textcolor{red}{(35,815)}          & \textbf{88,172} & 43,354  \textcolor{red}{(43,354)}      \\ \hline
    \textbf{Data Not Collected}     & 391                             & 431      & 796            & \textbf{4,359} \\ \hline
    \end{tabular}}
    \vspace{-2ex}
    \end{table*}
}
\newcommand{\conversionTable}[0]{
% \usepackage{booktabs}
% \usepackage{multirow}
% \usepackage{graphicx}
\begin{table}[t]
\centering
\caption{Deriving privacy label entries directly from segment annotations created using the Polisis framework.}
\label{tab:primary-conversion}
\resizebox{\columnwidth}{!}{%
\begin{tabular}{lcc}
\toprule
\multicolumn{1}{c}{\textbf{Apple Privacy Label}} &
  \multicolumn{2}{c}{\textbf{Polisis}} \\ \cmidrule{1-3} 
\multicolumn{1}{l}{\textbf{Privacy Type}} &
  \multicolumn{1}{c}{\textbf{\begin{tabular}[c]{@{}c@{}}High-level\\ Data Practice\end{tabular}}} &
  \multicolumn{1}{c}{\textbf{\begin{tabular}[c]{@{}c@{}}Identifiability\end{tabular}}} 
 \\ \cmidrule(r){1-3}
\cellcolor[HTML]{EFEFEF}\begin{tabular}[c]{@{}l@{}}Data Linked\\ to You  \end{tabular} &
\cellcolor[HTML]{EFEFEF}  \begin{tabular}[c]{@{}c@{}}First Party Collection/Use \\ Third Party Collection/Sharing \end{tabular} &
\cellcolor[HTML]{EFEFEF}  Identifiable  \\
\begin{tabular}[c]{@{}l@{}}Data Not Linked\\ to You  \end{tabular} &
  \begin{tabular}[c]{@{}c@{}}First Party Collection/Use  \\ Third Party Collection/Sharing \end{tabular} &
  Aggregated/anonymized  \\ \cmidrule{1-3}
\multicolumn{1}{l}{\textbf{Privacy Type}} &
  \multicolumn{1}{c}{\textbf{\begin{tabular}[c]{@{}c@{}}High-level\\ Data Practice\end{tabular}}} &
\multicolumn{1}{c}{\textbf{\begin{tabular}[c]{@{}c@{}}Does/Does Not\end{tabular}}} 
\\ \cmidrule{1-3}
\begin{tabular}[c]{@{}l@{}}Data Not\\ Collected \end{tabular} &
  \begin{tabular}[c]{@{}c@{}}First Party Collection/Use \\ Third Party Collection/Sharing \end{tabular} &
  Does Not \\ \cmidrule{1-3}
\textbf{Purpose} &
 \multicolumn{1}{c}{\textbf{\begin{tabular}[c]{@{}c@{}}High-level\\ Data Practice\end{tabular}}}   &
\multicolumn{1}{c}{\textbf{Purpose}}
\\ \cmidrule{1-3}
\cellcolor[HTML]{EFEFEF}\begin{tabular}[c]{@{}l@{}}App\\ Functionality \end{tabular} &
\cellcolor[HTML]{EFEFEF}  \begin{tabular}[c]{@{}c@{}}First Party Collection/Use \\ Third Party Collection/Sharing \end{tabular} &
\cellcolor[HTML]{EFEFEF}  \begin{tabular}[c]{@{}c@{}}Basic Service/feature \\ Additional Service/feature \\ Service operation \& security \end{tabular}  \\
Analytics  &
  \begin{tabular}[c]{@{}c@{}}First Party Collection/Use \\ Third Party Collection/Sharing \end{tabular} &
  Analytics/Research  
 \\
\cellcolor[HTML]{EFEFEF}\begin{tabular}[c]{@{}l@{}}Developers\\ Advertising \end{tabular} &
\cellcolor[HTML]{EFEFEF}  \begin{tabular}[c]{@{}c@{}}First Party Collection/Use \end{tabular} &
\cellcolor[HTML]{EFEFEF}  Advertising 
\\
\begin{tabular}[c]{@{}l@{}}Other\\ Purposes \end{tabular} &
  \begin{tabular}[c]{@{}c@{}}First Party Collection/Use \\ Third Party Collection/Sharing \end{tabular} &
  \begin{tabular}[c]{@{}c@{}}Merger/Acquisition \\ Legal requirement \\ Unspecified \end{tabular} 
\\
\cellcolor[HTML]{EFEFEF}\begin{tabular}[c]{@{}l@{}}Third Party\\ Advertising \end{tabular} &
\cellcolor[HTML]{EFEFEF}  Third Party Collection/Sharing &
\cellcolor[HTML]{EFEFEF}  Advertising  
\\
\begin{tabular}[c]{@{}l@{}}Product\\ Personalization \end{tabular} &
  \begin{tabular}[c]{@{}c@{}}First Party Collection/Use \\ Third Party Collection/Sharing \end{tabular} &
  \begin{tabular}[c]{@{}c@{}}Personalization/Customization \end{tabular} 
\\ \cmidrule(r){1-3}
\textbf{Data Category} &
 \multicolumn{1}{c}{\textbf{\begin{tabular}[c]{@{}c@{}}High-level\\ Data Practice\end{tabular}}}   &
\multicolumn{1}{c}{\textbf{\begin{tabular}[c]{@{}c@{}}Personal\\ Information Type\end{tabular}}}  
\\ \cmidrule{1-3}
\cellcolor[HTML]{EFEFEF} \begin{tabular}[c]{@{}l@{}}Contact\\ Info \end{tabular} &
\cellcolor[HTML]{EFEFEF}  \begin{tabular}[c]{@{}c@{}}First Party Collection/Use \\ Third Party Collection/Sharing \end{tabular} &
\cellcolor[HTML]{EFEFEF} Contact  
\\
Location  &
  \begin{tabular}[c]{@{}c@{}}First Party Collection/Use \\ Third Party Collection/Sharing \end{tabular} &
  Location 
\\
\cellcolor[HTML]{EFEFEF} \begin{tabular}[c]{@{}l@{}}Financial\\ Info \end{tabular} &
\cellcolor[HTML]{EFEFEF}  \begin{tabular}[c]{@{}c@{}}First Party Collection/Use \\ Third Party Collection/Sharing \end{tabular} &
\cellcolor[HTML]{EFEFEF} Financial 
\\
Identifiers  &
  \begin{tabular}[c]{@{}c@{}}First Party Collection/Use \\ Third Party Collection/Sharing \end{tabular} &
  \begin{tabular}[c]{@{}c@{}}Cookies \& Tracking Elements \\ IP address \& Device IDs \end{tabular} 
\\
\cellcolor[HTML]{EFEFEF} \begin{tabular}[c]{@{}l@{}}Usage\\ Data \end{tabular} &
\cellcolor[HTML]{EFEFEF}  \begin{tabular}[c]{@{}c@{}}First Party Collection/Use \\ Third Party Collection/Sharing \end{tabular} &
\cellcolor[HTML]{EFEFEF} User Online Activities  
\\
\begin{tabular}[c]{@{}l@{}}User\\ Content \end{tabular} &
  \begin{tabular}[c]{@{}c@{}}First Party Collection/Use \\ Third Party Collection/Sharing \end{tabular} &
  \begin{tabular}[c]{@{}c@{}}User Profile \\ Social Media Data \end{tabular} 
\\
\cellcolor[HTML]{EFEFEF} \begin{tabular}[c]{@{}l@{}}Health \&\\ Fitness \end{tabular} &
\cellcolor[HTML]{EFEFEF}  \begin{tabular}[c]{@{}c@{}}First Party Collection/Use \\ Third Party Collection/Sharing \end{tabular} &
\cellcolor[HTML]{EFEFEF}  Health 
\\ 
\begin{tabular}[c]{@{}l@{}}Browsing\\ History \end{tabular} &
  Third Party Collection/Sharing &
  User Online Activities 
 \\  
  \bottomrule
\end{tabular}%
}
\vspace{-3ex}
\end{table}
}

\newcommand{\inferentialConversionTable}[0]{
% \usepackage{booktabs}
% \usepackage{multirow}
% \usepackage{graphicx}
\begin{table}[ht]
\centering
\caption{Inferring privacy label entries from segment annotations created using the Polisis framework.}
\label{tab:inferential-conversion}
\resizebox{\columnwidth}{!}{%
\begin{tabular}{lcc}
\toprule
\multicolumn{1}{c}{\textbf{Apple Privacy Label}} &
  \multicolumn{2}{c}{\textbf{Polisis}} \\ \cmidrule{1-3} 
\multicolumn{1}{l}{\textbf{Privacy Type}} &
  \multicolumn{1}{c}{\textbf{\begin{tabular}[c]{@{}c@{}}High-level\\ Data Practice\end{tabular}}} &
  \multicolumn{1}{c}{\textbf{\begin{tabular}[c]{@{}c@{}}Purpose\end{tabular}}} 
 \\ \cmidrule{1-3}
\begin{tabular}[c]{@{}l@{}}Data Used to\\ Track You \end{tabular} &
\begin{tabular}[c]{@{}c@{}}Third Party Collection/Sharing \end{tabular} &
\cellcolor[HTML]{EFEFEF}  Advertising  
 \\ \cmidrule{1-3}
\textbf{Data Category} &
 \multicolumn{1}{c}{\textbf{\begin{tabular}[c]{@{}c@{}}High-level\\ Data Practice\end{tabular}}} &
 \multicolumn{1}{c}{\textbf{\begin{tabular}[c]{@{}c@{}}Personal\\ Information\\ Type\end{tabular}}}  
\\ \cmidrule{1-3}
\begin{tabular}[c]{@{}l@{}}Diagnostics \end{tabular} &
\begin{tabular}[c]{@{}c@{}}First Party Collection/Use \\ Third Party Collection/Sharing \end{tabular} &
\cellcolor[HTML]{EFEFEF}  \begin{tabular}[c]{@{}c@{}}Computer Information \\ IP address \& Device IDs \end{tabular}
\\ \cmidrule{2-3}
\multirow{2}{*}{Contacts} &
  \multirow{2}{*}{\begin{tabular}[c]{@{}c@{}}First Party Collection/Use \\ Third Party Collection/Sharing \end{tabular}} &
  \cellcolor[HTML]{EFEFEF} \begin{tabular}[c]{@{}c@{}}Social Media Data \\ `contact' , `friend'\end{tabular} 
    \\
     &
     &
  \begin{tabular}[c]{@{}c@{}}`address book', `phone book'\end{tabular} 
    \\ \cmidrule{2-3}
\begin{tabular}[c]{@{}l@{}}Purchases \end{tabular} &
\begin{tabular}[c]{@{}c@{}}First Party Collection/Use \\ Third Party Collection/Sharing \end{tabular} &
 \begin{tabular}[c]{@{}c@{}}Financial \\ User Online Activities \end{tabular} 
\\ \cmidrule{2-3}
\begin{tabular}[c]{@{}l@{}}Search\\ History \end{tabular} &
\begin{tabular}[c]{@{}c@{}}First Party Collection/Use \end{tabular} &
\cellcolor[HTML]{EFEFEF}  \begin{tabular}[c]{@{}c@{}}User Online Activities \\ `search'\end{tabular} 
\\ \cmidrule{2-3}
\multirow{-3}{*}{\begin{tabular}[c]{@{}l@{}}\\ \\ Sensitive\\ Info \end{tabular}} &
\multirow{-3}{*}{\begin{tabular}[c]{@{}c@{}}\\ \\ First Party Collection/Use \\ Third Party Collection/Sharing \end{tabular}} &
 \cellcolor[HTML]{EFEFEF} \begin{tabular}[c]{@{}c@{}}Demographic \\
 `race', `racial',`ethnic', `ethnicity', \\ 
 `sexual orientation', `sexual preference',\\
 `pregnancy', 'pregnant', `childbirth', \\
 `child birth', `child-birth', `disability',\\ 
 `religion', `religious', `religious belief', \\
 `trade union', `union member',`politics', \\\
 `political' `genetic',`genetic information', 
 \\'biometric'
 \end{tabular}
\\  
  \bottomrule
\end{tabular}%
}
\vspace{-2ex}
\end{table}
}

\newcommand{\templatesTable}[0]{
% \usepackage{booktabs}
% \usepackage{multirow}
% \usepackage{graphicx}
\begin{table*}[t]
\caption{Overview of privacy policy templates used to compare policies. Please note that matched policies are not mutually exclusive, i.e., policies can be found to be similar to more than one template.}
\label{tab:templates}
\resizebox{\textwidth}{!}{%
\begin{tabular}{@{}lrrrrrrrrrrrrrr@{}}
\toprule
\multicolumn{1}{c}{\multirow{2}{*}{\textbf{Template}}} &
  \multicolumn{1}{c}{\multirow{2}{*}{\textbf{\#Policies}}} &
  \multicolumn{1}{c}{\multirow{2}{*}{\textbf{\#Apps}}} &
  \multicolumn{3}{c}{\textbf{Data Used to Track You}} &
  \multicolumn{3}{c}{\textbf{Data Linked to You}} &
  \multicolumn{3}{c}{\textbf{Data Not Linked to You}} &
  \multicolumn{3}{c}{\textbf{Data Not Collected}} \\ \cmidrule(l){4-15} 
\multicolumn{1}{c}{} &
  \multicolumn{1}{c}{} &
  \multicolumn{1}{c}{} &
  \multicolumn{1}{c}{\textbf{Label Only}} &
  \multicolumn{1}{c}{\textbf{\begin{tabular}[c]{@{}c@{}}Policy Only\end{tabular}}} &
  \multicolumn{1}{c}{\textbf{Overlap}} &
  \multicolumn{1}{c}{\textbf{Label Only}} &
  \multicolumn{1}{c}{\textbf{\begin{tabular}[c]{@{}c@{}}Policy Only\end{tabular}}} &
  \multicolumn{1}{c}{\textbf{Overlap}} &
  \multicolumn{1}{c}{\textbf{Label Only}} &
  \multicolumn{1}{c}{\textbf{\begin{tabular}[c]{@{}c@{}}Policy Only\end{tabular}}} &
  \multicolumn{1}{c}{\textbf{Overlap}} &
  \multicolumn{1}{c}{\textbf{Label Only}} &
  \multicolumn{1}{c}{\textbf{\begin{tabular}[c]{@{}c@{}}Policy Only\end{tabular}}} &
  \multicolumn{1}{c}{\textbf{Overlap}} \\ \cmidrule(r){1-15}
\rowcolor[HTML]{EFEFEF}
\href{https://securiti.ai/privacy-center/ }{Securiti Privacy Center}           & 145759 & 221211 & 30877 & 51371 & 21619 & 4789 & 121108 & 94959 & 28267 & 85358 & 76230 & 78976 & 348 & 188 \\
\href{https://www.websiteprivacypolicygenerator.com/}{WebsitePrivacyPolicyGenerator.com} & 94170  & 131258 & 20363 & 23638 & 9555  & 4049 & 80955  & 45629 & 19539 & 49929 & 40513 & 56827 & 174 & 362 \\
\rowcolor[HTML]{EFEFEF}
\href{https://termly.io/products/privacy-policy-generator/}{Termly}                            & 89275  & 183697 & 25870 & 47995 & 21985 & 4909 & 94748  & 84013 & 27524 & 64195 & 61567 & 60331 & 335 & 166 \\
\href{https://www.termsfeed.com}{TermsFeed}                         & 86199  & 126844 & 19964 & 26507 & 9751  & 1947 & 74712  & 49945 & 14464 & 52646 & 42936 & 48868 & 57  & 68  \\
\rowcolor[HTML]{EFEFEF}
\href{https://www.enzuzo.com/privacy-policy-generator}{Enzuzo}                             & 80837  & 140996 & 21543 & 31389 & 14184 & 2531 & 76193  & 61901 & 19317 & 51293 & 47137 & 47615 & 75  & 95  \\
\href{https://www.freeprivacypolicy.com/free-privacy-policy-generator/}{FreePrivacyPolicy}              & 36224  & 88209  & 15513 & 18560 & 9921  & 2152 & 45845  & 40121 & 15949 & 28306 & 25494 & 29493 & 33  & 68  \\
\rowcolor[HTML]{EFEFEF}
\href{https://app-privacy-policy-generator.firebaseapp.com/}{App Privacy Policy Generator}      & 31786  & 39479  & 10787 & 1510  & 821   & 1015 & 27752  & 10895 & 4641  & 16892 & 13048 & 19115 & 20  & 59  \\
\href{https://www.privacypolicygenarator.info/}{PrivacyPolicyGenerator.info}       & 25504  & 53020  & 10479 & 8202  & 5205  & 3535 & 32314  & 17753 & 13379 & 14379 & 12686 & 24267 & 94  & 230 \\
\rowcolor[HTML]{EFEFEF}
\href{https://www.privacypolicies.com/}{PrivacyPolicies.com}              & 18292  & 41740  & 5880  & 9391  & 4610  & 1118 & 23125  & 17404 & 7321  & 12579 & 13402 & 14796 & 19  & 28  \\
\href{https://www.pandadoc.com/free-privacy-policy-template/}{PandaDoc}                             & 15306  & 30135  & 6729  & 4403  & 2189  & 2572 & 19933  & 8250  & 8698  & 7388  & 5834  & 15859 & 105 & 161 \\
\rowcolor[HTML]{EFEFEF}
\href{https://www.privacypolicygenerator.org}{PrivacyPolicyGenerator.org}        & 13429  & 24383  & 5900  & 3631  & 1798  & 2415 & 16028  & 6382  & 6940  & 6436  & 4600  & 13933 & 67  & 171 \\
\href{https://getterms.io/}{GetTerms}                         & 5274   & 11840  & 1757  & 1945  & 1395  & 319  & 6326   & 5089  & 2358  & 3567  & 3372  & 4250  & 1   & 6   \\
\rowcolor[HTML]{EFEFEF}
\href{https://www.iubenda.com/ }{iubenda}                           & 3718   & 6632   & 566   & 1465  & 699   & 22   & 4372   & 2176  & 1435  & 2168  & 2283  & 2041  & 0   & 0   \\
\href{https://websitepolicies.com}{WebsitePolicies.com}               & 1993   & 3302   & 132   & 1165  & 355   & 2    & 2130   & 1170  & 132   & 1628  & 1362  & 1347  & 0   & 0   \\
\rowcolor[HTML]{EFEFEF}
\href{https://www.shopify.com/tools/policy-generator}{Shopify}                           & 371    & 540    & 123   & 55    & 22    & 230  & 274    & 254   & 468   & 32    & 18    & 188   & 0   & 0   \\ \bottomrule
\end{tabular}%
}
\end{table*}
}

\newcommand{\trafficCollectionTable}[0]{
% \usepackage{graphicx}
% \usepackage[normalem]{ulem}
% \useunder{ine}{}{}
\begin{table*}[t]
\centering
\caption{An overview of network traffic collection for apps presented as case studies.}
\label{tab:traffic-collection}
\resizebox{0.82\textwidth}{!}{%
\begin{tabular}{llcccll}
\hline
\textbf{\#} &
  \multicolumn{1}{c}{\textbf{App Name}} &
  \multicolumn{1}{c}{\textbf{App Genre}} &
  \begin{tabular}[c]{@{}c@{}}\textbf{Declared in} \\ \textbf{Privacy Label} \end{tabular} &
  \begin{tabular}[c]{@{}c@{}}\textbf{Declared in} \\ \textbf{Privacy Policy} \end{tabular} &
  \multicolumn{1}{c}{\textbf{Trackers}} &
  \multicolumn{1}{c}{\textbf{Notes}}
   \\ \hline
\multicolumn{7}{l}{\textbf{Apps that do not declare tracking in their privacy label}} 
   \\
\rowcolor[HTML]{EFEFEF}
1 &
  \href{https://apps.apple.com/us/app/venmo/id351727428}{Venmo} &
  \multicolumn{1}{c}{Finance} &
  N &
  Y &
  \begin{tabular}[c]{@{}l@{}}Kochava, Optimizely\end{tabular} &
  \begin{tabular}[c]{@{}l@{}}Incomplete understanding of\\ App Store Requirements\end{tabular} \\
2 &
  \href{https://apps.apple.com/us/app/bible/id282935706}{Bible} &
  \multicolumn{1}{c}{Reference} &
  N &
  Y &
  \begin{tabular}[c]{@{}l@{}}Facebook, Google, Branch\end{tabular} &
  \begin{tabular}[c]{@{}l@{}}Incomplete understanding of\\ App Store's Definition of\\ tracking\end{tabular} \\
\rowcolor[HTML]{EFEFEF}
3 &
  \href{https://apps.apple.com/us/app/zillow-real-estate-rentals/id310738695}{Paypal} &
  \multicolumn{1}{c}{Finance} &
  N &
  Y &
  \begin{tabular}[c]{@{}l@{}}Adjust, Qualtrics\end{tabular} &
  \begin{tabular}[c]{@{}l@{}}Incomplete understanding of \\ third party tracking\end{tabular} \\
4 &
  \href{https://apps.apple.com/us/app/southwest-airlines/id344542975}{Southwest Airlines} &
  \multicolumn{1}{c}{Travel} &
  N &
  Y &
  \begin{tabular}[c]{@{}l@{}}Adobe, Qualtrics, Branch, Salesforce, Akamai \end{tabular} &
  \begin{tabular}[c]{@{}l@{}}Incomplete understanding of\\ App Store Requirements\end{tabular} \\
\rowcolor[HTML]{EFEFEF}
5 &
  \href{https://apps.apple.com/us/app/my-verizon/id416023011}{My Verizon} &
  \multicolumn{1}{c}{Utilities} &
  N &
  Y &
  \begin{tabular}[c]{@{}l@{}}Adobe, Google\end{tabular} &
  \begin{tabular}[c]{@{}l@{}}Incomplete understanding of\\ App Store's Definition of\\ tracking\end{tabular} \\
6 &
  \href{https://apps.apple.com/us/app/opentable/id296581815}{Open Table} &
  \multicolumn{1}{c}{Food \& Drink} &
  N &
  Y &
  \begin{tabular}[c]{@{}l@{}}Adjust, Mixpanel, Facebook\end{tabular} &
  \begin{tabular}[c]{@{}l@{}}Incomplete understanding of\\ App Store Requirements\end{tabular} \\
\rowcolor[HTML]{EFEFEF}
7 &
  \href{https://apps.apple.com/us/app/geico-mobile-car-insurance/id331763096}{Geico Mobile} &
  \multicolumn{1}{c}{Finance} &
  N &
  Y &
  \begin{tabular}[c]{@{}l@{}}Adobe, Airship, Branch, Google\end{tabular} &
  \begin{tabular}[c]{@{}l@{}}Incomplete understanding of\\ App Store's Definition of\\ tracking\end{tabular} \\
8 &
  \href{https://apps.apple.com/us/app/citi-mobile/id301724680}{Citi Mobile} &
  \multicolumn{1}{c}{Finance} &
  N &
  Y &
  \begin{tabular}[c]{@{}l@{}}Adobe, Google, Mixpanel\end{tabular} &
  \begin{tabular}[c]{@{}l@{}}Incomplete understanding of\\ App Store's Definition of\\ tracking\end{tabular} \\
\rowcolor[HTML]{EFEFEF}
9 &
  \href{https://apps.apple.com/us/app/crumbl/id1438166219}{Crumbl} &
  \multicolumn{1}{c}{Food \& Drink} &
  N &
  Y &
  \begin{tabular}[c]{@{}l@{}}Branch, Google, Facebook\end{tabular} &
  \begin{tabular}[c]{@{}l@{}}Incomplete understanding of \\ third party tracking\end{tabular} \\
10 &
  \href{https://apps.apple.com/us/app/classdojo/id552602056}{Class Dojo} &
  \multicolumn{1}{c}{Education} &
  N &
  Y &
  \begin{tabular}[c]{@{}l@{}}Datadog, Google, Zendesk\end{tabular} &
  \begin{tabular}[c]{@{}l@{}}Incomplete understanding of\\ App Store's Definition of\\ tracking\end{tabular} \\
\rowcolor[HTML]{EFEFEF}
11 &
  \href{https://apps.apple.com/us/app/indeed-job-search/id309735670}{Indeed Job Search} &
  \multicolumn{1}{c}{Business} &
  N &
  Y &
  \begin{tabular}[c]{@{}l@{}}AppsFlyer, iSpot, Google\end{tabular} &
  \begin{tabular}[c]{@{}l@{}}Incomplete understanding of\\ App Store requirements\end{tabular} \\
12 &
  \href{https://apps.apple.com/us/app/discord-chat-talk-hangout/id985746746}{Discord} &
  \multicolumn{1}{c}{Social Networking} &
  N &
  Y &
  \begin{tabular}[c]{@{}l@{}}Adjust, Google\end{tabular} &
  \begin{tabular}[c]{@{}l@{}}Incomplete understanding of\\ App Store's Definition of\\ tracking\end{tabular} \\
\rowcolor[HTML]{EFEFEF}
13 &
  \href{https://apps.apple.com/us/app/sams-club/id382497397}{Sam's Club} &
  \multicolumn{1}{c}{Shopping} &
  N &
  Y &
  \begin{tabular}[c]{@{}l@{}}Adobe, Branch, Google, PerimeterX, Moat Ads\end{tabular} &
  \begin{tabular}[c]{@{}l@{}}Policy Template Reuse\end{tabular} \\
14 &
  \href{https://apps.apple.com/us/app/lime-ridegreen/id1199780189}{Lime Ride} &
  \multicolumn{1}{c}{Travel} &
  N &
  Y &
  \begin{tabular}[c]{@{}l@{}}Amazon, Branch, Facebook, Google, Unity, Super Sonic Ads\end{tabular} &
  \begin{tabular}[c]{@{}l@{}}Incomplete understanding of\\ App Store requirements\end{tabular} \\
\rowcolor[HTML]{EFEFEF}
15 &
  \href{https://apps.apple.com/us/app/groupme/id392796698}{GroupMe} &
  \multicolumn{1}{c}{Social Networking} &
  N &
  Y &
  \begin{tabular}[c]{@{}l@{}}OneSignal, MixPanel\end{tabular} &
  \begin{tabular}[c]{@{}l@{}}Policy Template Reuse\end{tabular} \\
16 &
  \href{https://apps.apple.com/us/app/lexington-law-credit-repair/id593682112}{Lexington Law} &
  \multicolumn{1}{c}{Finance} &
  N &
  Y &
  \begin{tabular}[c]{@{}l@{}}Facebook, Google, Adobe, \\ TheTradeDesk, LiveIntent, \\ StackAdapt,  Bing, TikTok, \\ Taboola, Snapchat, Twitter\end{tabular} &
  Policy Template Reuse
   \\
\rowcolor[HTML]{EFEFEF}
17 &
  \href{https://apps.apple.com/us/app/creditrepair/id562091020}{CreditRepair} &
  \multicolumn{1}{c}{Finance} &
  N &
  Y &
  \begin{tabular}[c]{@{}l@{}}Facebook, Google, Adobe, \\ StackAdapt, TTD, Twitter, \\ Yahoo, LiveIntent, Taboola\end{tabular} &
  Policy Template Reuse
   \\
18 &
  \href{https://apps.apple.com/us/app/aldi-usa/id429396645}{Aldi} &
  \multicolumn{1}{c}{Shopping} &
  N &
  Y &
  Adobe, Google &
  \begin{tabular}[c]{@{}l@{}}Incomplete understanding of \\ App Store requirements \end{tabular}
   \\
% 4 &
%   \href{https://apps.apple.com/us/app/dish-cult-restaurant-bookings/id1141667303}{DishCult} &
%   \multicolumn{1}{c}{Food \& Drink} &
%   N &
%   Y &
%   Facebook, AppsFlyer &
%   \begin{tabular}[c]{@{}l@{}}Incomplete understanding of \\ third party collection\end{tabular}
%    \\
% \rowcolor[HTML]{EFEFEF}
% 5 &
%   \href{https://apps.apple.com/us/app/foodhub-online-takeaways/id1201845916}{FoodHub} &
%   \multicolumn{1}{c}{Food \& Drink} &
%   N &
%   Y &
%   Facebook, Google, MoEngage &
%   \begin{tabular}[c]{@{}l@{}}Incomplete understanding of \\ third party collection\end{tabular}
%    \\
% 6 &
%   \href{https://apps.apple.com/us/app/modern-milkman/id1448349321}{Modern Milkman} &
%   \multicolumn{1}{c}{Food \& Drink} &
%   N &
%   Y &
%   \begin{tabular}[c]{@{}l@{}}Facebook, Google, AppsFlyer, \\ LiveIntent, BidSwitch, ShareThis\end{tabular} &
%   \begin{tabular}[c]{@{}l@{}}Incomplete understanding of \\ App Store requirements \end{tabular}
%    \\
\rowcolor[HTML]{EFEFEF}   
19 &
  \href{https://apps.apple.com/us/app/axolochi/id1432184360}{Axolochi} &
  \multicolumn{1}{c}{Games} &
  N &
  Y &
  Google, SuperSonic, Unity &
  \begin{tabular}[c]{@{}l@{}}Incomplete understanding of \\ App Store's definition\\ of tracking\end{tabular}
   \\
20 &
  \href{https://apps.apple.com/us/app/hello-neighbor/id1386358600}{Hello Neighbor} &
  \multicolumn{1}{c}{Games} &
  N &
  Y &
  Google, SuperSonic &
  \begin{tabular}[c]{@{}l@{}}Incomplete understanding of \\ third party collection\end{tabular}
   \\
\rowcolor[HTML]{EFEFEF}   
% 9 &
%  \href{https://apps.apple.com/us/app/scripts-learn-chinese-writing/id1436303395}{\begin{tabular}[c]{@{}l@{}} Scripts - Learn\\ Chinese Writing\end{tabular}} &
%   \multicolumn{1}{c}{Education} &
%   N &
%   Y &
%   Facebook, Google &
%   \begin{tabular}[c]{@{}l@{}}Incomplete understanding of \\ App Store's requirements\end{tabular}
%    \\
% 10 &
%   \href{https://apps.apple.com/us/app/chemistry-periodic-table/id493558583}{\begin{tabular}[c]{@{}l@{}}Chemistry \&\\Periodic Table\end{tabular}} &
%   \multicolumn{1}{c}{Education} &
%   N &
%   Y &
%   Google &
%   \begin{tabular}[c]{@{}l@{}}Incomplete understanding of \\ third party collection\end{tabular}
%    \\
% \rowcolor[HTML]{EFEFEF}  
% 11 &
%  \href{https://apps.apple.com/us/app/demo-songwriting-studio/id1563264178}{\begin{tabular}[c]{@{}l@{}} Demo - Song-\\writing Studio\end{tabular}} &
%   \multicolumn{1}{c}{Music} &
%   N &
%   Y &
%   Facebook, Google, AppsFlyer &
%   \begin{tabular}[c]{@{}l@{}}Incomplete understanding of \\ third party collection\end{tabular}
%    \\
21 &
  \href{https://apps.apple.com/us/app/webmd-symptom-checker/id295076329}{WebMD} &
  \multicolumn{1}{c}{Medical} &
  N &
  Y &
  Adobe, Google &
  \begin{tabular}[c]{@{}l@{}}Incomplete understanding of \\ App Store's definition\\ of tracking\end{tabular}
   \\  
22 &
   \href{https://apps.apple.com/us/app/food-network-magazine-us/id503569987}{\begin{tabular}[c]{@{}l@{}}Food Network\\Magazine\end{tabular}} &
  \multicolumn{1}{c}{Food \& Drink} &
  N &
  Y &
  Facebook, Google &
  \begin{tabular}[c]{@{}l@{}}Incomplete understanding of \\ App Store's definition\\ of tracking\end{tabular}
   \\
\rowcolor[HTML]{EFEFEF}
23 &
  \href{https://apps.apple.com/us/app/best-buy/id314855255}{Best Buy} &
  \multicolumn{1}{c}{Shopping} &
  N &
  Y &
  Adobe, Google, Criteo &
  \begin{tabular}[c]{@{}l@{}}Incomplete understanding of \\ App Store's requirements\end{tabular}
   \\   
24 &
  \href{https://apps.apple.com/us/app/maya-my-period-tracker/id492534636}{Maya Period Tracker} &
  \multicolumn{1}{c}{Health \& Fitness} &
  Partial &
  Y &
  Facebook, Google &
  \begin{tabular}[c]{@{}l@{}}Incomplete understanding of \\ App Store's requirements\end{tabular}
   \\
\multicolumn{7}{l}{\textbf{Apps that declare tracking in their privacy label but have an unclear privacy policy}} 
   \\
\rowcolor[HTML]{EFEFEF}
25 &
  \href{https://apps.apple.com/us/app/shake-shack/id317279545}{Shake Shack} &
  Food \& Drink &
  Y &
  N &
  Facebook, Google &
  \begin{tabular}[c]{@{}l@{}}Not Stated in Policy\end{tabular} 
   \\  
26 &
  \href{https://apps.apple.com/us/app/kika-keyboard-for-iphone-ipad/id1035199024}{Kika Keyboard} &
  Utilities &
  Y &
  N &
 \begin{tabular}[c]{@{}l@{}}AppLovin, Facebook,\\ Google\end{tabular} &
  \begin{tabular}[c]{@{}l@{}}Not Stated in Policy\end{tabular} 
   \\
\rowcolor[HTML]{EFEFEF} 
27 &
  \href{https://apps.apple.com/us/app/photo-prints-now-cvs-photo/id1232461700}{Photo Prints CVS} &
  Photo \& Video &
  Y &
  N &
  Facebook, Google &
  \begin{tabular}[c]{@{}l@{}}Not Stated in Policy\end{tabular}  
   \\
28 &
  \href{https://apps.apple.com/us/app/everpix-cool-wallpapers-hd-4k/id921160527}{Everpix} &
  Entertainment &
  Y &
  N &
   \begin{tabular}[c]{@{}l@{}}AppLovin, Facebook,\\ Google, Liftoff\end{tabular} &
   \begin{tabular}[c]{@{}l@{}}Not Stated in Policy\end{tabular} 
   \\
\rowcolor[HTML]{EFEFEF}   
29 &
  \href{https://apps.apple.com/us/app/floatme-instant-cash/id1395667279}{FloatMe} &
  Finance &
  Y &
  N &
  \begin{tabular}[c]{@{}l@{}}Facebook, Google,\\ AppsFlyer\end{tabular} &
  \begin{tabular}[c]{@{}l@{}}Not Stated in Policy\end{tabular}
   \\
30 &
  \href{https://apps.apple.com/us/app/buffalo-wild-wings/id1031364004}{Buffalo Wild Wings} &
  Food \& Drink &
  Y &
  N &
  Google &
  Not Clearly Stated in Policy
   \\
\rowcolor[HTML]{EFEFEF}
31 &
  \href{https://apps.apple.com/us/app/the-general-auto-insurance-app/id1397958651}{\begin{tabular}[c]{@{}l@{}}The General Auto-\\Insurance App\end{tabular}} &
  Finance &
  Y &
  N &
  Facebook, Google &
  Not Clearly Stated in Policy
   \\
32 &
  \href{https://apps.apple.com/us/app/conservative-news/id1207514833}{Conservative News} &
  News &
  Y &
  N &
  \begin{tabular}[c]{@{}l@{}}Amazon, AppLovin,\\Flurry, Google\end{tabular} &
  Not Clearly Stated in Policy
   \\
\rowcolor[HTML]{EFEFEF}   
33 &
  \href{https://apps.apple.com/us/app/brainboom-word-brain-games/id1550734007}{BrainBoom} &
  Games &
  Y &
  Y &
  \begin{tabular}[c]{@{}l@{}}AppLovin, Facebook,\\ Google, Supersonic Ads,\\ InMobi, TapJoy, IronSource,\\ Vungle, AdColony\end{tabular} &
  \begin{tabular}[c]{@{}l@{}} Presented as an image,\\ difficult to parse\end{tabular}
   \\
34 &
  \href{https://apps.apple.com/us/app/stickman-boxing-battle-3d/id1491361807}{Stickman Boxing} &
  Games &
  Y &
  Y &
 \begin{tabular}[c]{@{}l@{}}Amazon, AppLovin,\\ Facebook, Google, IronSource,\\ Supersonic Ads, TapJoy,\\ Vungle, Yandex\end{tabular}  &
  \begin{tabular}[c]{@{}l@{}}Separate Declaration of\\ Data Collection and Purpose.\end{tabular}
   \\
\rowcolor[HTML]{EFEFEF} 
35 &
  \href{https://apps.apple.com/us/app/mcdonalds/id922103212}{McDonalds} &
  Food \& Drink &
  Y &
  Y &
  \begin{tabular}[c]{@{}l@{}}Adobe, Facebook,\\ Google, Kochava\end{tabular}&
  \begin{tabular}[c]{@{}l@{}}Policy segments linked\\ on landing page\end{tabular}
   \\  
36 &
  \href{https://apps.apple.com/us/app/episode-choose-your-story/id656971078}{\begin{tabular}[c]{@{}l@{}}Episode: Choose Your Story\end{tabular}} &
  Games &
  Y &
  Y &
  \begin{tabular}[c]{@{}l@{}}Adjust, Facebook,\\ Google\end{tabular} &
   \begin{tabular}[c]{@{}l@{}}Policy linked behind a link\\ on the landing page from\\ App Store\end{tabular} 
   \\
\rowcolor[HTML]{EFEFEF}  
37 &
  \href{https://apps.apple.com/us/app/jcpenney-shopping-coupons/id925338276}{JCPenney} &
  Shopping &
  Y &
  Y &
  \begin{tabular}[c]{@{}l@{}}Adobe, Facebook,\\ Google, UrbanAirship\end{tabular}  &
  \begin{tabular}[c]{@{}l@{}}Incorrect Policy Link.\\Different part of website\end{tabular} 
   \\ 
38 &
  \href{https://apps.apple.com/us/app/dosh-find-cash-back-deals/id1167047511}{Dosh} &
  Shopping &
  Y &
  Y &
  AppsFlyer, Google &
  \begin{tabular}[c]{@{}l@{}}Incorrect Policy Link.\\Different part of website \end{tabular} 
   \\
\rowcolor[HTML]{EFEFEF} 
39 &
  \href{https://apps.apple.com/us/app/cdl-prep-test-by-coco/id1527903479}{CDL Prep Test} &
  Reference &
  Y &
  N &
  Google &
  \begin{tabular}[c]{@{}l@{}}Incorrect Policy Link. Link broken.\end{tabular}
   \\
%  &
%    &
%    &
%    &
%    &
%    &
%    &
%    &
%    &
%    \\ 
\hline
\end{tabular}%
}
\end{table*}
}

%%%%%
% Data Categories Table: Browsing History
%%%%%

\newcommand{\dataCategoriesOverlap}[0]{
\onecolumn
    % \begin{longtable}[c]{@{}m{0.2\textwidth}p{0.2\textwidth}p{0.2\textwidth}p{0.2\textwidth}@{}}
    \begin{longtable}[c]{|m{0.18\textwidth}|C{0.14\textwidth}|C{0.14\textwidth}|C{0.14\textwidth}|C{0.14\textwidth}|C{0.14\textwidth}|}
        \caption{The number of apps with three of the privacy types associated with their collection of \textit{Data Categories}, as stated in privacy labels, against practices found in privacy policies. Please note that three of the \textit{Privacy Types} shown here, \textit{Data Used to Track You}, \textit{Data Linked to You} and \textit{Data Not Linked to You}, are not mutually exclusive. The \textbf{Not Mentioned} column indicates instances wherein the label or policy reports data collection, but not does not mention collecting the specific \textit{Data Category}. \textcolor{red}{(values)} indicate the number of apps that did \textit{not} also declare the corresponding privacy type found by the classifiers.} 
    \label{tab:data-categories-overlap-browsing-history}\\
    \toprule
    % \endfirsthead
    %
    \endhead
    \bottomrule
    \endfoot
    \endlastfoot
    % \begin{table*}[t]
    % \centering

    % \resizebox{\linewidth}{!}{
    % \begin{tabular}{|c|c|c|c|c|c|}
    % \hline
     \multicolumn{6}{|c|}{\textbf{Browsing History}} \\ \cmidrule{1-6} 
    \diagbox[width=8em]{\textbf{Policy}}{\textbf{Label}} & \textbf{Data Used to Track You} & \textbf{Data Linked to You} & \textbf{Data Not Linked to You} & \textbf{Data Not Collected} & \textbf{Not Mentioned} \\ \hline
    \textbf{Data Used to Track You} & \textbf{395}                   & 421 \textcolor{red}{(212)}     &   473 \textcolor{red}{(287)} &  23,394  \textcolor{red}{(23,394)} & 81,830 \textcolor{red}{(81,830)} \\ \hline
    \textbf{Data Linked to You}     & 624 \textcolor{red}{(331)} & \textbf{658} &     834 \textcolor{red}{(834)}     & 61,735 \textcolor{red}{(61,735)}   &  149,552 \textcolor{red}{(149,552)}  \\ \hline
    \textbf{Data Not Linked to You} & 272 \textcolor{red}{(119)}  & 324  \textcolor{red}{(324)}          & \textbf{467} & 27,573  \textcolor{red}{(27,573)}   & 83,745 \textcolor{red}{(83,745)}  \\ \hline
    \textbf{Data Not Collected}     & 1                             & 0      & 2            & \textbf{4,359}  & 0 \\ \hline
    \textbf{Not Mentioned}     & 301                             & 364      & 642            & 63,597  & \textbf{98,731} \\ \hline

    \multicolumn{6}{|c|}{\textbf{Contact Info}} \\ \cmidrule{1-6} 
    \diagbox[width=8em]{\textbf{Policy}}{\textbf{Label}} & \textbf{Data Used to Track You} & \textbf{Data Linked to You} & \textbf{Data Not Linked to You} & \textbf{Data Not Collected} & \textbf{Not Mentioned} \\ \hline
    \textbf{Data Used to Track You} & \textbf{1,450}                   & 13,095 \textcolor{red}{(11,710)}     &   2,808 \textcolor{red}{(2,643)} &  9,147  \textcolor{red}{(9,147)} & 10,467 \textcolor{red}{(10,467)} \\ \hline
    \textbf{Data Linked to You}     & 6,682 \textcolor{red}{(339)} & \textbf{78,411} &     18,954 \textcolor{red}{(13,501)}     & 64,845 \textcolor{red}{(64,845)}   &  66,580 \textcolor{red}{(66,580)}  \\ \hline
    \textbf{Data Not Linked to You} & 853 \textcolor{red}{(757)}  & 14,358  \textcolor{red}{(13,638)}          & \textbf{3,513} & 8,829  \textcolor{red}{(8,829)}   & 13,881 \textcolor{red}{(13,881)}  \\ \hline
    \textbf{Data Not Collected}     & 8                             & 175      & 81            & \textbf{4,359}  & 0 \\ \hline
    \textbf{Not Mentioned}     & 2,409                             & 33,552      & 9,555            &  66,690  & \textbf{64,109} \\ \hline

    \multicolumn{6}{|c|}{\textbf{Contacts}} \\ \cmidrule{1-6} 
    \diagbox[width=8em]{\textbf{Policy}}{\textbf{Label}} & \textbf{Data Used to Track You} & \textbf{Data Linked to You} & \textbf{Data Not Linked to You} & \textbf{Data Not Collected} & \textbf{Not Mentioned} \\ \hline
    \textbf{Data Used to Track You} & \textbf{0}                   & 51 \textcolor{red}{(51)}     &   15 \textcolor{red}{(15)} &  0   & 1,484 \textcolor{red}{(1,484)} \\ \hline
    \textbf{Data Linked to You}     & 11  & \textbf{630} &     458 \textcolor{red}{(458)}     & 4740 \textcolor{red}{(4740)}   &  21,182 \textcolor{red}{(21,182)}  \\ \hline
    \textbf{Data Not Linked to You} & 2 \textcolor{red}{(1)}  & 59  \textcolor{red}{(59)}          & \textbf{32} & 490  \textcolor{red}{(490)}   & 1,960 \textcolor{red}{(1,960)}  \\ \hline
    \textbf{Data Not Collected}     & 0                             & 18      & 9            & \textbf{4,359}  & 0 \\ \hline
    \textbf{Not Mentioned}     & 206                             & 4,995      & 2,698            &  128,332  & \textbf{237,427} \\ \hline

    \multicolumn{6}{|c|}{\textbf{Diagnostics}} \\ \cmidrule{1-6} 
    \diagbox[width=8em]{\textbf{Policy}}{\textbf{Label}} & \textbf{Data Used to Track You} & \textbf{Data Linked to You} & \textbf{Data Not Linked to You} & \textbf{Data Not Collected} & \textbf{Not Mentioned} \\ \hline
    \textbf{Data Used to Track You} & \textbf{6,725}                   & 10,928 \textcolor{red}{(6,346)}      &   26,069 \textcolor{red}{(22,465)} &  10,863  \textcolor{red}{(10,863)}  & 9,759 \textcolor{red}{(9,759)} \\ \hline
    \textbf{Data Linked to You}     & 19,745 \textcolor{red}{(5,818)} & \textbf{43,487} &     81,505 \textcolor{red}{(73,692)}     & 69,912 \textcolor{red}{(69,912)}   &  51,865 \textcolor{red}{(51,865)}  \\ \hline
    \textbf{Data Not Linked to You} & 5,257 \textcolor{red}{(2,221)}  & 9,877  \textcolor{red}{(7,692)}          & \textbf{31,190} & 17,018  \textcolor{red}{(17,018)}   & 14,905 \textcolor{red}{(14,905)}  \\ \hline
    \textbf{Data Not Collected}     & 169                             & 210      & 392            & \textbf{4,359}  & 0 \\ \hline
    \textbf{Not Mentioned}     & 7,301                             & 17,903      & 42,567            &  58,800  & \textbf{32,052} \\ 
    \hline

    \multicolumn{6}{|c|}{\textbf{Financial Info}} \\ \cmidrule{1-6} 
    \diagbox[width=8em]{\textbf{Policy}}{\textbf{Label}} & \textbf{Data Used to Track You} & \textbf{Data Linked to You} & \textbf{Data Not Linked to You} & \textbf{Data Not Collected} & \textbf{Not Mentioned} \\ \hline
    \textbf{Data Used to Track You} & \textbf{27}                   & 3,477 \textcolor{red}{(3,451)}     &   205 \textcolor{red}{(204)} &  2,387 \textcolor{red}{(2,387)} & 6,276 \textcolor{red}{(6,276)} \\ \hline
    \textbf{Data Linked to You}     & 194 \textcolor{red}{(13)} & \textbf{18,879} &     2,467 \textcolor{red}{(2,036)}     & 25,044 \textcolor{red}{(25,044)}   &  65,940 \textcolor{red}{(65,940)}  \\ \hline
    \textbf{Data Not Linked to You} & 23 \textcolor{red}{(22)}  & 5,055  \textcolor{red}{(5,050)}          & \textbf{183} & 2,570  \textcolor{red}{(8,829)}   & 7,067 \textcolor{red}{(7,067)}  \\ \hline
    \textbf{Data Not Collected}     & 1                             & 27      & 6            & \textbf{4,359}  & 0 \\ \hline
    \textbf{Not Mentioned}     & 335                             & 9,313      & 2,321            &  107,565  & \textbf{167,701} \\ 
    % \hline
\pagebreak
    \multicolumn{6}{|c|}{\textbf{Health}} \\ \cmidrule{1-6} 
    \diagbox[width=8em]{\textbf{Policy}}{\textbf{Label}} & \textbf{Data Used to Track You} & \textbf{Data Linked to You} & \textbf{Data Not Linked to You} & \textbf{Data Not Collected} & \textbf{Not Mentioned} \\ \hline
    \textbf{Data Used to Track You} & \textbf{0}                   & 52 \textcolor{red}{(52)}     &   18 \textcolor{red}{(18)} &  30 \textcolor{red}{30} & 133 \textcolor{red}{(133)} \\ \hline
    \textbf{Data Linked to You}     & 11 \textcolor{red}{(1)} & \textbf{898} &     299 \textcolor{red}{(286}     & 867 \textcolor{red}{(867)}   &  1,506 \textcolor{red}{(1,506)}  \\ \hline
    \textbf{Data Not Linked to You} & 1 \textcolor{red}{(1)}  & 61  \textcolor{red}{(61)}          & \textbf{25} & 55  \textcolor{red}{(55)}   & 72 \textcolor{red}{(72)}  \\ \hline
    \textbf{Data Not Collected}     & 0                             & 12      & 6            & \textbf{4,359}  & 0 \\ \hline
    \textbf{Not Mentioned}     & 92                            & 6,326      & 1,991            &  132,420  & \textbf{256,974} \\ \hline

    \multicolumn{6}{|c|}{\textbf{Identifiers}} \\ \cmidrule{1-6} 
    \diagbox[width=8em]{\textbf{Policy}}{\textbf{Label}} & \textbf{Data Used to Track You} & \textbf{Data Linked to You} & \textbf{Data Not Linked to You} & \textbf{Data Not Collected} & \textbf{Not Mentioned} \\ \hline
    \textbf{Data Used to Track You} & \textbf{28,152}                   & 41,171 \textcolor{red}{(22,171)}     &   22,056 \textcolor{red}{(10,954)} &  25,828 \textcolor{red}{25,828} & 28,866 \textcolor{red}{(28,866)} \\ \hline
    \textbf{Data Linked to You}     & 55,763 \textcolor{red}{(18,694)} & \textbf{86,392} &     47,009 \textcolor{red}{(40,280}     & 81,688 \textcolor{red}{(81,688)}   &  65,600 \textcolor{red}{(65,600)}  \\ \hline
    \textbf{Data Not Linked to You} & 21,097 \textcolor{red}{(12,635)}  & 30,657  \textcolor{red}{(28,207)}          & \textbf{19,027} & 25,844  \textcolor{red}{(25,844)}   & 30,043 \textcolor{red}{(30,043)}  \\ \hline
    \textbf{Data Not Collected}     & 276                             & 302      & 249            & \textbf{4,359}  & 0 \\ \hline
    \textbf{Not Mentioned}     & 13,361                            & 29,169      & 16,815            &  47,889  & \textbf{22,411} \\ \hline

    \multicolumn{6}{|c|}{\textbf{Location}} \\ \cmidrule{1-6} 
    \diagbox[width=8em]{\textbf{Policy}}{\textbf{Label}} & \textbf{Data Used to Track You} & \textbf{Data Linked to You} & \textbf{Data Not Linked to You} & \textbf{Data Not Collected} & \textbf{Not Mentioned} \\ \hline
    \textbf{Data Used to Track You} & \textbf{4,741}                   & 6,218 \textcolor{red}{(2,961)}     &   4,274 \textcolor{red}{(2,764)} &  5,711 \textcolor{red}{5,711} & 13,664 \textcolor{red}{(13,644)} \\ \hline
    \textbf{Data Linked to You}     & 10,861 \textcolor{red}{(3,370)} & \textbf{27,248} &     20,812 \textcolor{red}{(20,420}     & 30,410 \textcolor{red}{(30,410)}   &  67,749 \textcolor{red}{(67,749)}  \\ \hline
    \textbf{Data Not Linked to You} & 3,860 \textcolor{red}{(2,696)}  & 5,572  \textcolor{red}{(5,492)}          & \textbf{4,650} & 6,962  \textcolor{red}{(6,962)}   & 14,344 \textcolor{red}{(14,344)}  \\ \hline
    \textbf{Data Not Collected}     & 178                             & 204      & 137            & \textbf{4,359}  & 0 \\ \hline
    \textbf{Not Mentioned}     & 15,047                           & 24,228      & 24,818            &  100,638  & \textbf{97,879} \\ \hline

    \multicolumn{6}{|c|}{\textbf{Purchases}} \\ \cmidrule{1-6} 
    \diagbox[width=8em]{\textbf{Policy}}{\textbf{Label}} & \textbf{Data Used to Track You} & \textbf{Data Linked to You} & \textbf{Data Not Linked to You} & \textbf{Data Not Collected} & \textbf{Not Mentioned} \\ \hline
    \textbf{Data Used to Track You} & \textbf{136}                   & 2,664 \textcolor{red}{(2,552)}     &   275 \textcolor{red}{(251)} &  1,194 \textcolor{red}{1,194} & 3,255 \textcolor{red}{(3,255)} \\ \hline
    \textbf{Data Linked to You}     & 1,183 \textcolor{red}{(255)} & \textbf{12,790} &     1,339 \textcolor{red}{(1,339}     & 8,726 \textcolor{red}{(8,726)}   &  22,120 \textcolor{red}{(22,120)}  \\ \hline
    \textbf{Data Not Linked to You} & 194 \textcolor{red}{(142)}  & 4,878  \textcolor{red}{(4,878)}          & \textbf{172} & 1,853  \textcolor{red}{(1,853)}   & 4,151 \textcolor{red}{(4,151)}  \\ \hline
    \textbf{Data Not Collected}     & 13                             & 35      & 34            & \textbf{4,359}  & 0 \\ \hline
    \textbf{Not Mentioned}     & 8,629                           & 24,081      & 7,601            &  123,794  & \textbf{197,783} \\ \hline

    \multicolumn{6}{|c|}{\textbf{Search History}} \\ \cmidrule{1-6} 
    \diagbox[width=8em]{\textbf{Policy}}{\textbf{Label}} & \textbf{Data Used to Track You} & \textbf{Data Linked to You} & \textbf{Data Not Linked to You} & \textbf{Data Not Collected} & \textbf{Not Mentioned} \\ \hline
    \textbf{Data Used to Track You} & \textbf{47}                   & 233 \textcolor{red}{(211)}     &   123 \textcolor{red}{(98)} &  1,118 \textcolor{red}{1,118} & 3,708 \textcolor{red}{(3,708)} \\ \hline
    \textbf{Data Linked to You}     & 154 \textcolor{red}{(41)} & \textbf{920} &     1,303 \textcolor{red}{(1,303}     & 7,889 \textcolor{red}{(7,889}   &  30,287 \textcolor{red}{(30,287)}  \\ \hline
    \textbf{Data Not Linked to You} & 61 \textcolor{red}{(35)}  & 110  \textcolor{red}{(110)}          & \textbf{873} & 1,926  \textcolor{red}{(1,926)}   & 7,679 \textcolor{red}{(7,679)}  \\ \hline
    \textbf{Data Not Collected}     & 1                             & 6      & 8            & \textbf{4,359}  & 0 \\ \hline
    \textbf{Not Mentioned}     & 992                           & 4,253      & 4,600            &  124,521  & \textbf{222,281} \\ 
    % \hline

\pagebreak

    \multicolumn{6}{|c|}{\textbf{Sensitive Info}} \\ \cmidrule{1-6} 
    \diagbox[width=8em]{\textbf{Policy}}{\textbf{Label}} & \textbf{Data Used to Track You} & \textbf{Data Linked to You} & \textbf{Data Not Linked to You} & \textbf{Data Not Collected} & \textbf{Not Mentioned} \\ \hline
    \textbf{Data Used to Track You} & \textbf{32}                   & 335 \textcolor{red}{(303)}     &   75 \textcolor{red}{(75)} &  4,752 \textcolor{red}{4,752} & 17,912 \textcolor{red}{(17,912)} \\ \hline
    \textbf{Data Linked to You}     & 78 \textcolor{red}{(5)} & \textbf{2,144} &     480 \textcolor{red}{(480}     & 21,567 \textcolor{red}{(21,567}   &  74,790 \textcolor{red}{(74,790)}  \\ \hline
    \textbf{Data Not Linked to You} & 25 \textcolor{red}{(25)}  & 358  \textcolor{red}{(358)}          & \textbf{109} & 6,318  \textcolor{red}{(6,318)}   & 26,134 \textcolor{red}{(26,134)}  \\ \hline
    \textbf{Data Not Collected}     & 0                             & 8      & 5            & \textbf{4,359}  & 0 \\ \hline
    \textbf{Not Mentioned}     & 86                           & 3,123      & 530            &  108,606  & \textbf{177,261} \\ \hline

    \multicolumn{6}{|c|}{\textbf{Usage Data}} \\ \cmidrule{1-6} 
    \diagbox[width=8em]{\textbf{Policy}}{\textbf{Label}} & \textbf{Data Used to Track You} & \textbf{Data Linked to You} & \textbf{Data Not Linked to You} & \textbf{Data Not Collected} & \textbf{Not Mentioned} \\ \hline
    \textbf{Data Used to Track You} & \textbf{23,563}                   & 28,307 \textcolor{red}{(14,402)}     &   38,829 \textcolor{red}{(26,818)} &  23,394 \textcolor{red}{23,394} & 18,611 \textcolor{red}{(18,611)} \\ \hline
    \textbf{Data Linked to You}     & 48,575 \textcolor{red}{(20,829)} & \textbf{63,514} &     80,399 \textcolor{red}{(74,937}     & 87,767 \textcolor{red}{(87,767}   &  59,425 \textcolor{red}{(59,425)}  \\ \hline
    \textbf{Data Not Linked to You} & 21,458 \textcolor{red}{(9,950)}  & 31,074  \textcolor{red}{(28,120)}          & \textbf{44,701} & 32,792  \textcolor{red}{(32,792)}   & 25,493 \textcolor{red}{(25,493)}  \\ \hline
    \textbf{Data Not Collected}     & 307                             & 254      & 410            & \textbf{4,359}  & 0 \\ \hline
    \textbf{Not Mentioned}     & 9,930                           & 14,700      & 19,874            &  39,103  & \textbf{24,349} \\ \hline

    \multicolumn{6}{|c|}{\textbf{User Content}} \\ \cmidrule{1-6} 
    \diagbox[width=8em]{\textbf{Policy}}{\textbf{Label}} & \textbf{Data Used to Track You} & \textbf{Data Linked to You} & \textbf{Data Not Linked to You} & \textbf{Data Not Collected} & \textbf{Not Mentioned} \\ \hline
    \textbf{Data Used to Track You} & \textbf{226}                   & 1,632 \textcolor{red}{(1,456)}     &   1,089 \textcolor{red}{(1,031)} &  2,000 \textcolor{red}{2,000} & 4,915 \textcolor{red}{(4,915)} \\ \hline
    \textbf{Data Linked to You}     & 1,266 \textcolor{red}{(364)} & \textbf{22,550} &     8,669 \textcolor{red}{(6,556}     & 20,713 \textcolor{red}{(20,713}   &  41,193 \textcolor{red}{(41,193)}  \\ \hline
    \textbf{Data Not Linked to You} & 186 \textcolor{red}{(116)}  & 1,746  \textcolor{red}{(1,634)}          & \textbf{980} & 2,824  \textcolor{red}{(2,824)}   & 7,265 \textcolor{red}{(7,265)}  \\ \hline
    \textbf{Data Not Collected}     & 7                             & 75      & 72            & \textbf{4,359}  & 0 \\ \hline
    \textbf{Not Mentioned}     & 2,177                           & 34,233      & 18,898            &  111,707  & \textbf{144,204} \\ 
    % \hline
    \bottomrule
    % \end{tabular}}
    % \end{table*}
    \end{longtable}

\twocolumn
}

\newcommand{\manualVerificationTable}[0]{
    % For Extended Version
    % \usepackage{booktabs}
    % \usepackage[normalem]{ulem}
    \begin{table}[t]
    \centering
    \caption{Manual Verification of Mapping}
    \label{tab:mapping-verification}
    \resizebox{0.6\columnwidth}{!}{
    \begin{tabular}{@{}lr@{}}
    \toprule
    \textbf{Attribute} & \textbf{\begin{tabular}[c]{@{}r@{}}\# Accurate\\ (25 samples)\end{tabular}} \\ \midrule
    {\underline{\textbf{Privacy Types:}}} &    \\
    \rowcolor[HTML]{EFEFEF}
    Data Used to Track You        & 22  \\
    Data Linked to You            & 23 \\
    \rowcolor[HTML]{EFEFEF}
    Data Not Linked to You        & 21  \\
    Data Not Collected            & 25   \\
    {\underline{\textbf{Purposes:}}} &    \\
    \rowcolor[HTML]{EFEFEF}
    App Functionality             &  23  \\
    Analytics                       & 25  \\
    \rowcolor[HTML]{EFEFEF}
    Product Personalization            & 25   \\
    Third-party Advertising            &  24  \\
    \rowcolor[HTML]{EFEFEF}
    Developers Advertising            &  25  \\
    Other Purposes            &  25  \\
    {\underline{\textbf{Data Categories:}}} &    \\
    \rowcolor[HTML]{EFEFEF}
    Browsing History            &  24    \\
    Contact Info            &   25   \\
    \rowcolor[HTML]{EFEFEF}
    Financial Info            &   25   \\
    Health \& Fitness            &  25    \\
    \rowcolor[HTML]{EFEFEF}
    Identifiers            &  25    \\
    Location            &  25    \\
    \rowcolor[HTML]{EFEFEF}
    Usage Data            &  25    \\
    User Content            &  24    \\
    \rowcolor[HTML]{EFEFEF}
    Diagnostics            &   25   \\
    Contacts            &   25   \\
    \rowcolor[HTML]{EFEFEF}
    Purchases            &    25  \\
    Search History            &  24    \\
    \rowcolor[HTML]{EFEFEF}
    Sensitive Info            &  25    \\ \bottomrule
    \end{tabular}
    }
    \end{table}

}
%%%%%%%%%%%% Privacy Labels Anatomy %%%%%%%%%%%%
\newcommand{\privacyLabelsAnatomy}[0]{
\begin{figure}[t]
    \centering
    \includegraphics[width=0.8\columnwidth,keepaspectratio, page=1]{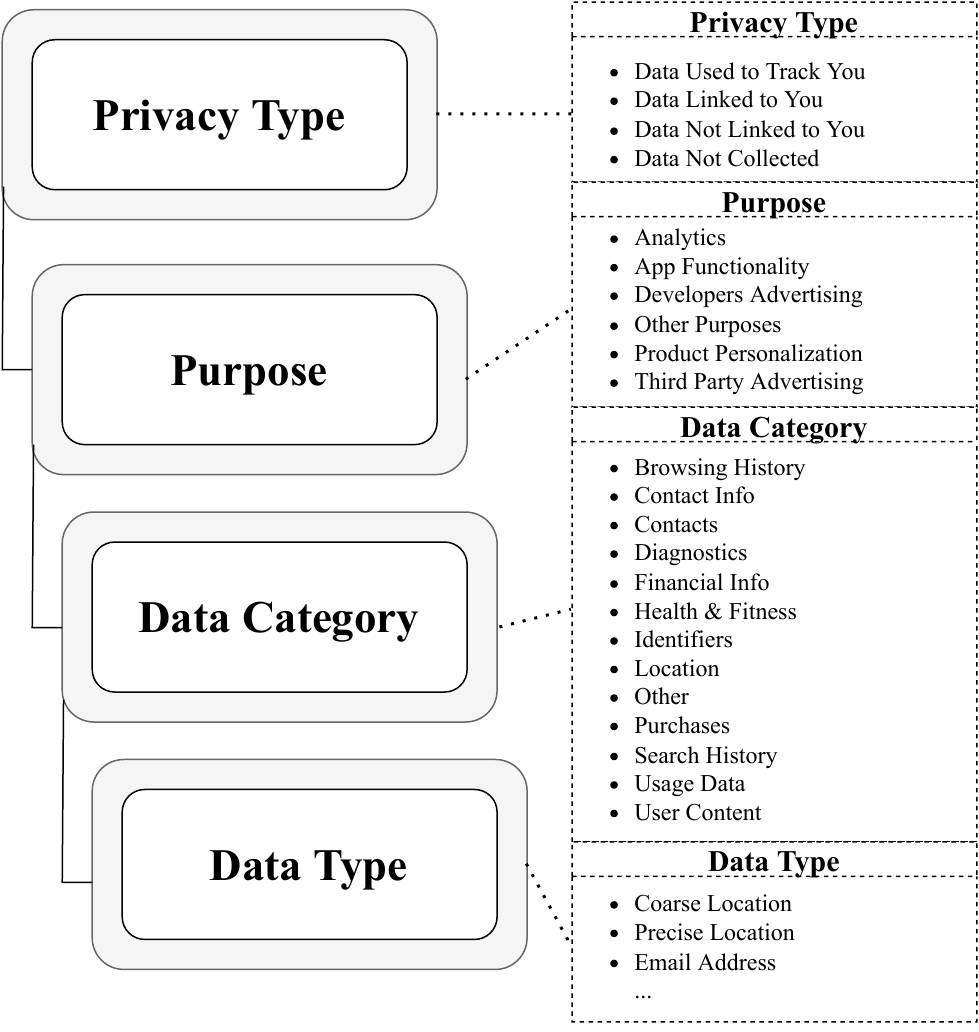}
    \vspace{-2ex}
    \caption{Anatomy of a Privacy Label.}
    \vspace{-3ex}
    \label{fig:privacy-labels-anatomy}
\end{figure}
}

%%%%%%%%%%%% Polisis Classifier Structure %%%%%%%%%%%%
\newcommand{\polisisStructure}[0]{
    \begin{figure}[t]
        \centering
        \includegraphics[width=\columnwidth,keepaspectratio, page=1]{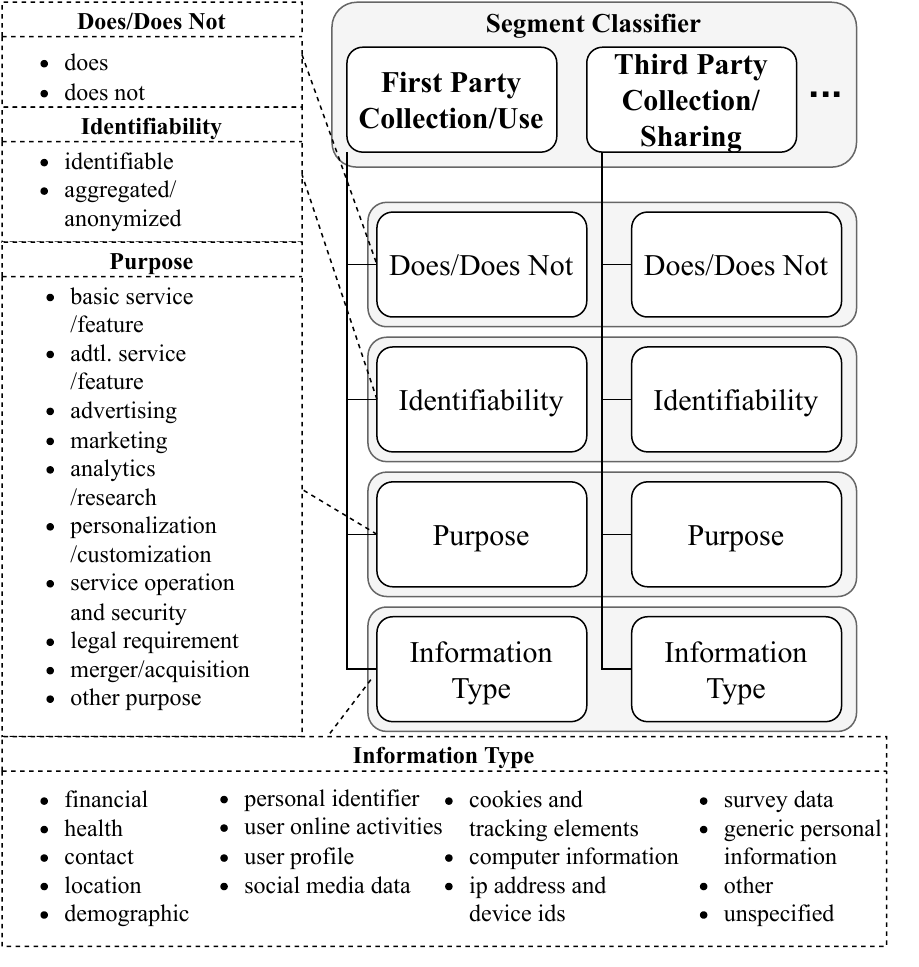}
        \caption{The hierarchical structure of the Polisis classifiers.}
        \label{fig:polisis-structure}
        \vspace{-3ex}
    \end{figure}
}
%%%%%%%%%%%% Methodology Figure %%%%%%%%%%%%
\newcommand{\methodologyFigure}[0]{
    \begin{figure}[t]
        \centering
        \includegraphics[width=0.7\columnwidth,keepaspectratio, page=1]{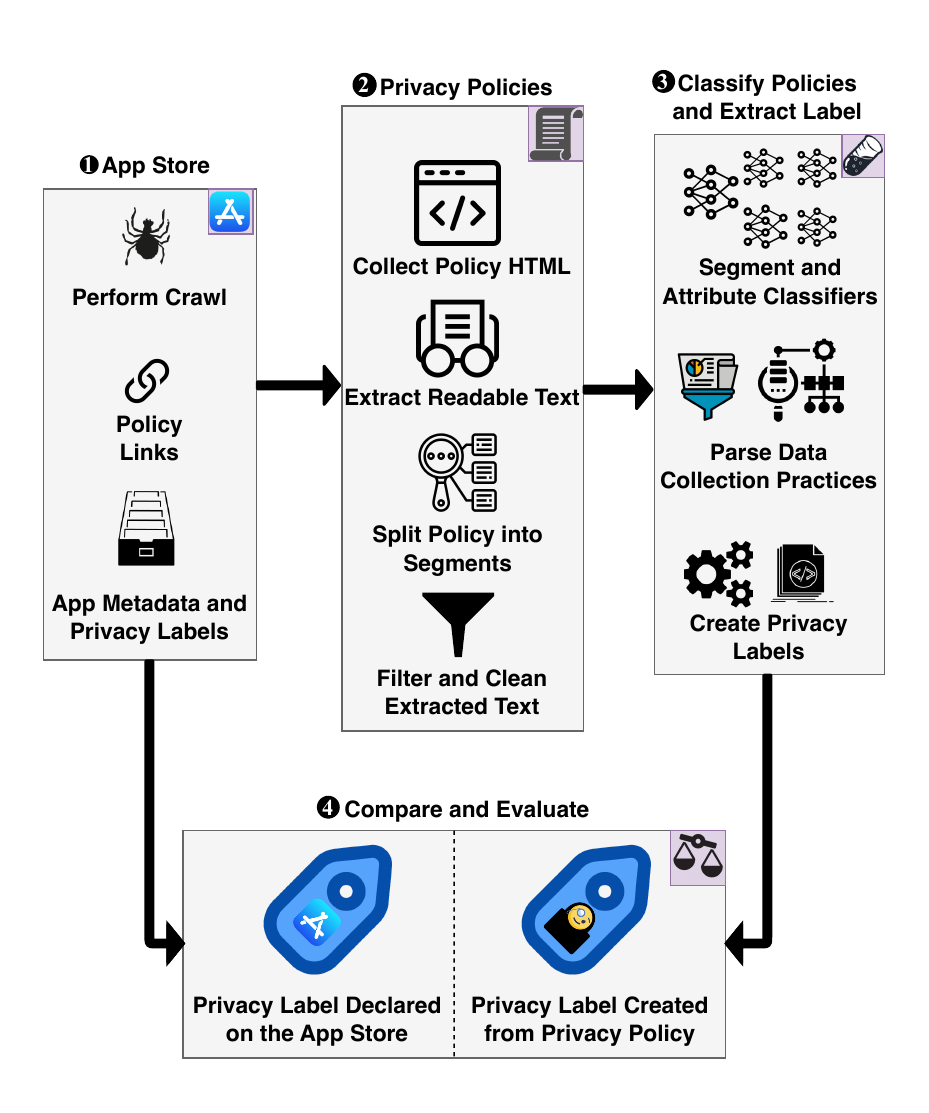}
        % \vspace{-1ex}
        \caption{An overview of the measurement workflow.}
        \vspace{-3ex}
        \label{fig:overview}
    \end{figure}
}

%%%%%%%%%%%% Privacy Types %%%%%%%%%%%%
\newcommand{\privacyTypes}[0]{
\begin{figure}[t]
    \centering
    \resizebox{0.8\linewidth}{!}{    \input{figures/privacy-types.pgf}}
    \vspace{-3ex}
    \caption{An overview of apps declaring data collection with corresponding \textit{Privacy Types} within their privacy policies (top) and on the App Store via privacy labels (bottom). The denominator is the total apps that we analyzed, i.e., 474,669 apps. Please note that the privacy types, except for \textit{Data Not Collected}, are \textit{not} mutually exclusive.}
    \label{fig:privacy-types}
    \vspace{-3ex}
\end{figure}
}

%%%%%%%%%%%% Purposes %%%%%%%%%%%%
\newcommand{\privacyTypesvsPurposesBar}[0]{
\begin{figure*}[t]
\centering
\input{figures/pt-vs-purpose.pgf}
\vspace{-3ex}
\caption[Privacy Types vs. Purposes]{The ratios of the six purposes for the \textit{Data Linked to You} and \textit{Data Not Linked to You} privacy types. The denominator is the number of apps with the designated privacy type either in their privacy label or their privacy policy, i.e., 419,504 apps with a \textit{Data Linked to You} label and 294,391 with a \textit{Data Not Linked to You} label. It is helpful to note here that privacy types shown here are \textit{not} mutually exclusive. Two other \textit{Privacy Types} are not shown here; the \textit{Data Used to Track You} privacy type refers to collection for the purpose of tracking, while the \textit{Data Not Collected} refers to the absence of \textit{any} data collection.\label{fig:pt-vs-purpose}}
\vspace{-2ex}
\end{figure*}
}

%%%%%%%%%%%% Data Categories %%%%%%%%%%%%
\newcommand{\privacyTypesvsDataCategoriesBar}[0]{
\begin{figure*}[t]
\centering
\resizebox{0.8\linewidth}{!}{\input{figures/data-categories-by-pt.pgf}}
\vspace{-2ex}
\caption[Privacy Types vs. Data Categories]{The ratios of data categories against privacy types. The denominator is the number of apps with the designated privacy type either in their privacy label or their privacy policy, i.e., 232,648 apps with \textit{Data Used to Track You}, 419,504 apps with \textit{Data Linked to You}, and 294,391 apps with \textit{Data Not Linked to You}. The three privacy types shown here are \textit{not} mutually exclusive.\label{fig:pt-vs-data-categories}}
\vspace{-1ex}
\end{figure*}
}

%%%%%%%%%%%% Apps that claim DNC: Purposes %%%%%%%%%%%%
\newcommand{\dncPrivacyTypesvsPurposesBar}[0]{
\begin{figure}[t]
\centering
\input{figures/pt-vs-purpose-dnc.pgf}
\caption[Data Not Collected -- Privacy Types vs. Purposes]{The ratios of the six \textit{Purposes} for the \textit{Data Linked to You} and \textit{Data Not Linked to You} Privacy Types, specific to apps that state on the App Store that they do not collect any data. The denominator is the total number of apps with a \textit{Data Not Collected} label on the App Store, i.e., 205,274 apps.\label{fig:dnc-pt-vs-purpose}}
\end{figure}
}

%%%%%%%%%%%% Apps that claim DNC: Data Categories %%%%%%%%%%%%
\newcommand{\dncPrivacyTypesvsDataCategoriesBar}[0]{
\begin{figure}[t]
\centering
\input{figures/data-categories-dnc.pgf}
\caption[Data Not Collected -- Privacy Types vs. Data Categories]{The ratios of the eight \textit{Data Categories} that were part of the \textit{Primary Conversion}, against the \textit{Data Linked to You} and \textit{Data Not Linked to You} Privacy Types, specific to apps that state on the App Store that they do not collect any data. The denominator is the total number of apps with a \textit{Data Not Collected} label on the App Store, i.e., 205,274 apps.\label{fig:dnc-pt-vs-data-categories}}
\end{figure}
}

%%%%%%%%%%%% DUTY: Data Categories %%%%%%%%%%%%
\newcommand{\dutyPrivacyTypesvsDataCategoriesBar}[0]{
\begin{figure}[t]
\centering
\input{figures/data-categories-duty.pgf}
\caption[Data Used to Track You -- Privacy Types vs. Data Categories]{The ratios of the eight \textit{Data Categories} that were part of the \textit{Inferential Conversion}, against the \textit{Data Used to Track You} Privacy Type. The denominator is the total number of apps with a \textit{Data Used to Track You} label, both, on the App Store and inferred from Polisis, i.e., 231,155 apps.\label{fig:duty-pt-vs-data-categories}}
\end{figure}
}

%%%%%%%%%%%% App Cost By Privacy Label %%%%%%%%%%%%
\newcommand{\appCostsByPrivacyLabel}[0]{
\begin{figure*}[ht]
\centering
\resizebox{0.7\textwidth}{!}{\input{figures/app_costs_by_privacy_label_type.pgf}}
\vspace{-3ex}
\caption[App Costs By Privacy Label Type Ratios]{The ratios of app costs for each of the four privacy types. The denominator is the number of apps with the designated app cost that have a privacy label. Free apps are more likely than paid apps to collect data, including data used to track and linked to users. Please note that privacy types shown here are \textit{not} mutually exclusive.\label{fig:app-costs-by-privacy}}
\vspace{-2ex}
\end{figure*}
}

%%%%%%%%%%%% App Size By Privacy Label %%%%%%%%%%%% 
\newcommand{\appSizeByPrivacyLabel}[0]{
\begin{figure*}[ht]
\centering
\input{figures/app_size_by_privacy_label_type.pgf}
\caption[App Size By Privacy Label Type Ratios]{The ratios of app sizes for each of the four \emph{Privacy Types}. The denominator is the number of apps with the designated app size that have a privacy label. Apps that are larger in size are more likely to collect data, including data used to track and linked to users. }
\end{figure*}\label{fig:app-size-by-privacy}
}

%%%%%%%%%%%% Content Rating By Privacy Label %%%%%%%%%%%%
\newcommand{\contentRatingByPrivacyLabel}[0]{
\begin{figure*}[t]
\centering
\resizebox{0.7\textwidth}{!}{\input{figures/content_rating_by_privacy_label_type.pgf}}
\caption[Content Rating By Privacy Label Type Ratios]{The ratios of content ratings for each of the four \emph{Privacy Types}. The denominator is the number of apps with the designated content rating that have a privacy label.\label{fig:content-rating-by-privacy}}
\end{figure*}
}

%%%%%%%%%%%% Ratings Count By Privacy Label %%%%%%%%%%%%
\newcommand{\ratingsCountByPrivacyLabel}[0]{
\begin{figure*}[t]
\centering
\input{figures/rating_counts_by_privacy_label_type.pgf}
\caption[Rating Counts By Privacy Label Type Ratios]{The ratios of the rating counts for each of the four \emph{Privacy Types}. The denominator is the number of apps with the designated rating counts that have a privacy label. Apps with a larger number of user ratings are more likely to collect data, including data used to track users. Ratings counts are not localized metadata and apps with low ratings counts in the US region may have higher counts elsewhere.\label{fig:ratings-counts-by-privacy}}
\end{figure*}
}

%%%%%%%%%%%% Release Date By Privacy Label %%%%%%%%%%%%
\newcommand{\releaseDateByPrivacyLabel}[0]{
\begin{figure*}[t]
\centering
\input{figures/release_date_by_privacy_label_type.pgf}
\caption[Release Date Year By Privacy Label Type Ratios]{The ratios of the release date year for each of the four \emph{Privacy Types}. The denominator is the number of apps with the designated release date year that have a privacy label. \label{fig:release-date-by-privacy}}
\end{figure*}
}

%%%%%%%%%%%% Version Date By Privacy Label %%%%%%%%%%%%
\newcommand{\versionDateByPrivacyLabel}[0]{
\begin{figure*}[t]
\centering
\input{figures/version_date_by_privacy_label_type.pgf}
\caption[Version Date Year By Privacy Label Type Ratios]{The ratios of the app latest version update year for each of the four \emph{Privacy Types}. The denominator is the number of apps with the designated version update year that have a privacy label. \label{fig:version-date-by-privacy}}
\end{figure*}
}

%%%%%%%%%%%% App Genre By Privacy Label %%%%%%%%%%%%
\newcommand{\appStoreGenreByPrivacyType}[0]{
\begin{figure*}[ht]
\centering
\resizebox{!}{0.95\textheight}{\input{figures/app_genre_by_privacy_label_type.pgf}}
\caption[App Genre By Privacy Label Type Ratios]{The ratios of app store genres for each of the four \emph{Privacy Types}. The denominator is the number of apps with the designated app store genre that have a privacy label.}\label{fig:app-genre-by-privacy} 
\end{figure*}
}

%%%%%%%%%%%% App Genre By Privacy Label %%%%%%%%%%%%
\newcommand{\appStoreGenreTopFiveByPrivacyType}[0]{
\begin{figure*}[ht]
\centering
\input{figures/app_genre_top5_by_privacy_label_type.pgf}
\caption[App Genre Top5 By Privacy Label Type Ratios]{The ratios of app store top five genres for each of the four \emph{Privacy Types}. The denominator is the number of apps with the designated app store genre that have a privacy label.}\label{fig:app-genre-top5-by-privacy} 
\end{figure*}
}

%%%%%%%%%%%% Children: Ratings Count By Privacy Label %%%%%%%%%%%%
\newcommand{\childrenRatingsCountByPrivacyLabel}[0]{
\begin{figure*}[t]
\centering
\resizebox{0.8\linewidth}{!}{\input{figures/children.pgf}}
\vspace{-3ex}
\caption[Children and User Content Rating By Privacy Label Type Ratios]{The ratios of the content ratings for each of the four privacy types, with an overlay (white bar) indicating the ratio of apps that also include a segment in their privacy policy, where they address privacy practices specific to children who engage with their services. The denominator is the number of apps with the designated content rating that have a privacy label. Please note that privacy types shown here are \textit{not} mutually exclusive. \label{fig:children}}
\vspace{-2ex}
\end{figure*}
}

%%%%%%%%%%%% Data Security %%%%%%%%%%%%
\newcommand{\dataSecurity}[0]{
\begin{figure}[ht]
    \centering
    \input{figures/security.pgf}
    \vspace{-7ex}
    \caption{The ratios of \textit{Privacy Types} associated with data collection on the App Store with an overlay (white bar) indicating the ratio of apps that address secure data storage/transfer practices in their privacy policies. The denominator is the total number of apps analyzed, i.e., 552,495.}
    \vspace{-2ex}
    \label{fig:security}
\end{figure}
}

%%%%%%%%%%%% Privacy Types %%%%%%%%%%%%
\newcommand{\templatePrivacyTypes}[0]{
\begin{figure}[t]
    \centering
    \resizebox{0.8\linewidth}{!}{\input{figures/templates-privacy-types.pgf}}
    \vspace{-3ex}
    \caption{An overview of the privacy types associated with data collection on the App Store, from privacy labels and privacy policies, specific to apps whose policies are similar to templates. The denominator is the total number of such apps, i.e., 300,535 apps. Please note that the privacy types, except for \textit{Data Not Collected}, are \textit{not} mutually exclusive.}
    \label{fig:templates-privacy-types}
    \vspace{-2ex}
\end{figure}
}

\section{Introduction}
Privacy policies are ubiquitous and required in many settings~\cite{FTC:FIPP:2000,FTC:Safeguards:2022,FTC:COPPA:2013,CCPA:2018}, and for better or worse, are an important tool for communicating about the behavior of systems. Natural language policies have many shortcomings and are full of technical details and jargon that significantly impact their usability as a tool to inform users clearly about the behaviors and data management practices~\cite{McDonald:2009,Cranor:2014}. %Even when users read privacy policies, they fail to completely understand the details, the impact on their privacy, and any appropriate countermeasures they should take.
\emph{Privacy nutrition labels}, or \emph{privacy labels}, offer an alternative to both simplify and standardize the communication of privacy behavior similar to food nutrition labels~\cite{Kelly:2009,nutrition-how-2022}.
% One proposal to improve privacy communication is to do away with natural language presentations of privacy behavior and instead use \emph{privacy nutrition labels}~\cite{Kelly:2009} or, more simply, \emph{privacy labels}. These are prescriptive labeling of applications, like food nutrition labels~\cite{nutrition-how-2022}, describing specific behaviors surrounding data collection and security. The key idea is that labels provide more clarity and transparency that is difficult to achieve via the privacy policy. 
In December 2020, Apple began requiring privacy labels~\cite{Apple:LabelsLive} for all new and updated apps in the App Store.  Apple's privacy labels ask developers to self-label (without verification) the data collection and sharing practices of their apps, the purposes, the types of data, and if that data is linked to user identities (see \autoref{fig:privacy-labels-anatomy} for more details). Essentially, privacy labels standardize the presentation of privacy behavior described in the privacy policy's natural language text.

In this paper, we answer the question: {\em How do privacy labels compare to the behavior described in the privacy policies?} 

We conducted a large-scale analysis of the Apple App Store by reviewing 474,669 apps' privacy policies and privacy labels using a validated implementation of PrivBERT~\cite{srinath-privbert-2021}, a transformer-based privacy policy language model. We fine-tuned PrivBERT with the OPP-115 corpus and mapped its features to Apple's privacy labels to identify discrepancies between the reported behavior of apps based on their labels compared to their privacy policies.

We find that there are large differences between privacy labels and privacy policies. Most prominently, according to our analysis of the privacy policies, nearly 228K {\em more} apps may be performing some amount of data linking than the number of apps that reported similar data collection in the labels. More alarming, 97\% of apps that report no data collection in their privacy label have statements in their privacy policy to the contrary. In many cases, mislabeling varies from the privacy policy regarding the kinds of data collected, particularly around app functionality and analytics or ``other'' functionality not prescribed by a privacy label.

We also compared free and paid apps. While paid apps use fewer privacy labels compared to free apps, the policies tell a different story: only 4\% of paid apps report collecting data that is linked to users, but the policies suggest that 76\%  paid apps perform such collection.
%Similarly, only 4\% collect data linked to users according to their labels, but the true number may be closer to 84\% according to their policies. 
%When observing the policies, the behavior of free and paid apps appear more in line with prior work~\cite{Han:2019,Han:2020}, in that there are limited differences in privacy behavior between these monetization schemes. 
%
We further analyzed privacy-relevant data practices that are not covered by privacy labels. We found that most apps (76\%) had a self-assigned content rating of 4+ on the App Store to indicate age appropriateness and enforce parental controls. Of these apps, only 50\% of such apps had a policy in place to handle data collected from children. Our case study further reveals that their policy might be to claim no responsibility for collecting and handling data collected from users under 13 years of age.
We also employ a similarity metric and identify that 65\% of evaluated apps potentially use templates, providing insight into a possible source of discrepancies. We further analyzed the network traffic from 30 apps, showing that their data collection practices diverged from those declared in privacy labels and privacy policies.

% \adam{maybe add a paragraph about children/content rating}

%According to the privacy policy 28\% of free apps are performing data tracking as compared to only 16\% reporting so in their labels. The difference is even greater for paid apps (without in-app purchases), where only 2\% report such behavior in their labels but 16\% indicate some form of data tracking in their privacy policy. %The truth of data tracking may be much worse. 

Our analysis indicates that privacy labels are likely misapplied in great numbers, even considering that classifiers are imperfect for analyzing privacy policies. 
% When reviewing specific examples, errors appear to emerge from benign misunderstandings. For example, many developers may mistake some actions as common business practices, such as collecting app performance and analytics, that are actually prescribed privacy labels. 
%as well as when this data could be associated with unique identifiers and thus "linked" to users. 
More guidance for developers would go a long way toward improving the accuracy of privacy labels. Still, there are also more concerning misapplications that could and should be addressed more broadly, such as the collection of data used to track users and apps falsely reporting that they do not collect any data. In these cases, the privacy policies are often explicit in this behavior, and the absence of a corresponding entry in the privacy label could lead to misunderstandings of the risks associated with using these apps and potentially violate Apple's App Store policies. First-level checks of the privacy policies when apps are submitted to the App Store could go a long way in highlighting and correcting some of the more common and egregious privacy label inaccuracies. In this work, we make the following contributions.
\begin{itemize}[leftmargin=*,nosep,noitemsep]
\item We build and validate a hierarchical framework that uses fine-tuned transformer models to extract multiple features from privacy policies.
\item We develop and validate a mapping between features extracted from classifiers and App Store privacy labels.
\item We collect and analyze the privacy labels of 474,669 apps against their policies and find large differences in their reported practices.
\item We use a similarity metric to compare policies against templates and find that their use might indicate a likely source of observed discrepancies. We also present examples from a case study of traffic collected 30 apps and show evidence of discrepancies.
\item We publicly release our code and dataset of app metadata and privacy policies to facilitate further research. The artifact associated with this paper can be accessed at \textcolor{blue}{\texttt{\urlstyle{tt}\url{https://github.com/masood/2024-pets-privacy-labels-policies}\urlstyle{rm}}}.
\end{itemize}

\section{Background and Related Work}

\privacyLabelsAnatomy

\paragraph{\textbf{Anatomy of a privacy label.}}
The Apple privacy labels are similar in style and content to the ``Privacy Facts'' label developed by Kelley et al. \cite{kelley-privacy-2013}.
The structure of a label is hierarchical (see
\autoref{fig:privacy-labels-anatomy}) and describes data collection practices under four levels:
\begin{enumerate*}[label=\textbf{(\arabic*)}]
    \item \textbf{Privacy Type:} Describes how the app handles collected data, which includes data collected for tracking users (with third parties), data collected and linked to users' identities, and data collected but aggregated/anonymized. An app’s privacy label may contain a combination of one, two, or all three types. An app may also report that data is not collected, which is mutually exclusive with the other types.
    \item \textbf{Purpose:} Discloses the intended reason for the data collection, e.g., for advertising, analytics, personalization.
    \item \textbf{Data Category:} Reports at a high level the category under which collected data falls.
    \item \textbf{Data Type:} Granular information that describes the data collected under the Data Category.
\end{enumerate*}

\paragraph{\textbf{Privacy nutrition labels.}} 
Privacy nutrition labels have been studied  from a variety of perspectives~\cite{FTC:2011,Cranor:2012,Balebako:2015,Kelly:2009,kelley-standardizing-2010,kelley-privacy-2013,emami-naeini-ask-2020,emami-naeini-which-2021,Schaub:2015}, but Apple's privacy label is the first wide-scale deployment~\cite{Apple:LabelsLive}. In an exploratory study, Li et al.~\cite{li-chi} observed the adoption of iOS privacy labels on the App Store and found that very few developers \textit{voluntarily} created privacy labels. Balash et al.~\cite{Balash:2022} performed a 66-week analysis of the privacy label adoption on the Apple App Store and 
identified a steady increase in the number of apps with privacy labels and likely under-reporting by developers forced to provide a label on a version update. 

Zhang et al.~\cite{Zhang:2022} conducted an in-depth interview study 
to determine the usability of iOS privacy labels from a user perspective. Most users found the privacy labels helpful 
despite misunderstandings that included unfamiliar terms and a confusing structure.
Garg et al.~\cite{Garg:2022} 
discovered that privacy label disclosures of sensitive information reduce app demand, and thus, the accuracy of the labels is important to help users make informed choices. 

Gardner et al. \cite{Gardner:2022} developed a tool to assist developers by prompting them while coding functionality that would potentially require a privacy label. 
Li et al.~\cite{li-understanding-2022} studied developers' creation of Apple's privacy nutrition labels and conducted semi-structured interviews. They found that errors and misunderstandings were prevalent in the privacy label generation process. These errors included under-reporting linked data, third-party data use, and missing data types. We observe the same when comparing the privacy policies and Li et al.'s findings regarding ``knowledge blindspots'' and misinterpreted Apple's definitions, likely leading to many of the misapplications we identified. 

\paragraph{\textbf{Privacy behavior of mobile apps.}} 
Numerous studies have  measured the privacy behaviors of mobile applications ~\cite{Andow:2019, Andow:2020, Breaux:2013, Bui:2021, Chen:2019, Slavin:2016, Yu:2016, Zimmeck:2019, Zimmeck:2017,Collins:2022}. One of the first approaches to automatically identify problems in privacy policies was PPChecker~\cite{Yu:2016}, which combined an NLP analysis of privacy policy text with bytecode analysis.
Andow et al.~\cite{Andow:2019} developed PolicyLint to identify contradictions within an individual policy.
Andow et al.~\cite{Andow:2020} also created PoliCheck, which considers third-party versus first-party entity access to personal data for an entity-sensitive consistency check.
Bui et al.~\cite{Bui:2021} extended PoliCheck to develop PurPliance that checks if data, entity, and purpose are equivalent to those extracted from data flows. In this paper, we choose Polisis~\cite{Harkous:2018} as the policy analysis tool as it produces output similar to the privacy labels. 

Zimmeck et al.~\cite{Zimmeck:2019} evaluated 1,035,853 Android apps using the Mobile App Privacy System (MAPS), a pipeline based on code analysis and supervised machine learning classifiers, to identify potential non-compliance with privacy standards.
Kollnig et al.~\cite{kollnig-goodbye-2022} analyzed 1,759 iOS apps using a combination of code analysis and network traffic monitoring, and they found that 80\% of the apps that claimed not to collect any data in the privacy labels contained at least one tracker library. We find that this discrepancy probably exists at scale.

Xiao et al.~\cite{Xiao:2022} analyzed 5,102 apps ($\sim$ 1\% of our dataset) by checking the privacy labels against actual data flows and focused on two levels of labels, \textit{Purposes} and \textit{Data Types}. They discovered that 67\% of those apps failed to accurately disclose their data collection practices, particularly around the use of \textit{User ID}, \textit{Device ID}, and \textit{Location} data. Our results complement their findings, where mentioning the collection of unique identifiers in an identifiable manner in the privacy policy is not reflected in the privacy labels. Further, our work analyzes apps at a much larger scale and covers \textit{Privacy Types}, \textit{Purposes}, and \textit{Data Categories}.

\paragraph{\textbf{Apple's deviations from recommendations.}} Although derived from Kelley et al.'s~\cite{kelley-privacy-2013} work, Apple's implementation deviates from its recommendations. While Kelley et al. noted, ``presenting \textit{[labels]} clearly and simply we could affect user decisions,'' Apple displays the nutrition label embedded down on the App Store, requiring interested users to scroll through details, where users may not see the labels before deciding to install an app. Additionally, Apple's labels do not give users choices or allow them to compare labels between apps. Further, recent user studies have found the labels to be confusing for  developers~\cite{li-understanding-2022}, showing the possibility that developers misapply labels. Finally, Kelley et al. highlighted the need for the labels to be accurate and noted, ``users believe this information is correct, is being verified, and will assume they misunderstand
something before they would believe the displays are incorrect.'' Since Apple's privacy labels are not vetted and are not trustworthy, this points to a serious concern about providing disinformation to end users. These factors further highlight the necessity to verify and demonstrate the discrepancies we present.

\section{Measurement Workflow}
\methodologyFigure
\polisisStructure

In \autoref{fig:overview}, we present the primary measurement workflow described in detail below.  
During all scans, we followed best practices of limiting the number of requests and respecting \texttt{403 Errors} by using exponential back-offs. 

\paragraph{\circlea{1} \textbf{Crawling the App Store.}}
We began by parsing the \texttt{XML} site map from Apple's App Store, which lists all apps currently published on the store, and then crawled each URL, parsing the privacy labels and associated metadata, such as the app name, version, size, type, user rating, genre, content rating, release date, seller name, and price. Notably, the metadata includes a link to the privacy policy. We also parsed the extended privacy label details, such as the purposes and data types, by performing an additional GET request to the Apple Catalog API~\cite{AppleCatalog}. In January 2024, there were 1.2M apps on the App Store. Of them, 995K apps had a privacy label, and we identified 993K apps with links to 669K unique policies (note that some apps link to the same policy).

\paragraph{\circlea{2} \textbf{Collecting Privacy Policies.}}
We extracted the HTML for each policy using a Python script. 
We leveraged the \textit{readability} library~\cite{Readability:2020,ReadabiliPy:2018}, a standalone version of the Firefox browser \textit{reader mode}. The library employs a complex set of heuristics to extract relevant text from web pages~\cite{FirefoxReadability:2022}, leaving us with de-cluttered HTML that we divided into segments based on the \texttt{<p>} tags. 
We then used a wrapper library on Google's language detection to discard non-English policies~\cite{LangDetect:2021}.
When policies included lists where each list entry was not self-contained, we merged these lists into the preceding text to provide relevant context. We scanned short lists, i.e., where each list item was composed of $<$20 words, and merged them into the preceding paragraph, thereby treating the entire list as a single segment. We then eliminated segments comprising $<$20 words. After cleaning, the classifiers individually processed each segment and mapped it back to the original policy. After excluding links that returned response errors, the \textit{readability} library successfully extracted relevant text from 286,717 policies, which we classified in the next stage.

\paragraph{\circlea{3} \textbf{Classifying Policies and Extracting Labels.}}
We analyzed policies with a similar approach to Polisis~\cite{Harkous:2018}, an  NLP framework that classifies data collection behavior from privacy policy text. Unfortunately, the prior published Polisis implementation is proprietary, and on reaching out, the authors informed us that their website can only take up to 30K policies. We completely re-implementated the classification framework to the same standards as prior work. We replaced their CNN-based approach with a state-of-the-art language model to improve classifier performance. We used PrivBERT~\cite{srinath-privbert-2021}, a transformer-based privacy policy language model, which was developed by pre-training the RoBERTa$_{BASE}$ model~\cite{roberta:2019} on 1M privacy policies. We fine-tuned PrivBERT on the OPP-115 corpus~\cite{Wilson:2016}. We present an overview of the framework structure in \autoref{fig:polisis-structure}. In~\autoref{tab:classifier-results} in the Appendix, we show that the PrivBERT classifiers perform better than CNNs. We provide more details about training and evaluating the models in \autoref{sec:framework}. 

We first passed each segment through the Segment Classifier to extract the high-level data practice. We passed any segments addressing \textit{First Party Collection/Use} or \textit{Third Party Collection/Sharing} through six Attribute Classifiers -- \textit{Does/Does Not}, \textit{Identifiability}, \textit{Purpose}, \textit{Personal Information Type}, \textit{Action First Party}, and \textit{Action Third Party} -- to extract annotations relevant to privacy labels. 
We used the \textit{Action First-Party} attribute to filter any segments explicitly addressing collection on websites (and not mobile apps). We used the \textit{Action Third-Party} attribute to eliminate instances wherein the third party only `sees` and does not collect data. We successfully detected segments addressing data collection in the policies of 474,966 apps ($n=280,767$ policies), which we then used to create privacy labels.

\conversionTable

\inferentialConversionTable

\paragraph{\circlea{4} \textbf{Compare and Evaluate.}}
The taxonomy of policy labeling does not always have a one-to-one mapping with Apple's privacy labels. So, we developed a grounded strategy based on qualitative coding to convert outputs from classifiers into equivalent privacy labels. Three researchers \textit{independently} coded the conversions and then discussed to reach an agreement on the mappings between OPP-115 and privacy labels. 
The coders completed three matching tasks:
\begin{itemize}[leftmargin=*,nosep,noitemsep]
    \item First, the coders determined which of the data practices found by the Segment Classifier, such as \textit{First Party Collection/Use} or \textit{Third Party Collection/Sharing}, that when combined with the Identifiability Attribute Classifier, such as ``Identifiable,'' ``Aggregated/Anonymized,'' ``Does'', or ``Does Not'', match to an appropriate Apple privacy label type,  such as \textit{Data Linked to You} or \textit{Data Not Collected}. For example, when the framework identifies a segment with a data practice of ``First Party Collection/Use'' and the data is ``Identifiable,'' that would associate with an Apple privacy label type of \textit{Data Linked to You}.

    \item Next, the coders matched the output of the Purpose Attribute Classifier against Apple's privacy label purposes. For example, a framework output of ``Basic Services/Features'' gets mapped to \textit{App Functionality} for privacy label purposes. 
    
    \item Finally, the coders matched the outputs of the Personal Information Type Attribute Classifier to the data categories provided in Apple's privacy label. For example, Polisis may identify that a segment discusses ``Contact,'' which then maps to the privacy label data category of \textit{Contact info}.
\end{itemize}
The combination of these three matching tasks provides a single privacy label entry for an app, according to the privacy policy, describing the privacy type (e.g., \textit{Data Linked to You}), the purpose (e.g., \textit{App Functionality}), and the data category collected (e.g., \textit{Contact Info}). We can find the full list of the direct conversions in this manner\autoref{tab:primary-conversion}. The coding process also revealed additional, inferred privacy labels from Polisis classification that included a combination of classifications and keywords relevant for \textit{Data Used to Track You} and remaining \textit{Data Categories}.~\autoref{tab:inferential-conversion} shows the inferred privacy labels. We further verified the mapping by randomly sampling labels generated from classifier outputs. In the Appendix, we present our evaluation in \autoref{tab:mapping-verification}.

\section{Limitations}
\label{sec:limitations}
Before proceeding, it is essential to note the limitations of our approach in comparing the privacy labels with the privacy policies.

\paragraph{\textbf{Ground truth.}} 
Foremost, we note that neither the labels nor the policies can provide comprehensive ground truth of app behavior, and even statistical and dynamic analysis has limitations. Here, we report only on observed discrepancies between the policies and the labels, but validating which is more in line with app behavior is beyond the scope of this paper. However, as these discrepancies occur at scale  (as reported in the next section), there are strong indications of prominent misapplication of privacy labels according to the privacy policies provided by app developers. Additionally, we present examples via case studies (\autoref{sec:case-studies} and \autoref{sec:case-studies-policies}) to show how such discrepancies occur with popular apps.

\paragraph{\textbf{Classifier Predictions.}} 
The outputs of language models introduce uncertainty that propagates further when combined. As a result of these inaccuracies, we can only report on the presence of statements addressing data collection practices in privacy policies and differences when compared with privacy labels.
However, the reported discrepancies are {\em much} larger than the associated uncertainties. Additionally, our framework analyzes privacy policies on a per-paragraph/per-segment basis, so it cannot detect explanations of app behaviors that span multiple segments.

\paragraph{\textbf{Train/Test Dataset.}} 
Without an updated corpus with equivalent robustness, we used the OPP-115 corpus to fine-tune language models~\cite{Wilson:2016}, an extensive dataset comprising manual annotations of 23k fine-grained data practices gathered from multiple graduate-level law students. However, the dataset includes old privacy policies that the researchers collected before the introduction of present-day privacy laws. We identify the limitations introduced by this dataset and recognize the need for an updated dataset. Additionally, specific annotations in the OPP-115 corpus do not directly map to the Apple privacy label taxonomy. As such, the independent annotators used a grounded approach to develop an inferential mapping to address this limitation (see \autoref{tab:inferential-conversion}). Finally, in \autoref{tab:classifier-results} performance, we manually evaluate the classifier outputs on new policies by randomly sampling segments from our dataset of app policies.

\paragraph{\textbf{Information Extraction.}} 
Privacy policies comprise varying formats, reducing the amount of information we can gather from our framework. As previously highlighted, our per-segment approach misses information that spans multiple, non-contiguous segments. Next, policies present information in various media formats (e.g., images) that we do not include in our analysis. Finally, many privacy policies contain links to third parties' privacy policies. We did not analyze the transitive closure of all privacy policies as part of this work. Apple's policy is for privacy labels to include all collection and tracking mechanisms, including third-party practices. Our analysis is a lower bound of data collection performed within an app, particularly related to third parties.

\begin{figure}[t]
    \centering
    \resizebox{0.8\linewidth}{!}{    \input{figures/privacy-types.pgf}}
    \vspace{-3ex}
    \caption{An overview of apps declaring data collection with corresponding \textit{Privacy Types} within their privacy policies (top) and on the App Store via privacy labels (bottom). The denominator is the total apps that we analyzed, i.e., 474,669 apps. Please note that the privacy types, except for \textit{Data Not Collected}, are \textit{not} mutually exclusive.}
    \label{fig:privacy-types}
    \vspace{-3ex}
\end{figure}

\privacyLabelPolicyOverlap
\section{Results}
\label{sec:results}
In this section, we directly compare developers' reported privacy labels to the output of language models following the hierarchical structure of the privacy labels (see \autoref{fig:privacy-labels-anatomy}).

\paragraph{\textbf{Privacy Types.}} We first consider the top level of privacy labels, the privacy types: \textit{Data Used to Track You}, \textit{Data Linked to You}, \textit{Data Not Linked to You}, and \textit{Data Not Collected}. We are primarily concerned with determining the number of apps with such a privacy type and if we can also find that privacy type in the policies. \autoref{fig:privacy-types} and  \autoref{tab:label-policy-overlap} provide a snapshot of the overlap of privacy types extracted from privacy policies and the privacy types declared in the privacy labels for the app on the App Store. As a helpful reminder while reading the numbers reported in this table, three of the privacy types, \textit{Data Used to Track You}, \textit{Data Linked to You}, and \textit{Data Not Linked to You}, are \textit{not} mutually exclusive. Apps may collect data linked to the user and aggregated/anonymized (i.e., not linked to the user), and they may also collect data to track the user.

The \textit{Data Linked to You} privacy type indicates that the app collects data linked to users, i.e., in an identifiable manner. Of the 190,965 apps indicated such collection on the App Store, our framework identified 88\% ($n=168,121$) (Fig.~\ref{fig:privacy-types}; lower half; yellow bar; hatches). More concerning, we observed an additional 228,539 apps that reported this practice in their policies but did not report it on the App Store (Fig.~\ref{fig:privacy-types}; top half; yellow bar; stripes).

We identified that 41\% ($n=88,172$) of the apps whose privacy labels stated that they collected data in an aggregated/anonymized manner, i.e., had \textit{Data Not Linked to You} privacy type, also said so in their policies (Fig.~\ref{fig:privacy-types}; lower half; blue bar; hatches). Of the remaining 59\% ($n=127,020$) apps that had the \textit{Data Not Linked to You} privacy type in their label but did not have a corresponding policy segment (Fig.~\ref{fig:privacy-types}; lower half; blue bar; stripes), 76\% ($n=97,029$) of those instead included segments in their privacy policy that indicated that they collect data linked to users (\autoref{tab:label-policy-overlap}; row2; col3). This difference may result from apps \textit{not} stating their aggregation practices in the same segment of the policy that addresses data collection. Despite factoring in uncertainty, there is a large gap between the practices declared in privacy labels and privacy policies.

Perhaps more problematic is apps that report they do not collect any data. Recall that the \textit{Data Not Collected} privacy type {\em is} mutually exclusive, i.e., developers only added this label to apps that claim \textit{not} to collect \textit{any} data from users. While 36\% ($n=172,924$) of the apps that we analyzed indicated in their privacy label that they did \textit{not} collect \textit{any} data, only 0.03\% ($n=4,359$) of these apps made similar statements in their policies (Fig.~\ref{fig:privacy-types}; lower half; green bar). More surprisingly, 84\% ($n=173,441$) of these apps stated in their policies that they collected data linked to users (\autoref{tab:label-policy-overlap}; row 4; col 2). 

Finally, of the 108,937 apps that stated on the App Store that they collected data to track users, our framework also reported similar practices in the privacy policies of 49\% ($n=53,359$) (Fig.~\ref{fig:privacy-types}; bottom half; red bar; hatches). We identified an additional 123,675 apps that did not declare this practice on the App Store (Fig.~\ref{fig:privacy-types}; top half; red bar; stripes).
Recall that the framework infers this privacy type, and we, therefore, partially report user tracking that apps engage in, presenting a lower bound of mislabeling. Our identification of apps that fail to report data collected for tracking indicates that many apps are under-reporting their tracking practices.

\textit{Takeaways.} Developers are very likely under-reporting their collection of identifiable data on the App Store. Most apps that indicate on the App Store that they do \textit{not} collect \textit{any} data state otherwise in their privacy policies.

\begin{figure*}[t]
\centering
\input{figures/pt-vs-purpose.pgf}
\vspace{-3ex}
\caption[Privacy Types vs. Purposes]{The ratios of the six purposes for the \textit{Data Linked to You} and \textit{Data Not Linked to You} privacy types. The denominator is the number of apps with the designated privacy type either in their privacy label or their privacy policy, i.e., 419,504 apps with a \textit{Data Linked to You} label and 294,391 with a \textit{Data Not Linked to You} label. It is helpful to note here that privacy types shown here are \textit{not} mutually exclusive. Two other \textit{Privacy Types} are not shown here; the \textit{Data Used to Track You} privacy type refers to collection for the purpose of tracking, while the \textit{Data Not Collected} refers to the absence of \textit{any} data collection.\label{fig:pt-vs-purpose}}
\vspace{-2ex}
\end{figure*}

\paragraph{\textbf{Purposes.}} We look at how apps claim to use the data they collect. \autoref{fig:pt-vs-purpose} presents a snapshot of the purposes associated with data collection, as identified from privacy labels and privacy policies. As a reminder, apps may collect both linked and not linked (anonymized) data. Additionally, apps may collect data for multiple purposes. For example, an app may collect your \textit{Location} in an anonymized manner to personalize user experience (\textit{Product Personalization}) and in an identifiable manner to help advertisers and agencies tailor the advertisements they display (\textit{Third Party Advertising}).

We find greater agreement between privacy labels and privacy policies for apps that collect data for \textit{App Functionality} and \textit{Analytics}. Of the 161,587 apps indicated in their privacy label that apps collect data linked to users for \textit{App Functionality}, 81\% ($n=130,108$) also included a corresponding statement in their privacy policy. Similarly, of the 105,729 apps that stated in their privacy label that they collect data linked to users for \textit{Analytics}, 68\% ($n=71,883$) also included a corresponding statement in their privacy policy (Fig.~\ref{fig:pt-vs-purpose}; bottom half; left plot; yellow bars 1 \& 2; hatches).

We find notable discrepancies in developers' reporting of \textit{Third-party Advertising} in their privacy policies and on the App Store. Considering data collection that is linked to users (Fig.~\ref{fig:pt-vs-purpose}; left plot; bar 4), 139,765 apps exclusively declare this purpose in their privacy policies (top half) and do not report this practice on the App Store.
Our findings are concerning since this is a lower bound. Privacy policies link to third-party policies instead of including details here. The results indicate that developers focus on their app's data collection practices when filling out privacy labels without considering third parties. We further highlight the problem of incomplete labeling with examples in~\S\ref{sec:case-studies} and \autoref{sec:case-studies-policies}.

Finally, we find that while 366,840 ($77\%$) apps stated in their privacy policies that they collected data in an identifiable manner for a purpose that does not fit into any of the options that Apple provides in their privacy label, only 17,487 ($5\%$) of these apps also addressed this on the App Store (Fig.~\ref{fig:pt-vs-purpose}; left plot; yellow bar 6).  
It appears that developers are less forthcoming about declaring data collection in their privacy labels for purposes beyond Apple's taxonomy, making limited use of the catch-all: \textit{Other Purposes}.

\textit{Takeaways.} Developers are \textit{more} likely to declare data collection for \textit{App Functionality} and \textit{Analytics} in either, privacy labels or privacy policies. Developers are also \textit{less} likely to declare data collection in their privacy labels for purposes beyond Apple's taxonomy, i.e., \textit{Other Purposes}.

\begin{figure*}[t]
\centering
\resizebox{0.8\linewidth}{!}{\input{figures/data-categories-by-pt.pgf}}
\vspace{-2ex}
\caption[Privacy Types vs. Data Categories]{The ratios of data categories against privacy types. The denominator is the number of apps with the designated privacy type either in their privacy label or their privacy policy, i.e., 232,648 apps with \textit{Data Used to Track You}, 419,504 apps with \textit{Data Linked to You}, and 294,391 apps with \textit{Data Not Linked to You}. The three privacy types shown here are \textit{not} mutually exclusive.\label{fig:pt-vs-data-categories}}
\vspace{-1ex}
\end{figure*}

\paragraph{\textbf{Data Categories.}} We additionally analyze the data categories collected by apps as stated in their privacy labels and policies.~\autoref{fig:pt-vs-data-categories} provides visual results of our findings, and we present additional details in \autoref{tab:data-categories-overlap-browsing-history} in \autoref{sec:additional-figs}.

We find that apps are more likely to declare in either their privacy policies or their privacy labels that they collect \textit{Contact Info} ($n=273,351$; $65\%$) and \textit{Identifiers} ($n=320,607$; $76\%$) linked to users (Fig.~\ref{fig:pt-vs-data-categories}; middle plot; yellow bars 2 \& 5). 
Apps that collect data to track users are more likely to use \textit{Browsing History} (46\%; $n=106,816$), \textit{Identifiers} (71\%; $n=164,732$), and \textit{Usage Data} (65\%; $n=150,651$) (Fig.~\ref{fig:pt-vs-data-categories}; upper plot; red bars 1, 5, \& 7). Our findings are in line with previous work that showed tracking activities target users with cookies and tracking pixels (\textit{Identifiers}) and monitor their browsing practices across sites and services (\textit{Browsing History} and \textit{Usage Data})~\cite{Acar:2014,Englehardt:2016}. 

However, we find that apps that state in their privacy policy that they collect \textit{Browsing History} (i.e., how users browse the Internet outside of the app)
and \textit{Sensitive Info} (such as racial/ethnic data, sexual orientation, etc.) linked to users are less likely to declare this collection in their privacy labels (Fig.~\ref{fig:pt-vs-data-categories}; middle plot; top half; yellow bars 1 \& 13). Surprisingly, of the 212,121 apps that stated in their privacy policy that they collect \textit{Browsing History} linked to users, only 658 (0.3\%) of these apps declared this practice in their privacy labels. 
While 96,837 apps indicated in their privacy policy that they collect some form of \textit{Sensitive Info}, only 2\% ($n=2,144$) apps also declared this collection in their privacy labels. Of notable concern, we find 22,171 apps and 11,710 apps mislabeling their collection of \textit{Identifiers} and \textit{Contact Info} respectively as being linked to users when their policies indicate that they use collected data to track users (see \autoref{tab:data-categories-overlap-browsing-history}). 

\textit{Takeaways.} Developers most commonly state that they collect \textit{Identifiers} and \textit{Contact Info} that are linked to users. Developers that state in their privacy policies that they collect \textit{Browsing History} or \textit{Sensitive Info} linked to users are less likely to declare this collection in their privacy labels. Apps that track users are more likely to use \textit{Browsing History}, \textit{Identifiers}, and \textit{Usage Data}, which is in line with prior findings about tracking practices.

\begin{figure*}[ht]
\centering
\resizebox{0.7\textwidth}{!}{\input{figures/app_costs_by_privacy_label_type.pgf}}
\vspace{-3ex}
\caption[App Costs By Privacy Label Type Ratios]{The ratios of app costs for each of the four privacy types. The denominator is the number of apps with the designated app cost that have a privacy label. Free apps are more likely than paid apps to collect data, including data used to track and linked to users. Please note that privacy types shown here are \textit{not} mutually exclusive.\label{fig:app-costs-by-privacy}}
\vspace{-2ex}
\end{figure*}

\paragraph{\textbf{Free vs. Paid Apps.}}%
The App Store has four pricing models: free apps, free apps with in-app purchases, paid apps, and paid apps with in-app purchases. 
Interestingly, when only observing privacy labels (Fig.~\ref{fig:app-costs-by-privacy}; all plots; bottom half), it would appear that paid apps have better privacy behaviors than their free counterparts. However, the altruism of paid apps compared to free apps disappears when considering the privacy policies (the top half of \autoref{fig:app-costs-by-privacy}). The privacy policy analysis better aligns with the observations of Han et al.~\cite{Han:2019, Han:2020}, who compared free and paid apps in the Android Play Store based on the inclusion of third-party advertising software, finding no differences between free and paid apps.

As a result of apparent under-reporting by paid apps, we find that they have the largest discrepancies of potentially under-reporting data collection practices in their privacy labels compared to the privacy policies. 
While the privacy policies suggest that 75\% ($n=21,330$) of paid apps collect data linked to users, only 4\% ($n=1,145$) paid apps have a privacy label of this type (Fig.~\ref{fig:app-costs-by-privacy}; second plot; yellow bar 3). More concerning, while the privacy policies of 21\% ($n=6,118$) paid apps report collecting data to track users, only 2\% ($n=643$) paid apps report this practice on the App Store (Fig.~\ref{fig:app-costs-by-privacy}; first plot; red bar 3).

\paragraph{\textbf{Content Rating.}}
Developers provide a \textit{Content Rating} as part of the app metadata to indicate the age appropriateness of their apps. 
These ratings are reviewed by Apple~\cite{Apple:Guidelines} and used to enforce parental control features that restrict children from accessing the app~\cite{Apple:AgeRatings}.
We find that most apps that have a 4+ content rating on the App Store (81\%; $n=419,762$), while fewer apps have 9+ (3\%; $n=16,687$), 12+ (9\%; $n=46,737$), or 17+ (13\%; $n=69,309$) content ratings. Since privacy labels do not indicate the app's data practices specific to children, users must review the privacy policy to learn this information. Given parental control settings, an app with a 4+, 9+, or 12+ rating could be used by minors, although they may not be the intended or target audience for the app. However, when an app specifically targets children, it is subject to additional regulations that may require parental consent.
We fine-tuned language models to identify policy segments that address \textit{International/Specific Audiences} and to identify further if the segment addresses \textit{Children}, then compare this output to the content rating.
Only 50\% ($n=179,168$) apps with a 4+ content rating also included a privacy policy segment that addresses data practices specific to children (Fig.~\ref{fig:children}; all plots; left-most bar). We were more likely to find similar policy segments for apps with different content ratings that can also be accessed by children, 9+ (65\%; $n=10,118$) and 12+ (59\%; $n=22,293$).

\begin{figure*}[t]
\centering
\resizebox{0.8\linewidth}{!}{\input{figures/children.pgf}}
\vspace{-3ex}
\caption[Children and User Content Rating By Privacy Label Type Ratios]{The ratios of the content ratings for each of the four privacy types, with an overlay (white bar) indicating the ratio of apps that also include a segment in their privacy policy, where they address privacy practices specific to children who engage with their services. The denominator is the number of apps with the designated content rating that have a privacy label. Please note that privacy types shown here are \textit{not} mutually exclusive. \label{fig:children}}
\vspace{-2ex}
\end{figure*}

We further looked at content ratings for different privacy types associated with data collection. 
Considering apps with a 4+ content rating, roughly half had a policy explicitly addressing children across privacy types. While 20\% ($n=74,320$), 37\% ($n=134,076$), and 44\% ($n=159,512$) of the apps with a 4+ content rating declare in their privacy label that they collect data used to track users, linked to users, and not linked to users respectively, only 58\% ($n=43,536$), 51\% ($n=68,715$), and 54\% ($n=86,743$) of those apps also addressed children in their privacy policies (Fig.~\ref{fig:children}; plots 1, 2, \& 3; bottom half; left-most bars; white overlay indicates addressing children).

While adding a 4+ content rating may help developers reach a wider audience, we only identified half of these apps consider data practices specific to children in the privacy policy. Additionally, even when apps address data collection from children in their privacy policies, these segments may absolve the developer of any responsibility. For example, ChowNow~\cite{ChowNow} is an app platform used by 3,182 different apps of local restaurants to receive online orders for takeout and delivery.
ChowNow adds a content rating of 4+ to its apps on the App Store, making it accessible for children. Recall that developers choose a content rating according to Apple's guidelines~\cite{Apple:Guidelines}; Apple does not assign this value. However, ChowNow's privacy policy absolves themselves of the responsibility of dealing with data collected from children. 

We acknowledge that our findings do not implicate the evaluated apps of violating COPPA~\cite{FTC:COPPA:2013}, which, for example, allows PII collection with specific restrictions (e.g., geolocation) provided that developers do not use data for targeting/profiling of minors and that they obtain informed parental or legal tutor consent. We highlight the lack of declaration of data practices in privacy policies, especially when considered optional, and the need to ensure transparency across platforms. Additionally, third-party libraries offer options to help applications comply with COPPA regulations, but prior work has shown that they are often misconfigured~\cite{Reyes:2018}.

\paragraph{\textbf{App Genre.}} We present an overview of our findings by app genre in~\autoref{fig:app-genre-by-privacy} in~\autoref{sec:additional-figs}. We find that \textit{Games} apps are most likely to collect data used to track users (60\%) and linked to users (59\%) (Fig.~\ref{fig:app-genre-by-privacy}; plots 1 \& 2; bar 8). Notably, while 83\% apps associated with the \textit{Stickers} genre stated on the App Store that they do not collect \textit{any} data, our analysis found that 66\% apps collected data linked to users (Fig.~\ref{fig:app-genre-by-privacy}; plots 2 \& 4; bar 23). Apps under the \textit{Stickers} genre are mostly lightweight apps made by smaller developers. They tend to have a 4+ content rating to reach a larger audience. They can include a few ad spaces and analytics libraries. Our intuition is that individual developers may not be aware of the data collection from third-party analytics and advertising libraries.

\paragraph{\textbf{App Popularity.}} Since the App Store does not reveal the number of downloads for an app, we instead rely on the number of user ratings as a proxy for app popularity. To better represent their disclosures, we bin rating counts within the same order of magnitude in a single category and present our findings in~\autoref{fig:ratings-counts-by-privacy} in ~\autoref{sec:additional-figs}. We find that with increased popularity, apps are more likely to declare data collection linked to users and used to track users. Our findings suggest that popular apps are more likely to be more thorough in their declaration of data collection practices because they receive more scrutiny.

\begin{figure}[t]
    \centering
    \resizebox{0.8\linewidth}{!}{\input{figures/templates-privacy-types.pgf}}
    \vspace{-3ex}
    \caption{An overview of the privacy types associated with data collection on the App Store, from privacy labels and privacy policies, specific to apps whose policies are similar to templates. The denominator is the total number of such apps, i.e., 300,535 apps. Please note that the privacy types, except for \textit{Data Not Collected}, are \textit{not} mutually exclusive.}
    \label{fig:templates-privacy-types}
    \vspace{-2ex}
\end{figure}

\paragraph{\textbf{Privacy Policy Templates.}}
Templates offer a valuable solution for creating privacy policies, as they provide a ready-made framework for organizations to establish clear guidelines regarding handling user data. These pre-designed templates serve as a starting point that developers can customize to meet specific requirements and legal obligations. By utilizing templates, businesses can save time and effort by avoiding the need to create privacy policies from scratch.
Additionally, templates help ensure compliance with privacy regulations by incorporating standard clauses and disclosures, ensuring that the privacy policy aligns with applicable laws such as GDPR or CCPA. However, it is essential for organizations to carefully review and tailor the template's content to accurately reflect their unique practices, guaranteeing transparency in communicating their privacy practices to users.

We evaluated the policies in our dataset to identify the use of templates. We searched for privacy policy templates and generators and gathered a list of services. We then visited each service and signed up, if required. We collected a set of 15 privacy policy templates, which we cleaned and divided into individual sentences.
We represented the text in both the templates and the policies using in-domain word embeddings derived from privacy policies shared by Harkous et al.~\cite{Harkous:2018}.
For each policy in our dataset, we conducted a comprehensive sentence-level comparison. We compared each sentence in a policy against every sentence in a template. We employed the cosine similarity metric to measure the semantic resemblance between two sentences. We deemed sentences similar if their cosine similarity exceeded a threshold of 0.8. We established a criterion to determine if a policy derived from a template: if over half of the sentences in a policy were similar to over half of the sentences in the template, we identified the policy as template-like. 

We find that the privacy policies of 65\% ($n=306,404$) apps potentially use templates. We looked at the privacy labels these apps have declared on the App Store. 
Considering privacy types, 23\%, 45\%, 46\%, and 31\% of these apps declare \textit{Data Used to Track You}, \textit{Data Linked to You}, \textit{Data Not Linked to You}, and \textit{Data Not Collected} privacy types in their labels on the App Store (Fig.~\ref{fig:templates-privacy-types}; bottom half; all bars). These findings align with all evaluated apps (see \autoref{fig:privacy-types}).
A majority of evaluated apps use template-like privacy policies. The use of templates possibly affects the discrepancies between the declaration of data collection practices in privacy labels and privacy policies. 
Templates often use generators, which offer significant value by ensuring developers thoroughly consider various data collection and sharing practices.
These generators are similar to creating privacy labels on the App Store.  
However, it is essential to recognize that templates are not one-size-fits-all solutions. Developers must review and tailor policies derived from templates to accurately reflect individual apps' unique data collection practices. By carefully reviewing and customizing policies, developers can ensure the accuracy of their disclosures.

\section{Case Studies}
\label{sec:case-studies}

Without Apple verifying privacy labels (and policies), their contents may not wholly clarify actual app practices. We present case studies of app behavior to shed light on the potential disparities between stated data collection practices and real-world app behavior. We use network requests captured from app usage to behavior developers report in labels and policies.

We used an iPhone running iOS 17.3.1 (released Feb 2024) with a man-in-the-middle (MiTM) proxy~\cite{mitmproxy} to gather outgoing traffic to determine domains that apps accessed. We evaluated each app in the following manner: (1) We installed the app directly from the App Store. (2) We established a connection between the iPhone and the proxy. (3) Upon opening the app, the proxy captured and stored any outgoing requests made by the app. (4) After closing the app and terminating the proxy connection, we deleted the app before evaluating the next app in the sequence.

We included 39 apps in the analysis, split between (a) 24 apps that declare data collection for advertising purposes in their privacy policies but not on their privacy labels and (b) 15 apps that declare a ``Data Used to Track You'' privacy type in their label on the App Store, but we could not infer such a practice from their privacy policies. We then compared the domains in the captured network requests against EasyList, EasyPrivacy, and WhoTracks.Me to identify trackers~\cite{EasyList,WhoTrackMe:Site,Whotracks:2019}. We provide an overview of our findings in~\autoref{tab:traffic-collection} in~\autoref{sec:traffic-collection}. The analysis presented in this study is an exploratory case study of 30 apps' network behavior. It should not be considered representative of the practices of all apps on the App Store.

The evaluated apps contact numerous tracking domains, with Facebook and Google being the most prominent. Further, developers often do not include analytics libraries within their purview of tracking, but guides from these libraries show that their practices are more nuanced~\cite{Appsflyer:Guidelines,GoogleAnalytics:Disclosures}. Additionally, inconsistencies between privacy disclosures and network traffic persist across different app categories. When privacy policies mention third-party libraries, they refer to third-party policies, resulting in incomplete inferences from an automated approach like the one presented in this work. We elaborate on potential explanations for our observations below.

\paragraph{\textbf{Policy Reuse.}} Developers with multiple apps on the App Store reuse the privacy policies linked with individual apps. While this practice may result from using generic templates for some developers, organizations can also reuse these templates with multiple services. For example, different developer accounts publish Lexington Law and CreditRepair (\#1 \& \#2 in~\autoref{tab:traffic-collection}), and the apps link to different privacy policies on the App Store. However, their privacy labels and privacy policies are identical. They are subsidiaries of the same organization, PGX Holdings Inc., and reuse declaration statements even if these statements apply to those subsidiaries. Developers must update templates to ensure accurate data collection practices, which can then reflect the accuracy of privacy labels.

\paragraph{\textbf{Understanding Third Party Collection.}} When applications state in their privacy policies that they do not share data with third parties except to provide certain services (not including targeted advertising), it is possible that developers do not clearly understand or parse the nuances of data collection and sharing performed by integrated third parties. For example, Paypal, Crumbl, and Discord (\#3, \#9, \#12 in~\autoref{tab:traffic-collection}) have policies covering data collection and sharing from third parties. To their credit, third-party libraries provide guidelines and disclosure links for developers to review before filling out their privacy labels and privacy policies (examples,~\cite{Appsflyer:Guidelines,Unity:Guidelines,Meta:Guidelines,GoogleAnalytics:Disclosures}). However, these guides include multiple caveats that can further complicate developers' understanding, requiring them to process against their use cases and translate into Apple's data collection definitions and requirements.

\paragraph{\textbf{Understanding App Store requirements.}} Apple requires that developers declare all data collected in the app, including the practices of third-party partners, except for certain scenarios wherein disclosure is deemed optional~\cite{Apple:PrivacyLabelDetails}. While apps like Venmo, Southwest Airlines, Open Table, and Indeed (\#1, \#4, \#6, \#11 in~\autoref{tab:traffic-collection}) fill their privacy labels with multiple data categories under the \textit{Data Linked to You} and \textit{Data Not Linked to You} privacy types, they fail to do the same while declaring \textit{Data Used to Track You}. Their privacy policies include statements highlighting third-party data collection and sharing for advertising and measurement purposes, indicating the developers' understanding of such activity. However, despite the App Store requiring the disclosure of all data collection practices, the developers' interpretation of optional caveats may affect their creation of privacy labels. For example, the period tracking app, Maya (\#24 in~\autoref{tab:traffic-collection}), declared the sharing of \textit{Usage Data} for tracking users, but the third-party libraries that it uses additionally collect and use identifiers and device information to track users~\cite{Meta:Guidelines,GoogleAnalytics:Disclosures}.

\paragraph{\textbf{Understanding Apple's Definition of Tracking.}} Apple details practices that it considers to fall under \textit{Tracking}, along with examples and caveats~\cite{Apple:PrivacyLabelDetails}. However, recent work has found that developers find it difficult to understand this definition and correctly declare data collection used to track users~\cite{li-understanding-2022}. Apps like Axolochi, WebMD, and Food Network Magazine (\#19, \#21, and \#22 in~\autoref{tab:traffic-collection}) acknowledge the use of tracking technologies in their privacy policies. However, the absence of similar declarations in privacy labels can stem from confusion around their understanding of Apple's definition of tracking. A recent study by Li et al.~\cite{li-understanding-2022} showed that developers find it difficult to correctly identify data linked to users and data used to track users.

Next, we present possible reasons for discrepancies for apps with a \textit{Data Used to Track You} privacy type in the App Store label but prove it challenging to automatically capture tracking practices from their privacy policies.

\paragraph{\textbf{Non-exhaustive Policies.}} The privacy policies of Shake Shack, Kika Keyboard, Photo Prints CVS, Everpix, and FloatMe (\#25, \#26, \#27, \#28, \#29 in~\autoref{tab:traffic-collection}) mention third party collection and sharing in terms of legal compliance and mergers/acquisitions. These privacy policies do not comprehensively cover all practices and data collection scenarios, making it difficult to identify such practices without ground truth.

\paragraph{\textbf{Unclear Policy Statements.}} Even when developers declare third-party data collection and sharing in their privacy policies, such declaration is not explicit or clear to enable automatic detection and inference. The policies of Buffalo Wild Wings, The General Auto Insurance App, Conservative News (\#30, \#31, \#32 in~\autoref{tab:traffic-collection}) include statements of sharing of information with ``non-affiliated third parties'', ``vendors'', ``third party code and libraries'', but do not make explicit the specific data categories collected and the use of this data for tracking, advertising, or advertising measurement.

\paragraph{\textbf{Complex Formats.}} Being free-form documents, privacy policies do not need to be presented in standard, machine-parsable formats. While developers provide correct links to their policies on the App Store, we can only access the content of the policy behind a further link(s), as is the case with apps like McDonalds, Episode (\#35, \#36 in~\autoref{tab:traffic-collection}). Additionally, the policy for BrainBoom (\#33 in~\autoref{tab:traffic-collection}) presents information in mixed formats, i.e., text and images, further complicating our ability to identify all practices. Finally, apps like JCPenney, Dosh, and CDL Prep Test (\#37, \#38, \#39 in~\autoref{tab:traffic-collection}) provide incorrect or broken links on the App Store, resulting in the extraction of incorrect from automated crawls.

\section{Discussion and Conclusions}

We analyzed 474,669 apps on the App Store, comparing the practices reported in privacy policies to those reported in privacy labels by performing automated NLP classification of the privacy policies.
We find that most apps are likely under-reporting data collection practices in their privacy labels compared to their privacy policies. 
We find that almost all (97\%) apps that indicate in their privacy labels that they do \textit{not} collect \textit{any} data engage in some form of data collection according to their privacy policy. Additionally, the privacy labels of 84\% of paid apps indicate that they do not collect any data. In contrast, privacy policies suggest that the actual number may be closer to only 6.4\% paid apps. Privacy policy analysis also reveals additional information about data practices not captured in privacy labels, including that most apps (81\%) selected a 4+ content rating, but only 50\% of these apps mention data collected from children in their privacy policies. 

\paragraph{\textbf{Ethics.}} The analysis and findings we present are based on publicly available data. We only mention popular apps (determined from rating counts) associated with large companies or developed by services with numerous associated apps. We reached out to Apple and shared our paper before publication. We encourage communication from developers and researchers to make use of our code and data to verify privacy labels.

In the remainder of this section, we discuss some of the implications of this analysis, such as the ground truth of privacy behavior when considering privacy labels or privacy policies. We also consider what factors likely lead to the misapplication of labels and recommendations for improving the current state. 

\paragraph{\textbf{Privacy Behavior Ground Truth.}} Since Apple's labels are not validated, we considered the privacy policies a reasonable reference point of comparison. However, it isn't easy to know the actual ground truth of privacy behavior, even if we fully dynamically and statically analyze every app. In this paper, we compare privacy labels against privacy policies as a point of comparison of the declaration of data practices across platforms. Privacy policies do not serve as ground truth for actual app behavior.
While there are limitations to the approach we take in analyzing privacy policies using classifiers, the NLP methods of extracting free-form text levels get us closer to a viable understanding of data collection practices than the privacy labels, as currently used. We believe that this is the case for two reasons. 
First, classifier outputs introduce uncertainties that stem from the fact that policies are analyzed on a per-segment basis, so discussions of data aggregation or anonymization that occurs in one segment, separate from the data that is collected, might appear as data linking when it is, in fact, not linked. However, even with these statements, the app's behavior remains ambiguous according to the privacy policy regarding which specific data categories are aggregated or anonymized. Apps could often link data based on unique identifiers stated in other policy segments. Our observations suggest that developers mislabel many apps even after considering uncertainties from classier outputs. 
Second, there are also significant cases of under-reporting from classifiers due to how Apple links to privacy policies and the use of secondary privacy policies from third-party libraries. Many privacy policies link to other policies that we did not analyze. The App Store links also point to the developers' and not the specific apps' privacy policies. These policies usually address all services provided by the developer. For example, Subsplash~\cite{Subsplash:Policy} and ChowNow~\cite{ChowNow} affect thousands of apps, and it is unknown how the eventual customer uses that data and if policies reflect such scenarios. 

\textit{Takeaway.} We need improved notions of ground truth, which can dynamically identify data collection within apps at scale. However, even with their shortcomings, privacy policies provide a first-level check to identify discrepancies in privacy labels.

\paragraph{\textbf{Source of Confusion Around Privacy Labels.}} 
It may also be that the processes for generating a privacy policy, including legal staff, are quite different from those selecting the labels, leaving the onus on the development team to make an accurate submission to the App Store. This split in responsibilities could confuse the kinds of data covered by the privacy label (as compared to what is in the policy) and what Apple would consider linked or not linked to users. For example, a recent study by Li et al.~\cite{li-understanding-2022} showed that developers find it difficult to correctly identify data linked to users and data used to track users. Our results suggest that there is a large amount of mismatch in both data linked and not linked regarding the {\em Purposes}, where {\em App Functionality} and {\em Analytics} are particularly confusing, especially when apps may collect unique identifiers, as well as collecting {\em other} kinds of data that this should match to the {\em Other Purposes} category. 

\textit{Takeaway.} We argue that inaccurate labels are not necessarily the developers' fault but that better guidance and education are required to help them match app practices to labels.

\paragraph{\textbf{Divergent Incentive Models}}
Privacy policies have become a standard and accepted part of notice and consent laws, and failure to provide an accurate and comprehensive privacy policy could lead to serious legal consequences. Companies are well incentivized to provide broad privacy policies that provide legal cover for their data collection practices in a way that protects them from any jeopardy, including hiring lawyers and other policy experts to craft and review them. Given their length and legal jargon, research shows that privacy policies are neither well understood~\cite{reidenberg:2015} nor actively reviewed by most users~\cite{jensen:2004}.
In contrast, privacy labels are now forward-facing and published directly on the App Store without needing to follow any links to review. Recent results by Garg et al.~\cite{Garg:2022} have even suggested that privacy labels can reduce app demand in cases of collecting sensitive information. The incentive for privacy labels may be an economic rather than a legal one, and these diverging incentive models may help explain some of the large differences we observed between privacy policies and privacy labels.
This setup may change, and it is reasonable to consider that privacy labels should face the same regulatory scrutiny as privacy policies due to their role. One could also argue that Apple can expand privacy labels to include more explicit details about data collection behaviors, some of which may indeed be crucial to users for making meaningful and informed decisions about whether to install an app on their computing devices. However, we need balance as adding too much information contradicts the goal of privacy labels to provide a succinct and readable description of the app behavior without needing to read the privacy policy. 

\textit{Takeaway.} Unfortunately, privacy labels appear to suffer from the transparency paradox~\cite{Nissenbaum:2011}: the inherent conflict between the transparency of textual meaning and the transparency of data-handling practices.

\paragraph{\textbf{Improved NLP Models for Privacy Labels.}} Classification approaches~\cite{Harkous:2018,Zimmeck:2019,Wilson:2016} offer much promise in helping to verify additional labeling of apps, like privacy labels. However, these approaches have several shortcomings as researchers did not design them for this task. Foremost, the analysis process is on a per-segment basis, which is helpful in inferring practices that policies completely describe in individual segments. However, policies often describe practices in parts that automated frameworks do not correctly capture across multiple segments. This shortcoming is partly due to the models' design and training data (OPP-115 dataset~\cite{Wilson:2016}), which researchers labeled on a per-segment basis.
Additionally, given that services, including Google in Android~\cite{Google:2022}, are adopting privacy labels more broadly, it may be time to update the models and training data to reflect privacy labels as the outcome. For example, the OPP-115 dataset could be re-annotated with privacy labels, forming the basis for new NLP models and more reliable tools to assist developers, researchers, and regulators better.

\textit{Takeaway.} The community needs new datasets that align with the taxonomies used by Apple and Google. We also need stronger NLP approaches that can consider cross-segment contexts in privacy policies and thus comprehensively extract the nuances of data collection practices highlighted within the free-form text. 

\paragraph{\textbf{Regulation and Legal Compliance.}} Apple requires developers to create a single privacy label for all regions and all users of an app. The App Store does not allow developers to explicitly comply with region-specific (GDPR, CCPA) and age-specific (COPPA) laws. Instead, it encourages developers to create a single, universal label that is either too extensive or too sparse --- neither version accurately represents a user's experience. Further, in the absence of vetting from Apple, the responsibility for accuracy solely lies with app developers. The existing structure of the ecosystem helps the App Store appear to care about user privacy but absolves Apple of responsibility for inaccuracy and disinformation. 

\paragraph{\textbf{Recommendations for Apple.}} With recent studies highlighting that privacy labels are hard to understand~\cite{li-understanding-2022,Zhang:2022}, Apple could reconsider the taxonomy and descriptions of privacy labels.
Additionally, Apple's lack of obvious vetting or regulation of the privacy labels may not incentivize the creation of accurate labels, particularly without any feedback to developers. Our imperfect framework can provide a first-level check for developers to consider more comprehensive arrays of labels for their apps. With Apple imposing a short embargo to review new apps before posting to the store, the platform could also incorporate some form of policy-based analysis into the review process.

% \input{conclusion}

%-------------------------------------------------------------------------------
\section*{Acknowledgments}
%-------------------------------------------------------------------------------

We thank the anonymous reviewers for their helpful comments. This material is based upon work supported by the United States National Science Foundation under Grant Nos. 2247952 and 2247953.

% %-------------------------------------------------------------------------------
% \section*{Availability}
% %-------------------------------------------------------------------------------

% USENIX program committees give extra points to submissions that are
% backed by artifacts that are publicly available. If you made your code
% or data available, it's worth mentioning this fact in a dedicated
% section.

%-------------------------------------------------------------------------------
\begin{footnotesize}
\bibliographystyle{ACM-Reference-Format}
\bibliography{main}


\begin{thebibliography}{83}

%%% the Web address.

\ifx \showCODEN    \undefined \def \showCODEN     #1{\unskip}     \fi
\ifx \showDOI      \undefined \def \showDOI       #1{#1}\fi
\ifx \showISBNx    \undefined \def \showISBNx     #1{\unskip}     \fi
\ifx \showISBNxiii \undefined \def \showISBNxiii  #1{\unskip}     \fi
\ifx \showISSN     \undefined \def \showISSN      #1{\unskip}     \fi
\ifx \showLCCN     \undefined \def \showLCCN      #1{\unskip}     \fi
\ifx \shownote     \undefined \def \shownote      #1{#1}          \fi
\ifx \showarticletitle \undefined \def \showarticletitle #1{#1}   \fi
\ifx \showURL      \undefined \def \showURL       {\relax}        \fi
% invisible to TeX
\providecommand\bibfield[2]{#2}
\providecommand\bibinfo[2]{#2}
\providecommand\natexlab[1]{#1}
\providecommand\showeprint[2][]{arXiv:#2}

\bibitem[Sub(2020a)]%
        {Subsplash}
 \bibinfo{year}{2020}\natexlab{a}.
\newblock \bibinfo{title}{Subsplash {\textbar} {Church} mobile apps, websites,
  media, giving \& more}.
\newblock
\newblock
\urldef\tempurl%
\url{https://www.subsplash.com/}
\showURL{%
\tempurl}


\bibitem[Sub(2020b)]%
        {Subsplash:Policy}
 \bibinfo{year}{2020}\natexlab{b}.
\newblock \bibinfo{title}{{Subsplash Privacy Policy}}.
\newblock
\newblock
\urldef\tempurl%
\url{https://www.subsplash.com/legal/privacy}
\showURL{%
\tempurl}


\bibitem[Sub(2022)]%
        {Subsplash:DappyTKeys}
 \bibinfo{year}{2022}\natexlab{}.
\newblock \bibinfo{title}{{DappyTKeys}}.
\newblock
\newblock
\urldef\tempurl%
\url{https://apps.apple.com/us/app/dappytkeys/id1536818077}
\showURL{%
\tempurl}


\bibitem[Eas(2023)]%
        {EasyList}
 \bibinfo{year}{2023}\natexlab{}.
\newblock \bibinfo{title}{{EasyList}}.
\newblock
\newblock
\urldef\tempurl%
\url{https://easylist.to/}
\showURL{%
\tempurl}


\bibitem[Acar et~al\mbox{.}(2014)]%
        {Acar:2014}
\bibfield{author}{\bibinfo{person}{Gunes Acar}, \bibinfo{person}{Christian
  Eubank}, \bibinfo{person}{Steven Englehardt}, \bibinfo{person}{Marc Juarez},
  \bibinfo{person}{Arvind Narayanan}, {and} \bibinfo{person}{Claudia Diaz}.}
  \bibinfo{year}{2014}\natexlab{}.
\newblock \showarticletitle{The Web Never Forgets: Persistent Tracking
  Mechanisms in the Wild}. In \bibinfo{booktitle}{\emph{Proceedings of the 2014
  ACM SIGSAC Conference on Computer and Communications Security}} (Scottsdale,
  Arizona, USA) \emph{(\bibinfo{series}{CCS '14})}.
  \bibinfo{publisher}{Association for Computing Machinery},
  \bibinfo{address}{New York, NY, USA}, \bibinfo{pages}{674–689}.
\newblock
\showISBNx{9781450329576}
\urldef\tempurl%
\url{https://doi.org/10.1145/2660267.2660347}
\showDOI{\tempurl}


\bibitem[{ALDI International Services GmbH \& Co. oHG}(2023)]%
        {Aldi}
\bibfield{author}{\bibinfo{person}{{ALDI International Services GmbH \& Co.
  oHG}}.} \bibinfo{year}{2023}\natexlab{}.
\newblock \bibinfo{title}{{ALDI USA}}.
\newblock
\newblock
\urldef\tempurl%
\url{https://apps.apple.com/us/app/aldi-usa/id429396645}
\showURL{%
\tempurl}


\bibitem[{Aldi US}(2023)]%
        {Aldi:Policy}
\bibfield{author}{\bibinfo{person}{{Aldi US}}.}
  \bibinfo{year}{2023}\natexlab{}.
\newblock \bibinfo{title}{Aldi: {U.S.} {Privacy} {Policy}}.
\newblock
\newblock
\urldef\tempurl%
\url{https://www.aldi.us/en/online-privacy-notice/}
\showURL{%
\tempurl}


\bibitem[Andow et~al\mbox{.}(2019)]%
        {Andow:2019}
\bibfield{author}{\bibinfo{person}{Benjamin Andow},
  \bibinfo{person}{Samin~Yaseer Mahmud}, \bibinfo{person}{Wenyu Wang},
  \bibinfo{person}{Justin Whitaker}, \bibinfo{person}{William Enck},
  \bibinfo{person}{Bradley Reaves}, \bibinfo{person}{Kapil Singh}, {and}
  \bibinfo{person}{Tao Xie}.} \bibinfo{year}{2019}\natexlab{}.
\newblock \showarticletitle{{PolicyLint}: Investigating Internal Privacy Policy
  Contradictions on Google Play}. In \bibinfo{booktitle}{\emph{28th USENIX
  Security Symposium (USENIX Security 19)}}. \bibinfo{publisher}{USENIX
  Association}, \bibinfo{address}{Santa Clara, CA}, \bibinfo{pages}{585--602}.
\newblock
\showISBNx{978-1-939133-06-9}
\urldef\tempurl%
\url{https://www.usenix.org/conference/usenixsecurity19/presentation/andow}
\showURL{%
\tempurl}


\bibitem[Andow et~al\mbox{.}(2020)]%
        {Andow:2020}
\bibfield{author}{\bibinfo{person}{Benjamin Andow},
  \bibinfo{person}{Samin~Yaseer Mahmud}, \bibinfo{person}{Justin Whitaker},
  \bibinfo{person}{William Enck}, \bibinfo{person}{Bradley Reaves},
  \bibinfo{person}{Kapil Singh}, {and} \bibinfo{person}{Serge Egelman}.}
  \bibinfo{year}{2020}\natexlab{}.
\newblock \showarticletitle{Actions Speak Louder than Words: {Entity-Sensitive}
  Privacy Policy and Data Flow Analysis with {PoliCheck}}. In
  \bibinfo{booktitle}{\emph{29th USENIX Security Symposium (USENIX Security
  20)}}. \bibinfo{publisher}{USENIX Association}, \bibinfo{pages}{985--1002}.
\newblock
\showISBNx{978-1-939133-17-5}
\urldef\tempurl%
\url{https://www.usenix.org/conference/usenixsecurity20/presentation/andow}
\showURL{%
\tempurl}


\bibitem[Apple(2020)]%
        {Apple:Guidelines}
\bibfield{author}{\bibinfo{person}{Apple}.} \bibinfo{year}{2020}\natexlab{}.
\newblock \bibinfo{title}{App {Store} {Review} {Guidelines} - {Apple}
  {Developer}}.
\newblock
\newblock
\urldef\tempurl%
\url{https://developer.apple.com/app-store/review/guidelines/}
\showURL{%
Retrieved 2022-05-15 from \tempurl}


\bibitem[Apple(2022a)]%
        {Apple:AgeRatings}
\bibfield{author}{\bibinfo{person}{Apple}.} \bibinfo{year}{2022}\natexlab{a}.
\newblock \bibinfo{title}{Age {Ratings} - {Apple} {Developer}}.
\newblock
\newblock
\urldef\tempurl%
\url{https://developer.apple.com/help/app-store-connect/reference/age-ratings/}
\showURL{%
Retrieved 2023-05-23 from \tempurl}


\bibitem[Apple(2022b)]%
        {AppleCatalog}
\bibfield{author}{\bibinfo{person}{Apple}.} \bibinfo{year}{2022}\natexlab{b}.
\newblock \bibinfo{title}{{Apple Catalog API}}.
\newblock
\newblock
\urldef\tempurl%
\url{https://amp-api.apps.apple.com/v1/catalog/}
\showURL{%
Retrieved 2022-10-11 from \tempurl}


\bibitem[Apps(2022)]%
        {Subsplash:FamilyLife}
\bibfield{author}{\bibinfo{person}{FamilyLife Apps}.}
  \bibinfo{year}{2022}\natexlab{}.
\newblock \bibinfo{title}{{FamilyLife}®}.
\newblock
\newblock
\urldef\tempurl%
\url{https://apps.apple.com/us/app/familylife/id903170704}
\showURL{%
\tempurl}


\bibitem[AppsFlyer(2023)]%
        {Appsflyer:Guidelines}
\bibfield{author}{\bibinfo{person}{AppsFlyer}.}
  \bibinfo{year}{2023}\natexlab{}.
\newblock \bibinfo{title}{Preparing for the {App} {Store} review—nutrition
  labels}.
\newblock
\newblock
\urldef\tempurl%
\url{https://support.appsflyer.com/hc/en-us/articles/207032086-Preparing-for-the-App-Store-review-nutrition-labels}
\showURL{%
\tempurl}


\bibitem[Balash et~al\mbox{.}(2022)]%
        {Balash:2022}
\bibfield{author}{\bibinfo{person}{David~G Balash}, \bibinfo{person}{Mir~Masood
  Ali}, \bibinfo{person}{Xiaoyuan Wu}, \bibinfo{person}{Chris Kanich}, {and}
  \bibinfo{person}{Adam~J Aviv}.} \bibinfo{year}{2022}\natexlab{}.
\newblock \showarticletitle{Longitudinal Analysis of Privacy Labels in the
  Apple App Store}.
\newblock \bibinfo{journal}{\emph{arXiv preprint arXiv:2206.02658}}
  (\bibinfo{year}{2022}).
\newblock
\urldef\tempurl%
\url{https://doi.org/10.48550/arXiv.2206.02658}
\showURL{%
\tempurl}


\bibitem[Balebako et~al\mbox{.}(2015)]%
        {Balebako:2015}
\bibfield{author}{\bibinfo{person}{Rebecca Balebako}, \bibinfo{person}{Florian
  Schaub}, \bibinfo{person}{Idris Adjerid}, \bibinfo{person}{Alessandro
  Acquisti}, {and} \bibinfo{person}{Lorrie Cranor}.}
  \bibinfo{year}{2015}\natexlab{}.
\newblock \showarticletitle{The Impact of Timing on the Salience of Smartphone
  App Privacy Notices}. In \bibinfo{booktitle}{\emph{Proceedings of the 5th
  Annual ACM CCS Workshop on Security and Privacy in Smartphones and Mobile
  Devices}} (Denver, Colorado, USA) \emph{(\bibinfo{series}{SPSM '15})}.
  \bibinfo{publisher}{Association for Computing Machinery},
  \bibinfo{address}{New York, NY, USA}, \bibinfo{pages}{63–74}.
\newblock
\showISBNx{9781450338196}
\urldef\tempurl%
\url{https://doi.org/10.1145/2808117.2808119}
\showDOI{\tempurl}


\bibitem[{Braincake}(2023)]%
        {PregnancyTracker:Policy}
\bibfield{author}{\bibinfo{person}{{Braincake}}.}
  \bibinfo{year}{2023}\natexlab{}.
\newblock \bibinfo{title}{{Privacy} {Policy}}.
\newblock
\newblock
\urldef\tempurl%
\url{http://braincake.net/pregnancytracker_privacy.html}
\showURL{%
\tempurl}


\bibitem[Breaux and Rao(2013)]%
        {Breaux:2013}
\bibfield{author}{\bibinfo{person}{Travis~D. Breaux} {and}
  \bibinfo{person}{Ashwini Rao}.} \bibinfo{year}{2013}\natexlab{}.
\newblock \showarticletitle{Formal analysis of privacy requirements
  specifications for multi-tier applications}. In
  \bibinfo{booktitle}{\emph{2013 21st IEEE International Requirements
  Engineering Conference (RE)}}. \bibinfo{pages}{14--23}.
\newblock
\urldef\tempurl%
\url{https://doi.org/10.1109/RE.2013.6636701}
\showDOI{\tempurl}


\bibitem[Bui et~al\mbox{.}(2021)]%
        {Bui:2021}
\bibfield{author}{\bibinfo{person}{Duc Bui}, \bibinfo{person}{Yuan Yao},
  \bibinfo{person}{Kang~G. Shin}, \bibinfo{person}{Jong-Min Choi}, {and}
  \bibinfo{person}{Junbum Shin}.} \bibinfo{year}{2021}\natexlab{}.
\newblock \showarticletitle{Consistency Analysis of Data-Usage Purposes in
  Mobile Apps}. In \bibinfo{booktitle}{\emph{Proceedings of the 2021 ACM SIGSAC
  Conference on Computer and Communications Security}} (Virtual Event, Republic
  of Korea) \emph{(\bibinfo{series}{CCS '21})}. \bibinfo{publisher}{Association
  for Computing Machinery}, \bibinfo{address}{New York, NY, USA},
  \bibinfo{pages}{2824–2843}.
\newblock
\showISBNx{9781450384544}
\urldef\tempurl%
\url{https://doi.org/10.1145/3460120.3484536}
\showDOI{\tempurl}


\bibitem[{Center for Food Safety and Applied Nutrition}(2022)]%
        {nutrition-how-2022}
\bibfield{author}{\bibinfo{person}{{Center for Food Safety and Applied
  Nutrition}}.} \bibinfo{year}{2022}\natexlab{}.
\newblock \bibinfo{booktitle}{\emph{How to {Understand} and {Use} the
  {Nutrition} {Facts} {Label}}}.
\newblock U.S. Food and Drug Administration.
\newblock
\urldef\tempurl%
\url{https://www.fda.gov/food/new-nutrition-facts-label/how-understand-and-use-nutrition-facts-label}
\showURL{%
\tempurl}
\newblock
\shownote{Publisher: FDA}.


\bibitem[Chen et~al\mbox{.}(2019)]%
        {Chen:2019}
\bibfield{author}{\bibinfo{person}{Yi Chen}, \bibinfo{person}{Mingming Zha},
  \bibinfo{person}{Nan Zhang}, \bibinfo{person}{Dandan Xu},
  \bibinfo{person}{Qianqian Zhao}, \bibinfo{person}{Xuan Feng},
  \bibinfo{person}{Kan Yuan}, \bibinfo{person}{Fnu Suya}, \bibinfo{person}{Yuan
  Tian}, \bibinfo{person}{Kai Chen}, \bibinfo{person}{XiaoFeng Wang}, {and}
  \bibinfo{person}{Wei Zou}.} \bibinfo{year}{2019}\natexlab{}.
\newblock \showarticletitle{Demystifying Hidden Privacy Settings in Mobile
  Apps}. In \bibinfo{booktitle}{\emph{2019 IEEE Symposium on Security and
  Privacy (SP)}}. \bibinfo{pages}{570--586}.
\newblock
\urldef\tempurl%
\url{https://doi.org/10.1109/SP.2019.00054}
\showDOI{\tempurl}


\bibitem[ChowNow(2022)]%
        {ChowNow:Bagelman}
\bibfield{author}{\bibinfo{person}{ChowNow}.} \bibinfo{year}{2022}\natexlab{}.
\newblock \bibinfo{title}{{Bagelman}}.
\newblock
\newblock
\urldef\tempurl%
\url{https://apps.apple.com/us/app/bagelman/id1254203081}
\showURL{%
\tempurl}


\bibitem[{ChowNow}(2022)]%
        {ChowNow:ElCharrito}
\bibfield{author}{\bibinfo{person}{{ChowNow}}.}
  \bibinfo{year}{2022}\natexlab{}.
\newblock \bibinfo{title}{{El Charrito}}.
\newblock
\newblock
\urldef\tempurl%
\url{https://apps.apple.com/us/app/el-charrito/id1080457082}
\showURL{%
\tempurl}


\bibitem[Chownow(2022a)]%
        {ChowNow}
\bibfield{author}{\bibinfo{person}{Chownow}.} \bibinfo{year}{2022}\natexlab{a}.
\newblock \bibinfo{title}{Online Food Ordering System for Restaurants}.
\newblock
\newblock
\urldef\tempurl%
\url{https://get.chownow.com/}
\showURL{%
\tempurl}


\bibitem[Chownow(2022b)]%
        {ChowNow:Policy}
\bibfield{author}{\bibinfo{person}{Chownow}.} \bibinfo{year}{2022}\natexlab{b}.
\newblock \bibinfo{title}{Privacy Policy - ChowNow}.
\newblock
\newblock
\urldef\tempurl%
\url{https://get.chownow.com/privacy-policy}
\showURL{%
\tempurl}


\bibitem[Cortesi et~al\mbox{.}(10  )]%
        {mitmproxy}
\bibfield{author}{\bibinfo{person}{Aldo Cortesi}, \bibinfo{person}{Maximilian
  Hils}, \bibinfo{person}{Thomas Kriechbaumer}, {and}
  \bibinfo{person}{contributors}.} \bibinfo{year}{2010--}\natexlab{}.
\newblock \bibinfo{title}{{mitmproxy}: A free and open source interactive
  {HTTPS} proxy}.
\newblock
\newblock
\urldef\tempurl%
\url{https://mitmproxy.org/}
\showURL{%
\tempurl}
\newblock
\shownote{[Version 9.0]}.


\bibitem[Cranor(2012)]%
        {Cranor:2012}
\bibfield{author}{\bibinfo{person}{Lorrie~Faith Cranor}.}
  \bibinfo{year}{2012}\natexlab{}.
\newblock \showarticletitle{Necessary but not sufficient: Standardized
  mechanisms for privacy notice and choice}.
\newblock \bibinfo{journal}{\emph{J. on Telecomm. \& High Tech. L.}}
  \bibinfo{volume}{10} (\bibinfo{year}{2012}), \bibinfo{pages}{273}.
\newblock
\urldef\tempurl%
\url{http://jthtl.org/content/articles/V10I2/JTHTLv10i2_Cranor.PDF}
\showURL{%
\tempurl}


\bibitem[Cranor et~al\mbox{.}(2014)]%
        {Cranor:2014}
\bibfield{author}{\bibinfo{person}{Lorrie~Faith Cranor},
  \bibinfo{person}{Candice Hoke}, \bibinfo{person}{Pedro Leon}, {and}
  \bibinfo{person}{Alyssa Au}.} \bibinfo{year}{2014}\natexlab{}.
\newblock \bibinfo{booktitle}{\emph{Are {They} {Worth} {Reading}? {An}
  {In}-{Depth} {Analysis} of {Online} {Advertising} {Companies}’ {Privacy}
  {Policies}}}.
\newblock \bibinfo{type}{{SSRN} {Scholarly} {Paper}} ID 2418590.
  \bibinfo{institution}{Social Science Research Network},
  \bibinfo{address}{Rochester, NY}.
\newblock
\urldef\tempurl%
\url{https://papers.ssrn.com/abstract=2418590}
\showURL{%
\tempurl}


\bibitem[{Credit Karma, Inc.}(2022)]%
        {CreditKarma}
\bibfield{author}{\bibinfo{person}{{Credit Karma, Inc.}}}
  \bibinfo{year}{2022}\natexlab{}.
\newblock \bibinfo{title}{{Credit Karma on the App Store}}.
\newblock
\newblock
\urldef\tempurl%
\url{https://apps.apple.com/us/app/credit-karma/id519817714}
\showURL{%
\tempurl}


\bibitem[Developer(2022a)]%
        {Apple:PrivacyLabelDetails}
\bibfield{author}{\bibinfo{person}{Apple Developer}.}
  \bibinfo{year}{2022}\natexlab{a}.
\newblock \bibinfo{title}{App {Privacy} {Details} - {App} {Store}}.
\newblock
\newblock
\urldef\tempurl%
\url{https://developer.apple.com/app-store/app-privacy-details/}
\showURL{%
\tempurl}


\bibitem[Developer(2022b)]%
        {Apple:LabelsLive}
\bibfield{author}{\bibinfo{person}{Apple Developer}.}
  \bibinfo{year}{2022}\natexlab{b}.
\newblock \bibinfo{title}{App privacy labels now live on the {App} {Store} -
  {Latest} {News} - {Apple} {Developer}}.
\newblock
\newblock
\urldef\tempurl%
\url{https://developer.apple.com/news/?id=3wann9gh}
\showURL{%
\tempurl}


\bibitem[Emami-Naeini et~al\mbox{.}(2020)]%
        {emami-naeini-ask-2020}
\bibfield{author}{\bibinfo{person}{Pardis Emami-Naeini},
  \bibinfo{person}{Yuvraj Agarwal}, \bibinfo{person}{Lorrie Faith~Cranor},
  {and} \bibinfo{person}{Hanan Hibshi}.} \bibinfo{year}{2020}\natexlab{}.
\newblock \showarticletitle{Ask the {Experts}: {What} {Should} {Be} on an {IoT}
  {Privacy} and {Security} {Label}?}. In \bibinfo{booktitle}{\emph{2020 {IEEE}
  {Symposium} on {Security} and {Privacy} ({SP})}}. \bibinfo{publisher}{IEEE},
  \bibinfo{address}{San Jose, CA, USA}, \bibinfo{pages}{447--464}.
\newblock
\urldef\tempurl%
\url{https://doi.org/10.1109/SP40000.2020.00043}
\showDOI{\tempurl}
\newblock
\shownote{ISSN: 2375-1207}.


\bibitem[Emami-Naeini et~al\mbox{.}(2021)]%
        {emami-naeini-which-2021}
\bibfield{author}{\bibinfo{person}{Pardis Emami-Naeini},
  \bibinfo{person}{Janarth Dheenadhayalan}, \bibinfo{person}{Yuvraj Agarwal},
  {and} \bibinfo{person}{Lorrie~Faith Cranor}.}
  \bibinfo{year}{2021}\natexlab{}.
\newblock \showarticletitle{Which {Privacy} and {Security} {Attributes} {Most}
  {Impact} {Consumers}’ {Risk} {Perception} and {Willingness} to {Purchase}
  {IoT} {Devices}?}. In \bibinfo{booktitle}{\emph{2021 {IEEE} {Symposium} on
  {Security} and {Privacy} ({SP})}}. \bibinfo{publisher}{IEEE},
  \bibinfo{address}{San Francisco, CA, USA}, \bibinfo{pages}{519--536}.
\newblock
\showISBNx{978-1-72818-934-5}
\urldef\tempurl%
\url{https://doi.org/10.1109/SP40001.2021.00112}
\showDOI{\tempurl}


\bibitem[Englehardt and Narayanan(2016)]%
        {Englehardt:2016}
\bibfield{author}{\bibinfo{person}{Steven Englehardt} {and}
  \bibinfo{person}{Arvind Narayanan}.} \bibinfo{year}{2016}\natexlab{}.
\newblock \showarticletitle{Online Tracking: A 1-Million-Site Measurement and
  Analysis}. In \bibinfo{booktitle}{\emph{Proceedings of the 2016 ACM SIGSAC
  Conference on Computer and Communications Security}} (Vienna, Austria)
  \emph{(\bibinfo{series}{CCS '16})}. \bibinfo{publisher}{Association for
  Computing Machinery}, \bibinfo{address}{New York, NY, USA},
  \bibinfo{pages}{1388–1401}.
\newblock
\showISBNx{9781450341394}
\urldef\tempurl%
\url{https://doi.org/10.1145/2976749.2978313}
\showDOI{\tempurl}


\bibitem[{Federal Trade Commission}(2000)]%
        {FTC:FIPP:2000}
\bibfield{author}{\bibinfo{person}{{Federal Trade Commission}}.}
  \bibinfo{year}{2000}\natexlab{}.
\newblock \bibinfo{title}{Privacy {Online}: {Fair} {Information} {Practices} in
  the {Electronic} {Marketplace}: {A} {Federal} {Trade} {Commission} {Report}
  to {Congress}}.
\newblock
\newblock
\urldef\tempurl%
\url{https://www.ftc.gov/sites/default/files/documents/reports/privacy-online-report-congress/priv-23a.pdf}
\showURL{%
\tempurl}


\bibitem[{Federal Trade Commission}(2013)]%
        {FTC:COPPA:2013}
\bibfield{author}{\bibinfo{person}{{Federal Trade Commission}}.}
  \bibinfo{year}{2013}\natexlab{}.
\newblock \bibinfo{title}{{Children's Online Privacy Protection Rule (COPPA)}}.
\newblock
\newblock
\urldef\tempurl%
\url{https://www.ftc.gov/legal-library/browse/rules/childrens-online-privacy-protection-rule-coppa}
\showURL{%
\tempurl}


\bibitem[{Federal Trade Commission}(2022)]%
        {FTC:Safeguards:2022}
\bibfield{author}{\bibinfo{person}{{Federal Trade Commission}}.}
  \bibinfo{year}{2022}\natexlab{}.
\newblock \bibinfo{title}{{FTC} {Safeguards} {Rule}: {What} {Your} {Business}
  {Needs} to {Know}}.
\newblock
\newblock
\urldef\tempurl%
\url{https://www.ftc.gov/business-guidance/resources/ftc-safeguards-rule-what-your-business-needs-know}
\showURL{%
\tempurl}


\bibitem[{Fitness Labs}(2023)]%
        {PregnancyTracker}
\bibfield{author}{\bibinfo{person}{{Fitness Labs}}.}
  \bibinfo{year}{2023}\natexlab{}.
\newblock \bibinfo{title}{{Pregnancy} {Tracker}: {Baby} {Bump}}.
\newblock
\newblock
\urldef\tempurl%
\url{https://apps.apple.com/us/app/pregnancy-tracker-baby-bump/id1453373942}
\showURL{%
\tempurl}


\bibitem[Gardner et~al\mbox{.}(2022)]%
        {Gardner:2022}
\bibfield{author}{\bibinfo{person}{Jack Gardner}, \bibinfo{person}{Yuanyuan
  Feng}, \bibinfo{person}{Kayla Reiman}, \bibinfo{person}{Zhi Lin},
  \bibinfo{person}{Akshath Jain}, {and} \bibinfo{person}{Norman Sadeh}.}
  \bibinfo{year}{2022}\natexlab{}.
\newblock \showarticletitle{Helping Mobile Application Developers Create
  Accurate Privacy Labels}. In \bibinfo{booktitle}{\emph{2022 IEEE European
  Symposium on Security and Privacy Workshops (EuroS\&PW)}}.
  \bibinfo{pages}{212--230}.
\newblock
\urldef\tempurl%
\url{https://doi.org/10.1109/EuroSPW55150.2022.00028}
\showDOI{\tempurl}


\bibitem[Garg and Telangb(2022)]%
        {Garg:2022}
\bibfield{author}{\bibinfo{person}{Rajiv Garg} {and} \bibinfo{person}{Rahul
  Telangb}.} \bibinfo{year}{2022}\natexlab{}.
\newblock \showarticletitle{Impact of App Privacy Label Disclosure on Demand:
  An Empirical Analysis}.
\newblock \bibinfo{journal}{\emph{Workshop on the Economics of Information
  Security (WEIS)}} (\bibinfo{year}{2022}).
\newblock
\urldef\tempurl%
\url{https://weis2022.econinfosec.org/wp-content/uploads/sites/10/2022/06/weis22-telang.pdf}
\showURL{%
\tempurl}


\bibitem[Ghostery(2023)]%
        {WhoTrackMe:Site}
\bibfield{author}{\bibinfo{person}{Ghostery}.} \bibinfo{year}{2023}\natexlab{}.
\newblock \bibinfo{title}{{WhoTracks.me - Bringing Transparency to Online
  Tracking}}.
\newblock
\newblock
\urldef\tempurl%
\url{https://whotracks.me/}
\showURL{%
\tempurl}


\bibitem[Google(2020)]%
        {Google:2022}
\bibfield{author}{\bibinfo{person}{Google}.} \bibinfo{year}{2020}\natexlab{}.
\newblock \bibinfo{title}{Provide information for {Google} {Play}'s {Data}
  safety section - {Play} {Console} {Help}}.
\newblock
\newblock
\urldef\tempurl%
\url{https://support.google.com/googleplay/android-developer/answer/10787469}
\showURL{%
Retrieved 2022-05-15 from \tempurl}


\bibitem[Google(2023)]%
        {GoogleAnalytics:Disclosures}
\bibfield{author}{\bibinfo{person}{Google}.} \bibinfo{year}{2023}\natexlab{}.
\newblock \bibinfo{title}{Privacy {Disclosures} {Policy}}.
\newblock
\newblock
\urldef\tempurl%
\url{https://support.google.com/analytics/answer/7318509}
\showURL{%
\tempurl}


\bibitem[Han et~al\mbox{.}(2019)]%
        {Han:2019}
\bibfield{author}{\bibinfo{person}{Catherine Han}, \bibinfo{person}{Irwin
  Reyes}, \bibinfo{person}{Amit Elazari Bar~On}, \bibinfo{person}{Joel
  Reardon}, \bibinfo{person}{{\'A}lvaro Feal}, \bibinfo{person}{Serge Egelman},
  \bibinfo{person}{Narseo Vallina-Rodriguez}, {et~al\mbox{.}}}
  \bibinfo{year}{2019}\natexlab{}.
\newblock \showarticletitle{Do you get what you pay for? comparing the privacy
  behaviors of free vs. paid apps}. In \bibinfo{booktitle}{\emph{Workshop on
  Technology and Consumer Protection (ConPro 2019), in conjunction with the
  39th IEEE Symposium on Security and Privacy, 23 May 2019, San Francisco, CA,
  USA.}} \bibinfo{publisher}{IEEE}, \bibinfo{address}{San Francisco, CA, USA},
  \bibinfo{numpages}{7}~pages.
\newblock
\urldef\tempurl%
\url{https://www.ieee-security.org/TC/SPW2019/ConPro/papers/han-conpro19.pdf}
\showURL{%
\tempurl}


\bibitem[Han et~al\mbox{.}(2020)]%
        {Han:2020}
\bibfield{author}{\bibinfo{person}{Catherine Han}, \bibinfo{person}{Irwin
  Reyes}, \bibinfo{person}{{\'{A}}lvaro Feal}, \bibinfo{person}{Joel Reardon},
  \bibinfo{person}{Primal Wijesekera}, \bibinfo{person}{Narseo
  Vallina{-}Rodriguez}, \bibinfo{person}{Amit Elazari~Bar On},
  \bibinfo{person}{Kenneth~A. Bamberger}, {and} \bibinfo{person}{Serge
  Egelman}.} \bibinfo{year}{2020}\natexlab{}.
\newblock \showarticletitle{The Price is (Not) Right: Comparing Privacy in Free
  and Paid Apps}.
\newblock \bibinfo{journal}{\emph{Proc. Priv. Enhancing Technol.}}
  \bibinfo{volume}{2020}, \bibinfo{number}{3} (\bibinfo{year}{2020}),
  \bibinfo{pages}{222--242}.
\newblock
\urldef\tempurl%
\url{https://doi.org/10.2478/popets-2020-0050}
\showDOI{\tempurl}


\bibitem[Harkous et~al\mbox{.}(2018)]%
        {Harkous:2018}
\bibfield{author}{\bibinfo{person}{Hamza Harkous}, \bibinfo{person}{Kassem
  Fawaz}, \bibinfo{person}{R{\'e}mi Lebret}, \bibinfo{person}{Florian Schaub},
  \bibinfo{person}{Kang~G. Shin}, {and} \bibinfo{person}{Karl Aberer}.}
  \bibinfo{year}{2018}\natexlab{}.
\newblock \showarticletitle{Polisis: Automated Analysis and Presentation of
  Privacy Policies Using Deep Learning}. In \bibinfo{booktitle}{\emph{27th
  USENIX Security Symposium (USENIX Security 18)}}. \bibinfo{publisher}{USENIX
  Association}, \bibinfo{address}{Baltimore, MD}, \bibinfo{pages}{531--548}.
\newblock
\showISBNx{978-1-939133-04-5}
\urldef\tempurl%
\url{https://www.usenix.org/conference/usenixsecurity18/presentation/harkous}
\showURL{%
\tempurl}


\bibitem[{HyperBeard Games}(2023)]%
        {Axolochi:Policy}
\bibfield{author}{\bibinfo{person}{{HyperBeard Games}}.}
  \bibinfo{year}{2023}\natexlab{}.
\newblock \bibinfo{title}{Privacy {Policy}: {HyperBeard} {Games}}.
\newblock
\newblock
\urldef\tempurl%
\url{https://docs.google.com/document/d/1bJpQtjxoxmf1leATVgnRbZWkV9W-iDngLx_pTsvu8Vk/edit}
\showURL{%
\tempurl}


\bibitem[Inc.(2023)]%
        {Axolochi}
\bibfield{author}{\bibinfo{person}{HyperBeard Inc.}}
  \bibinfo{year}{2023}\natexlab{}.
\newblock \bibinfo{title}{Axolochi}.
\newblock
\newblock
\urldef\tempurl%
\url{https://apps.apple.com/us/app/axolochi/id1432184360}
\showURL{%
\tempurl}


\bibitem[Jensen and Potts(2004)]%
        {jensen:2004}
\bibfield{author}{\bibinfo{person}{Carlos Jensen} {and} \bibinfo{person}{Colin
  Potts}.} \bibinfo{year}{2004}\natexlab{}.
\newblock \showarticletitle{Privacy policies as decision-making tools: an
  evaluation of online privacy notices}. In
  \bibinfo{booktitle}{\emph{Proceedings of the SIGCHI conference on Human
  Factors in Computing Systems}}. \bibinfo{pages}{471--478}.
\newblock


\bibitem[Karaj et~al\mbox{.}(2019)]%
        {Whotracks:2019}
\bibfield{author}{\bibinfo{person}{Arjaldo Karaj}, \bibinfo{person}{Sam
  Macbeth}, \bibinfo{person}{Rémi Berson}, {and} \bibinfo{person}{Josep~M.
  Pujol}.} \bibinfo{year}{2019}\natexlab{}.
\newblock \bibinfo{title}{{WhoTracks} .{Me}: {Shedding} light on the opaque
  world of online tracking}.
\newblock
\newblock
\urldef\tempurl%
\url{https://doi.org/10.48550/arXiv.1804.08959}
\showDOI{\tempurl}
\newblock
\shownote{arXiv:1804.08959 [cs]}.


\bibitem[Kelley et~al\mbox{.}(2009)]%
        {Kelly:2009}
\bibfield{author}{\bibinfo{person}{Patrick~Gage Kelley},
  \bibinfo{person}{Joanna Bresee}, \bibinfo{person}{Lorrie~Faith Cranor}, {and}
  \bibinfo{person}{Robert~W Reeder}.} \bibinfo{year}{2009}\natexlab{}.
\newblock \showarticletitle{A ``nutrition label'' for privacy}. In
  \bibinfo{booktitle}{\emph{Proceedings of the 5th Symposium on Usable Privacy
  and Security}}. \bibinfo{pages}{1--12}.
\newblock


\bibitem[Kelley et~al\mbox{.}(2010)]%
        {kelley-standardizing-2010}
\bibfield{author}{\bibinfo{person}{Patrick~Gage Kelley},
  \bibinfo{person}{Lucian Cesca}, \bibinfo{person}{Joanna Bresee}, {and}
  \bibinfo{person}{Lorrie~Faith Cranor}.} \bibinfo{year}{2010}\natexlab{}.
\newblock \showarticletitle{Standardizing privacy notices: an online study of
  the nutrition label approach}. In \bibinfo{booktitle}{\emph{Proceedings of
  the {SIGCHI} {Conference} on {Human} {Factors} in {Computing} {Systems}}}
  \emph{(\bibinfo{series}{{CHI} '10})}. \bibinfo{publisher}{Association for
  Computing Machinery}, \bibinfo{address}{New York, NY, USA},
  \bibinfo{pages}{1573--1582}.
\newblock
\showISBNx{978-1-60558-929-9}
\urldef\tempurl%
\url{https://doi.org/10.1145/1753326.1753561}
\showDOI{\tempurl}


\bibitem[Kelley et~al\mbox{.}(2013)]%
        {kelley-privacy-2013}
\bibfield{author}{\bibinfo{person}{Patrick~Gage Kelley},
  \bibinfo{person}{Lorrie~Faith Cranor}, {and} \bibinfo{person}{Norman Sadeh}.}
  \bibinfo{year}{2013}\natexlab{}.
\newblock \showarticletitle{Privacy as part of the app decision-making
  process}. In \bibinfo{booktitle}{\emph{Proceedings of the {SIGCHI}
  {Conference} on {Human} {Factors} in {Computing} {Systems}}}.
  \bibinfo{publisher}{ACM}, \bibinfo{address}{Paris France},
  \bibinfo{pages}{3393--3402}.
\newblock
\showISBNx{978-1-4503-1899-0}
\urldef\tempurl%
\url{https://doi.org/10.1145/2470654.2466466}
\showDOI{\tempurl}


\bibitem[Kollnig et~al\mbox{.}(2022)]%
        {kollnig-goodbye-2022}
\bibfield{author}{\bibinfo{person}{Konrad Kollnig}, \bibinfo{person}{Anastasia
  Shuba}, \bibinfo{person}{Max Van~Kleek}, \bibinfo{person}{Reuben Binns},
  {and} \bibinfo{person}{Nigel Shadbolt}.} \bibinfo{year}{2022}\natexlab{}.
\newblock \showarticletitle{Goodbye Tracking? Impact of IOS App Tracking
  Transparency and Privacy Labels}. In \bibinfo{booktitle}{\emph{2022 ACM
  Conference on Fairness, Accountability, and Transparency}} (Seoul, Republic
  of Korea) \emph{(\bibinfo{series}{FAccT '22})}.
  \bibinfo{publisher}{Association for Computing Machinery},
  \bibinfo{address}{New York, NY, USA}, \bibinfo{pages}{508–520}.
\newblock
\showISBNx{9781450393522}
\urldef\tempurl%
\url{https://doi.org/10.1145/3531146.3533116}
\showDOI{\tempurl}


\bibitem[Li et~al\mbox{.}(2022b)]%
        {li-understanding-2022}
\bibfield{author}{\bibinfo{person}{Tianshi Li}, \bibinfo{person}{Kayla Reiman},
  \bibinfo{person}{Yuvraj Agarwal}, \bibinfo{person}{Lorrie~Faith Cranor},
  {and} \bibinfo{person}{Jason~I. Hong}.} \bibinfo{year}{2022}\natexlab{b}.
\newblock \showarticletitle{Understanding Challenges for Developers to Create
  Accurate Privacy Nutrition Labels}. In \bibinfo{booktitle}{\emph{CHI
  Conference on Human Factors in Computing Systems}} (New Orleans, LA, USA)
  \emph{(\bibinfo{series}{CHI '22})}. \bibinfo{publisher}{Association for
  Computing Machinery}, \bibinfo{address}{New York, NY, USA}, Article
  \bibinfo{articleno}{588}, \bibinfo{numpages}{24}~pages.
\newblock
\showISBNx{9781450391573}
\urldef\tempurl%
\url{https://doi.org/10.1145/3491102.3502012}
\showDOI{\tempurl}


\bibitem[Li et~al\mbox{.}(2022a)]%
        {li-chi}
\bibfield{author}{\bibinfo{person}{Yucheng Li}, \bibinfo{person}{Deyuan Chen},
  \bibinfo{person}{Tianshi Li}, \bibinfo{person}{Yuvraj Agarwal},
  \bibinfo{person}{Lorrie~Faith Cranor}, {and} \bibinfo{person}{Jason~I.
  Hong}.} \bibinfo{year}{2022}\natexlab{a}.
\newblock \showarticletitle{Understanding iOS Privacy Nutrition Labels: An
  Exploratory Large-Scale Analysis of App Store Data}. In
  \bibinfo{booktitle}{\emph{Extended Abstracts of the 2022 CHI Conference on
  Human Factors in Computing Systems}} (New Orleans, LA, USA)
  \emph{(\bibinfo{series}{CHI EA '22})}. \bibinfo{publisher}{Association for
  Computing Machinery}, \bibinfo{address}{New York, NY, USA}, Article
  \bibinfo{articleno}{356}, \bibinfo{numpages}{7}~pages.
\newblock
\showISBNx{9781450391566}
\urldef\tempurl%
\url{https://doi.org/10.1145/3491101.3519739}
\showDOI{\tempurl}


\bibitem[Liu et~al\mbox{.}(2019)]%
        {roberta:2019}
\bibfield{author}{\bibinfo{person}{Yinhan Liu}, \bibinfo{person}{Myle Ott},
  \bibinfo{person}{Naman Goyal}, \bibinfo{person}{Jingfei Du},
  \bibinfo{person}{Mandar Joshi}, \bibinfo{person}{Danqi Chen},
  \bibinfo{person}{Omer Levy}, \bibinfo{person}{Mike Lewis},
  \bibinfo{person}{Luke Zettlemoyer}, {and} \bibinfo{person}{Veselin
  Stoyanov}.} \bibinfo{year}{2019}\natexlab{}.
\newblock \bibinfo{title}{{RoBERTa}: {A} {Robustly} {Optimized} {BERT}
  {Pretraining} {Approach}}.
\newblock
\newblock
\urldef\tempurl%
\url{https://doi.org/10.48550/arXiv.1907.11692}
\showDOI{\tempurl}
\newblock
\shownote{arXiv:1907.11692 [cs]}.


\bibitem[McDonald and Cranor(2009)]%
        {McDonald:2009}
\bibfield{author}{\bibinfo{person}{Aleecia~M. McDonald} {and}
  \bibinfo{person}{Lorrie~Faith Cranor}.} \bibinfo{year}{2009}\natexlab{}.
\newblock \showarticletitle{The {Cost} of {Reading} {Privacy} {Policies}}.
\newblock \bibinfo{journal}{\emph{HeinOnline}} \bibinfo{volume}{4},
  \bibinfo{number}{3} (\bibinfo{year}{2009}), \bibinfo{pages}{543--568}.
\newblock
\urldef\tempurl%
\url{https://heinonline.org/HOL/P?h=hein.journals/isjlpsoc4\&i=563}
\showURL{%
\tempurl}


\bibitem[Meta(2022)]%
        {Meta:Guidelines}
\bibfield{author}{\bibinfo{person}{Meta}.} \bibinfo{year}{2022}\natexlab{}.
\newblock \bibinfo{title}{Resources for {Completing} {App} {Store} {Data}
  {Practice} {Questionnaires} for {Apps} {That} {Include} the {Facebook} or
  {Audience} {Network} {SDK}}.
\newblock
\newblock
\urldef\tempurl%
\url{https://developers.facebook.com/blog/post/2022/07/18/resources-for-completing-app-store-data-practice-questionnaires-apps-facebook-or-audience-network-sdk/}
\showURL{%
\tempurl}
\newblock
\shownote{publisher: Meta}.


\bibitem[Mozilla(2020)]%
        {Readability:2020}
\bibfield{author}{\bibinfo{person}{Mozilla}.} \bibinfo{year}{2020}\natexlab{}.
\newblock \bibinfo{title}{{Readability.js}}.
\newblock
\newblock
\urldef\tempurl%
\url{https://github.com/mozilla/readability}
\showURL{%
\tempurl}


\bibitem[Munyendo et~al\mbox{.}(2022)]%
        {Collins:2022}
\bibfield{author}{\bibinfo{person}{Collins~W. Munyendo},
  \bibinfo{person}{Yasemin Acar}, {and} \bibinfo{person}{Adam~J. Aviv}.}
  \bibinfo{year}{2022}\natexlab{}.
\newblock \showarticletitle{Desperate Times Call for Desperate Measures: User
  Concerns with Mobile Loan Apps in Kenya}. In \bibinfo{booktitle}{\emph{2022
  IEEE Symposium on Security and Privacy (SP)}}. \bibinfo{pages}{2304--2319}.
\newblock
\urldef\tempurl%
\url{https://doi.org/10.1109/SP46214.2022.9833779}
\showDOI{\tempurl}


\bibitem[Nissenbaum(2011)]%
        {Nissenbaum:2011}
\bibfield{author}{\bibinfo{person}{Helen Nissenbaum}.}
  \bibinfo{year}{2011}\natexlab{}.
\newblock \showarticletitle{A contextual approach to privacy online}.
\newblock \bibinfo{journal}{\emph{Daedalus}} \bibinfo{volume}{140},
  \bibinfo{number}{4} (\bibinfo{year}{2011}), \bibinfo{pages}{32--48}.
\newblock


\bibitem[of~Google's language-detection library~to Python.(2021)]%
        {LangDetect:2021}
\bibfield{author}{\bibinfo{person}{Port of~Google's language-detection
  library~to Python.}} \bibinfo{year}{2021}\natexlab{}.
\newblock \bibinfo{title}{langdetect}.
\newblock
\newblock
\urldef\tempurl%
\url{https://github.com/Mimino666/langdetect}
\showURL{%
\tempurl}


\bibitem[{Office of the Attorney General of California}(2018)]%
        {CCPA:2018}
\bibfield{author}{\bibinfo{person}{{Office of the Attorney General of
  California}}.} \bibinfo{year}{2018}\natexlab{}.
\newblock \bibinfo{title}{California {Consumer} {Privacy} {Act} ({CCPA})}.
\newblock
\newblock


\bibitem[{Paytronix Systems Inc}(2022)]%
        {ChowNow:HomeState}
\bibfield{author}{\bibinfo{person}{{Paytronix Systems Inc}}.}
  \bibinfo{year}{2022}\natexlab{}.
\newblock \bibinfo{title}{{HomeState, A Texas Kitchen}}.
\newblock
\newblock
\urldef\tempurl%
\url{https://apps.apple.com/us/app/homestate-a-texas-kitchen/id925093380}
\showURL{%
\tempurl}


\bibitem[Reidenberg et~al\mbox{.}(2015)]%
        {reidenberg:2015}
\bibfield{author}{\bibinfo{person}{Joel~R Reidenberg}, \bibinfo{person}{Travis
  Breaux}, \bibinfo{person}{Lorrie~Faith Cranor}, \bibinfo{person}{Brian
  French}, \bibinfo{person}{Amanda Grannis}, \bibinfo{person}{James~T Graves},
  \bibinfo{person}{Fei Liu}, \bibinfo{person}{Aleecia McDonald},
  \bibinfo{person}{Thomas~B Norton}, {and} \bibinfo{person}{Rohan Ramanath}.}
  \bibinfo{year}{2015}\natexlab{}.
\newblock \showarticletitle{Disagreeable privacy policies: Mismatches between
  meaning and users' understanding}.
\newblock \bibinfo{journal}{\emph{Berkeley Tech. LJ}}  \bibinfo{volume}{30}
  (\bibinfo{year}{2015}), \bibinfo{pages}{39}.
\newblock


\bibitem[Reyes et~al\mbox{.}(2018)]%
        {Reyes:2018}
\bibfield{author}{\bibinfo{person}{Irwin Reyes}, \bibinfo{person}{Primal
  Wijesekera}, \bibinfo{person}{Joel Reardon}, \bibinfo{person}{Amit
  Elazari~Bar On}, \bibinfo{person}{Abbas Razaghpanah}, \bibinfo{person}{Narseo
  Vallina-Rodriguez}, {and} \bibinfo{person}{Serge Egelman}.}
  \bibinfo{year}{2018}\natexlab{}.
\newblock \showarticletitle{“Won’t Somebody Think of the Children?”
  Examining COPPA Compliance at Scale}.
\newblock \bibinfo{journal}{\emph{Proceedings on Privacy Enhancing
  Technologies}}  \bibinfo{volume}{3} (\bibinfo{year}{2018}),
  \bibinfo{pages}{63--83}.
\newblock
\urldef\tempurl%
\url{https://petsymposium.org/2018/files/papers/issue3/popets-2018-0021.pdf}
\showURL{%
\tempurl}


\bibitem[Schaub et~al\mbox{.}(2015)]%
        {Schaub:2015}
\bibfield{author}{\bibinfo{person}{Florian Schaub}, \bibinfo{person}{Rebecca
  Balebako}, \bibinfo{person}{Adam~L. Durity}, {and}
  \bibinfo{person}{Lorrie~Faith Cranor}.} \bibinfo{year}{2015}\natexlab{}.
\newblock \showarticletitle{A Design Space for Effective Privacy Notices}. In
  \bibinfo{booktitle}{\emph{Eleventh Symposium On Usable Privacy and Security
  (SOUPS 2015)}}. \bibinfo{publisher}{USENIX Association},
  \bibinfo{address}{Ottawa}, \bibinfo{pages}{1--17}.
\newblock
\showISBNx{978-1-931971-249}
\urldef\tempurl%
\url{https://www.usenix.org/conference/soups2015/proceedings/presentation/schaub}
\showURL{%
\tempurl}


\bibitem[Slavin et~al\mbox{.}(2016)]%
        {Slavin:2016}
\bibfield{author}{\bibinfo{person}{Rocky Slavin}, \bibinfo{person}{Xiaoyin
  Wang}, \bibinfo{person}{Mitra~Bokaei Hosseini}, \bibinfo{person}{James
  Hester}, \bibinfo{person}{Ram Krishnan}, \bibinfo{person}{Jaspreet Bhatia},
  \bibinfo{person}{Travis~D. Breaux}, {and} \bibinfo{person}{Jianwei Niu}.}
  \bibinfo{year}{2016}\natexlab{}.
\newblock \showarticletitle{Toward a Framework for Detecting Privacy Policy
  Violations in Android Application Code}. In \bibinfo{booktitle}{\emph{2016
  IEEE/ACM 38th International Conference on Software Engineering (ICSE)}}.
  \bibinfo{pages}{25--36}.
\newblock
\urldef\tempurl%
\url{https://doi.org/10.1145/2884781.2884855}
\showDOI{\tempurl}


\bibitem[Srinath et~al\mbox{.}(2021)]%
        {srinath-privbert-2021}
\bibfield{author}{\bibinfo{person}{Mukund Srinath}, \bibinfo{person}{Shomir
  Wilson}, {and} \bibinfo{person}{C~Lee Giles}.}
  \bibinfo{year}{2021}\natexlab{}.
\newblock \showarticletitle{Privacy at Scale: Introducing the {P}riva{S}eer
  Corpus of Web Privacy Policies}. In \bibinfo{booktitle}{\emph{Proceedings of
  the 59th Annual Meeting of the Association for Computational Linguistics and
  the 11th International Joint Conference on Natural Language Processing
  (Volume 1: Long Papers)}}, \bibfield{editor}{\bibinfo{person}{Chengqing
  Zong}, \bibinfo{person}{Fei Xia}, \bibinfo{person}{Wenjie Li}, {and}
  \bibinfo{person}{Roberto Navigli}} (Eds.). \bibinfo{publisher}{Association
  for Computational Linguistics}, \bibinfo{address}{Online},
  \bibinfo{pages}{6829--6839}.
\newblock
\urldef\tempurl%
\url{https://doi.org/10.18653/v1/2021.acl-long.532}
\showDOI{\tempurl}


\bibitem[{Staff FTC}(2011)]%
        {FTC:2011}
\bibfield{author}{\bibinfo{person}{{Staff FTC}}.}
  \bibinfo{year}{2011}\natexlab{}.
\newblock \showarticletitle{Protecting consumer privacy in an era of rapid
  change--A proposed framework for businesses and policymakers}.
\newblock \bibinfo{journal}{\emph{Journal of Privacy and Confidentiality}}
  \bibinfo{volume}{3}, \bibinfo{number}{1} (\bibinfo{year}{2011}),
  \bibinfo{numpages}{112}~pages.
\newblock
\urldef\tempurl%
\url{https://www.ftc.gov/reports/protecting-consumer-privacy-era-rapid-change-recommendations-businesses-policymakers}
\showURL{%
\tempurl}


\bibitem[Support(2022)]%
        {FirefoxReadability:2022}
\bibfield{author}{\bibinfo{person}{Mozilla Support}.}
  \bibinfo{year}{2022}\natexlab{}.
\newblock \bibinfo{title}{Firefox {Reader} {View} for clutter-free web pages
  {\textbar} {Firefox} {Help}}.
\newblock
\newblock
\urldef\tempurl%
\url{https://support.mozilla.org/en-US/kb/firefox-reader-view-clutter-free-web-pages}
\showURL{%
\tempurl}


\bibitem[{The Alan Turing Institute}(2018)]%
        {ReadabiliPy:2018}
\bibfield{author}{\bibinfo{person}{{The Alan Turing Institute}}.}
  \bibinfo{year}{2018}\natexlab{}.
\newblock \bibinfo{title}{{ReadabiliPy}}.
\newblock
\newblock
\urldef\tempurl%
\url{https://github.com/alan-turing-institute/ReadabiliPy}
\showURL{%
\tempurl}


\bibitem[Unity(2023)]%
        {Unity:Guidelines}
\bibfield{author}{\bibinfo{person}{Unity}.} \bibinfo{year}{2023}\natexlab{}.
\newblock \bibinfo{title}{Apple privacy survey for {Unity} {Ads}}.
\newblock
\newblock
\urldef\tempurl%
\url{https://docs.unity.com/ads/en/manual/ApplePrivacySurvey}
\showURL{%
\tempurl}


\bibitem[Walmart(2022)]%
        {Walmart}
\bibfield{author}{\bibinfo{person}{Walmart}.} \bibinfo{year}{2022}\natexlab{}.
\newblock \bibinfo{title}{{Walmart - Shopping \& Grocery}}.
\newblock
\newblock
\urldef\tempurl%
\url{https://apps.apple.com/us/app/walmart-shopping-grocery/id338137227}
\showURL{%
\tempurl}


\bibitem[{WebMD}(2023a)]%
        {WebMD:Policy}
\bibfield{author}{\bibinfo{person}{{WebMD}}.} \bibinfo{year}{2023}\natexlab{a}.
\newblock \bibinfo{title}{{WebMD} {Privacy} {Policy}}.
\newblock
\newblock
\urldef\tempurl%
\url{https://www.webmd.com/about-webmd-policies/about-privacy-policy}
\showURL{%
\tempurl}


\bibitem[{WebMD}(2023b)]%
        {WebMD}
\bibfield{author}{\bibinfo{person}{{WebMD}}.} \bibinfo{year}{2023}\natexlab{b}.
\newblock \bibinfo{title}{{WebMD}: {Symptom} {Checker}}.
\newblock
\newblock
\urldef\tempurl%
\url{https://apps.apple.com/us/app/webmd-symptom-checker/id295076329}
\showURL{%
\tempurl}


\bibitem[Wilson et~al\mbox{.}(2016)]%
        {Wilson:2016}
\bibfield{author}{\bibinfo{person}{Shomir Wilson}, \bibinfo{person}{Florian
  Schaub}, \bibinfo{person}{Aswarth~Abhilash Dara}, \bibinfo{person}{Frederick
  Liu}, \bibinfo{person}{Sushain Cherivirala}, \bibinfo{person}{Pedro
  Giovanni~Leon}, \bibinfo{person}{Mads Schaarup~Andersen},
  \bibinfo{person}{Sebastian Zimmeck}, \bibinfo{person}{Kanthashree~Mysore
  Sathyendra}, \bibinfo{person}{N.~Cameron Russell}, \bibinfo{person}{Thomas~B.
  Norton}, \bibinfo{person}{Eduard Hovy}, \bibinfo{person}{Joel Reidenberg},
  {and} \bibinfo{person}{Norman Sadeh}.} \bibinfo{year}{2016}\natexlab{}.
\newblock \showarticletitle{The Creation and Analysis of a Website Privacy
  Policy Corpus}. In \bibinfo{booktitle}{\emph{Proceedings of the 54th Annual
  Meeting of the Association for Computational Linguistics (Volume 1: Long
  Papers)}}. \bibinfo{publisher}{Association for Computational Linguistics},
  \bibinfo{address}{Berlin, Germany}, \bibinfo{pages}{1330--1340}.
\newblock
\urldef\tempurl%
\url{https://doi.org/10.18653/v1/P16-1126}
\showDOI{\tempurl}


\bibitem[Xiao et~al\mbox{.}(2022)]%
        {Xiao:2022}
\bibfield{author}{\bibinfo{person}{Yue Xiao}, \bibinfo{person}{Zhengyi Li},
  \bibinfo{person}{Yue Qin}, \bibinfo{person}{Xiaolong Bai},
  \bibinfo{person}{Jiale Guan}, \bibinfo{person}{Xiaojing Liao}, {and}
  \bibinfo{person}{Luyi Xing}.} \bibinfo{year}{2022}\natexlab{}.
\newblock \showarticletitle{Lalaine: Measuring and Characterizing
  Non-Compliance of Apple Privacy Labels at Scale}.
\newblock \bibinfo{journal}{\emph{CoRR}}  \bibinfo{volume}{abs/2206.06274}
  (\bibinfo{year}{2022}).
\newblock
\urldef\tempurl%
\url{https://doi.org/10.48550/arXiv.2206.06274}
\showDOI{\tempurl}
\showeprint[arXiv]{2206.06274}


\bibitem[Yu et~al\mbox{.}(2016)]%
        {Yu:2016}
\bibfield{author}{\bibinfo{person}{Le Yu}, \bibinfo{person}{Xiapu Luo},
  \bibinfo{person}{Xule Liu}, {and} \bibinfo{person}{Tao Zhang}.}
  \bibinfo{year}{2016}\natexlab{}.
\newblock \showarticletitle{Can We Trust the Privacy Policies of Android
  Apps?}. In \bibinfo{booktitle}{\emph{2016 46th Annual IEEE/IFIP International
  Conference on Dependable Systems and Networks (DSN)}}.
  \bibinfo{pages}{538--549}.
\newblock
\urldef\tempurl%
\url{https://doi.org/10.1109/DSN.2016.55}
\showDOI{\tempurl}


\bibitem[Zhang et~al\mbox{.}(2022)]%
        {Zhang:2022}
\bibfield{author}{\bibinfo{person}{Shikun Zhang}, \bibinfo{person}{Yuanyuan
  Feng}, \bibinfo{person}{Yaxing Yao}, \bibinfo{person}{Lorrie~Faith Cranor},
  {and} \bibinfo{person}{Norman Sadeh}.} \bibinfo{year}{2022}\natexlab{}.
\newblock \showarticletitle{How Usable Are iOS App Privacy Labels?}
\newblock \bibinfo{journal}{\emph{Proceedings on Privacy Enhancing
  Technologies}}  \bibinfo{volume}{4} (\bibinfo{year}{2022}),
  \bibinfo{pages}{204--228}.
\newblock


\bibitem[Zimmeck et~al\mbox{.}(2019)]%
        {Zimmeck:2019}
\bibfield{author}{\bibinfo{person}{Sebastian Zimmeck}, \bibinfo{person}{Peter
  Story}, \bibinfo{person}{Daniel Smullen}, \bibinfo{person}{Abhilasha
  Ravichander}, \bibinfo{person}{Ziqi Wang}, \bibinfo{person}{Joel~R.
  Reidenberg}, \bibinfo{person}{N.~Cameron Russell}, {and}
  \bibinfo{person}{Norman~M. Sadeh}.} \bibinfo{year}{2019}\natexlab{}.
\newblock \showarticletitle{MAPS: Scaling Privacy Compliance Analysis to a
  Million Apps}.
\newblock \bibinfo{journal}{\emph{Proceedings on Privacy Enhancing
  Technologies}}  \bibinfo{volume}{2019} (\bibinfo{year}{2019}),
  \bibinfo{pages}{66 -- 86}.
\newblock


\bibitem[Zimmeck et~al\mbox{.}(2017)]%
        {Zimmeck:2017}
\bibfield{author}{\bibinfo{person}{Sebastian Zimmeck}, \bibinfo{person}{Ziqi
  Wang}, \bibinfo{person}{Lieyong Zou}, \bibinfo{person}{Roger Iyengar},
  \bibinfo{person}{Bin Liu}, \bibinfo{person}{Florian Schaub},
  \bibinfo{person}{Shomir Wilson}, \bibinfo{person}{Norman Sadeh},
  \bibinfo{person}{{Steven M.} Bellovin}, {and} \bibinfo{person}{Joel
  Reidenberg}.} \bibinfo{year}{2017}\natexlab{}.
\newblock \showarticletitle{Automated Analysis of Privacy Requirements for
  Mobile Apps}. In \bibinfo{booktitle}{\emph{Proceedings 2017 Network and
  Distributed System Security Symposium}}.
\newblock
\urldef\tempurl%
\url{https://doi.org/10.14722/ndss.2017.23034}
\showDOI{\tempurl}


\end{thebibliography}
\end{footnotesize}

 \appendix

%\section{Methodology}

%In this section we both describe the re implementation of Polisis using the same training data as described by Harkous et al.~\cite{Harkous:2018}, followed by the measurement and analysis workflow. 
\section{Reimplementing and Training The Classification Framework}
\label{sec:framework}

%Polisis is a NLP tool for analyzing privacy policies. There is a publicly available front-end interface at \url{pribot.org}, however, after consulting with the providers, it was inappropriate to make many hundred of thousands requests to this interface. Additionally, the Polisis implementation at \url{pribot.org} is, unfortunately, proprietary, but the implementation details and the datasets used to train the algorithm are public, and there is an  open-sourced version of the implementation from the authors' other work~\cite{Windl:2022}. We further reached out to the Harkous et al.~\cite{Harkous:2018} and obtained an additional corpus that they  used to create in-domain word embeddings, which we describe later in this section.  Based on these details, we fully re-implemented Polisis to the same standards as in prior work to perform large-scale analysis. Below, we provide an overview of the Polisis framework and describe the procedure we adopted to recreate and train the NLP classifiers at its core. 

\paragraph{\textbf{Hierarchical Structure.}} Our implementation of the framework closely follows that used for Polisis~\cite{Harkous:2018}, which in-turn relies on the OPP-115 corpus~\cite{Wilson:2016}. It comprises a \textit{hierarchical, multi-level} set of classifiers.
The framework takes a paragraph-length segment of text as input, and passes it to a Segment Classifier to first determine one or more high-level data practices addressed in the segment. These data practices may look like, \textit{First Party Collection/Use}, \textit{Data Security}, \textit{International/Specific Audiences}, etc. The framework further passes the segment through multiple Attribute Classifiers, each of which determine one or more attribute values relevant to the data practice determined by the Segment Classifier. For example, if the segment addresses \textit{First Party Collection/Use}, the \textit{Does/Does Not}  Attribute Classifier determines if the policy claims to engage in data collection, the \textit{Identifiability}  Attribute Classifier determines if the data collection can be linked to the user, the \textit{Purpose}  Attribute Classifier determines the stated reason for data collection, and the \textit{Personal Information Type}  Attribute Classifier determines the data categories addressed in the segment.  The framework classifies one segment of the policy at a time, and the data practices addressed in the entire policy are determined by collating results from all segments. An overview of this structure is provided in \autoref{fig:polisis-structure}.

%\paragraph{Training Classifiers to Interpret Policies.}
% TODO: Explain Polisis & OPP-115

\paragraph{\textbf{Training Dataset.}} The Online Privacy Policies (OPP-115) dataset, created by Wilson et al.~\cite{Wilson:2016}, is an annotated dataset of 115 privacy policies. Each policy is divided into paragraph-length \textit{segments}, and manually annotated by law school students. Each segment was annotated at two levels -- first, the annotator chose one or more high-level data practices that the segment addresses (e.g., First Party Collection/Use, Third Party Collection/Sharing); then, depending on the initial selections, they annotated segments with multiple attribute-value pairs (e.g., information\_type: financial, purpose:advertising, etc.). Overall, the task covered 10 data practices and 20 associated attributes, with 138 distinct values across attributes. We developed one classifier to determine high-level data practices addressed in a segment, followed by a classifier each for the different attributes associated with the identified data practice. 
% Of these, 4 attributes were relevant to the creation of privacy labels -- we expand on their use later in this section.

\paragraph{\textbf{Train-Test Split.}} For each attribute, we collected all segments that had a relevant annotation for the attribute in the OPP-115 dataset. We then performed a separate 80-20 train-test split for each collection of segments belonging to an attribute. In this aspect, we differed from Harkous et al.~\cite{Harkous:2018}, who instead set 65 of the 115 policies aside for training, and used relevant segments from these 65 policies to train all attribute classifiers -- a choice that would have resulted in varied amounts of training data being used for each attribute.

% \paragraph{\textbf{Word Embeddings.}} Text classifiers deal with text by representing their features as building blocks. A simple example of this would consider the \textit{frequency} of occurrence of each word as a feature used to train classifiers. This approach is limited in its ability to interpret words outside of the dataset used to train classifiers, hence limiting its ability to generalize.

% Word embeddings offer a different approach by extracting vector representations of words, in an unsupervised manner, from a large corpus of text. These representations can directly be used as features to classifiers, and also account for words that the classifier has not observed in its training phase.

% General purpose embeddings like GloVe~\cite{Pennington:Glove:2014} and Word2vec~\cite{Mikolov:Word2Vec:2013}, have been trained on large text corpuses and provide useful word representations for most use cases. However, domain-specific embeddings generates better classification results~\cite{Tang:2014}. We therefore reached out to Harkous et al.~\cite{Harkous:2018} and requested a copy of their corpus of 130K privacy policies of apps on the Google Play Store. We then trained a word-embeddings model using \textit{fastText}~\cite{Bojanowski:fastText:2017}, which also helped create representations for subwords, and account for words outside of the policies in the training corpus.

\manualVerificationTable

\paragraph{\textbf{Evaluation Metrics.}} The authors of PrivBERT~\cite{srinath-privbert-2021} presented an example of fine-tuning a segment classifier using the OPP-115 corpus, in which they manually tuned the hyperparameters used to train the model. We followed a similar approach to develop each classifier.~\autoref{tab:classifier-results} presents the evaluation reports for the classifier's precision, recall, and F1 scores on an unseen test set. Following the practice highlighted by Harkous et al.~\cite{Harkous:2018}, we evaluate each classifier's ability to detect both the \textit{presence} and \textit{absence} of an attribute in a given text segment. 
Additionally, since the OPP-115 corpus is old, we additionally manually evaluated classifier outputs on randomly sampled segments of Apple App Store policies, which we also report in~\autoref{tab:classifier-results}. For each attribute, we randomly sampled 25 segments for which the classifier reported the presence of an attribute and also sampled 25 segments for which it reported the absence of an attribute. In this manner, we cover 50 segments each for 35 attributes across privacy policies. \autoref{tab:classifier-results} also compares the classification reports for implementing the Polisis CNN-based approach against the performance of the fine-tuned BERT-based models. Finally, to verify our mapping of classifier outputs to privacy label attributes, two researchers randomly sampled and manually evaluated the outputs for 25 instances of each label output (see~\autoref{tab:mapping-verification}).

% \paragraph{\textbf{Hyperparameters and Evaluation Metrics.}} The hyperparameters for each classifier were determined using a randomized grid-search. We adopted a similar approach to that of Harkous et al.~\cite{Harkous:2018}, and evaluated classifiers based on the precision, recall, and F1 scores, macro-averaged per label. We present the evaluation results in \autoref{tab:classifier-results}, where the numbers show the classifiers' ability to detect both, the \textit{presence} and \textit{absence} of a label in a given text segment. We also report the accuracy and confidence intervals for each classifier result after performing bootstrap sampling on the test set~\cite{BootstrappingTestSet:2022,Sanchez:2019}. 

% \methodologyFigure

% \usepackage{graphicx}
\begin{table*}[t]
\centering
\caption{Classification results for the attributes that were used in the creation of Privacy Labels.}
\label{tab:classifier-results}
% \resizebox{0.8\textwidth}{!}{%
\resizebox*{!}{0.95\textheight}{%
\begin{tabular}{lrrrrrrrrr}
\\
\toprule
&
  \multicolumn{4}{|c|}{\textbf{PrivBERT Classification Report}} &
  \multicolumn{2}{c}{\textbf{\begin{tabular}[c]{@{}c@{}}Manual Check on\\ New Segments\end{tabular}}} &
  \multicolumn{3}{|c}{\textbf{Polisis Classification Report}} \\\cline{2-10}
 \multicolumn{1}{c}{\multirow{-4}{*}{\textbf{Classifier Output}}}&
  \multicolumn{1}{|c}{\multirow{-1}{*}{\textbf{Precision}}} &
  \multicolumn{1}{c}{\multirow{-1}{*}{\textbf{Recall}}} &
  \multicolumn{1}{c}{\multirow{-1}{*}{\textbf{F1}}} &
  % \multicolumn{1}{c|}{{\textbf{\begin{tabular}[c]{@{}c@{}}Support\\(Presence$|$Absence)\end{tabular}}} &
  \multicolumn{1}{c|}{\textbf{\begin{tabular}[c]{@{}c@{}}Support\\ (Presence$|$Absence)\end{tabular}}} &
  \multicolumn{1}{c}{\multirow{-1}{*}{\textbf{Presence}}} &
  \multicolumn{1}{c}{\multirow{-1}{*}{\textbf{Absence}}} &
  \multicolumn{1}{|c}{\multirow{-1}{*}{\textbf{Precision}}} &
  \multicolumn{1}{c}{\multirow{-1}{*}{\textbf{Recall}}} &
  \multicolumn{1}{c}{\multirow{-1}{*}{\textbf{F1}}} \\
\hline
 % &
 %  \multicolumn{1}{l}{} &
 %  \multicolumn{1}{l}{} &
 %  \multicolumn{1}{l}{} &
 %  \multicolumn{1}{l}{} &
 %  \multicolumn{1}{l}{} &
 %  \multicolumn{1}{l}{} &
 %  \multicolumn{1}{l}{} \\
\textbf{Segment Classifier} &
  \multicolumn{1}{l}{} &
  \multicolumn{1}{l}{} &
  \multicolumn{1}{l}{} &
  \multicolumn{1}{l}{} &
  \multicolumn{1}{l}{} &
  \multicolumn{1}{l}{} &
  \multicolumn{1}{l}{} &
  \multicolumn{1}{l}{} &
  \multicolumn{1}{l}{} \\
First Party Collection/Use &
  0.89 &
  0.88 &
  0.89 &
  298$|$460 &
  25/25 &
  24/25 &
  0.80 &
  0.79 &
  0.80 \\
Third Party Collection/Sharing &
  0.92 &
  0.92 &
  0.92 &
  258$|$500 &
  24/25 &
  23/25 &
  0.88 &
  0.85 &
  0.86 \\
International and Specific Audiences &
  0.97 &
  0.97 &
  0.97 &
  59$|$699 &
  25/25 &
  23/25 &
  0.97 &
  0.95 &
  0.96 \\
  \hline
\multicolumn{1}{r}{Average} &
  0.93 &
  0.92 &
  0.93 &
  \multicolumn{1}{l}{} &
   &
   &
  0.88 &
  0.86 &
  0.87 \\
  \hline
 &
  \multicolumn{1}{l}{} &
  \multicolumn{1}{l}{} &
  \multicolumn{1}{l}{} &
  \multicolumn{1}{l}{} &
  \multicolumn{1}{l}{} &
  \multicolumn{1}{l}{} &
  \multicolumn{1}{l}{} &
  \multicolumn{1}{l}{} &
  \multicolumn{1}{l}{} \\
\textbf{Identifiability} &
  \multicolumn{1}{l}{} &
  \multicolumn{1}{l}{} &
  \multicolumn{1}{l}{} &
  \multicolumn{1}{l}{} &
  \multicolumn{1}{l}{} &
  \multicolumn{1}{l}{} &
  \multicolumn{1}{l}{} &
  \multicolumn{1}{l}{} &
  \multicolumn{1}{l}{} \\
Identifiable &
  0.94 &
  0.91 &
  0.92 &
  115$|$54 &
  25/25 &
  20/25 &
  0.75 &
  0.76 &
  0.75 \\
Aggregated or Anonymized &
  0.96 &
  0.97 &
  0.96 &
  59$|$110 &
  20/25 &
  25/25 &
  0.85 &
  0.85 &
  0.80 \\
   \hline
\multicolumn{1}{r}{Average} &
  0.95 &
  0.94 &
  0.94 &
  \multicolumn{1}{l}{} &
   &
   &
   0.80 &
   0.80 &
   0.80 \\ \hline
 &
  \multicolumn{1}{l}{} &
  \multicolumn{1}{l}{} &
  \multicolumn{1}{l}{} &
  \multicolumn{1}{l}{} &
  \multicolumn{1}{l}{} &
  \multicolumn{1}{l}{} &
  \multicolumn{1}{l}{} &
  \multicolumn{1}{l}{} &
  \multicolumn{1}{l}{} \\
\textbf{Does/Does Not} &
  \multicolumn{1}{l}{} &
  \multicolumn{1}{l}{} &
  \multicolumn{1}{l}{} &
  \multicolumn{1}{l}{} &
  \multicolumn{1}{l}{} &
  \multicolumn{1}{l}{} &
  \multicolumn{1}{l}{} &
  \multicolumn{1}{l}{} &
  \multicolumn{1}{l}{} \\
Does Not &
  0.87 &
  0.86 &
  0.86 &
  47$|$393 &
  25/25 &
  25/25 &
  0.91 &
  0.80 &
  0.84 \\
Does &
  0.78 &
  0.83 &
  0.81 &
  428$|$12 &
  25/25 &
  25/25 &
  0.74 &
  0.66 &
  0.70 \\ \hline
\multicolumn{1}{r}{Average} &
  0.83 &
  0.85 &
  0.84 &
  \multicolumn{1}{l}{} &
   &
   &
  0.82 &
  0.73 &
  0.77\\ \hline
\multicolumn{1}{r}{} &
  \multicolumn{1}{l}{} &
  \multicolumn{1}{l}{} &
  \multicolumn{1}{l}{} &
  \multicolumn{1}{l}{} &
  \multicolumn{1}{l}{} &
  \multicolumn{1}{l}{} \\
\textbf{Purpose} &
  \multicolumn{1}{l}{} &
  \multicolumn{1}{l}{} &
  \multicolumn{1}{l}{} &
  \multicolumn{1}{l}{} &
  \multicolumn{1}{l}{} &
  \multicolumn{1}{l}{} \\
Additional Service/Feature &
  0.83 &
  0.86 &
  0.84 &
  103$|$399 &
  25/25 &
  24/25 &
  0.82 &
  0.79 &
  0.80 \\
Advertising &
  0.95 &
  0.93 &
  0.94 &
  125$|$377 &
  24/25 &
  25/25 &
  0.87 &
  0.84 &
  0.86 \\
Analytics/Research &
  0.89 &
  0.90 &
  0.89 &
  88$|$414 &
  24/25 &
  25/25 &
  0.86 &
  0.85 &
  0.85 \\
Basic Service/Feature &
  0.84 &
  0.84 &
  0.84 &
  135$|$367 &
  25/25 &
  22/25 &
  0.80 &
  0.80 &
  0.80 \\
Legal Requirement &
  0.90 &
  0.87 &
  0.89 &
  35$|$467 &
  25/25 &
  25/25 &
  0.92 &
  0.83 & 
  0.87 \\
Marketing &
  0.86 &
  0.85 &
  0.86 &
  123$|$379 &
  25/25 &
  25/25 &
  0.84 &
  0.82 &
  0.83 \\
Merger &
  0.94 &
  1.00 &
  0.97 &
  13$|$489 &
  25/25 &
  25/25 &
  1.00 &
  0.88 &
  0.93 \\
Personalization &
  0.88 &
  0.85 &
  0.86 &
  70$|$432 &
  25/25 &
  23/25 & 
  0.86 &
  0.80 &
  0.82 \\
Service Operation and Security &
  0.90 &
  0.90 &
  0.90 &
  37$|$465 &
  25/25 &
  23/25 &
  0.86 &
  0.81 &
  0.83 \\
Unspecified &
  0.79 &
  0.79 &
  0.78 &
  227$|$275 &
  23/25 &
  25/25 &
  0.81 &
  0.73 &
  0.76 \\ \hline
\multicolumn{1}{r}{Average} &
  0.88 &
  0.88 &
  0.87 &
  \multicolumn{1}{l}{} &
   &
   &
  0.86 &
  0.81 &
  0.83 \\ \hline
\multicolumn{1}{r}{} &
  \multicolumn{1}{l}{} &
  \multicolumn{1}{l}{} &
  \multicolumn{1}{l}{} &
  \multicolumn{1}{l}{} &
  \multicolumn{1}{l}{} &
  \multicolumn{1}{l}{} &
  \multicolumn{1}{l}{} &
  \multicolumn{1}{l}{} &
  \multicolumn{1}{l}{} \\
\textbf{Personal Information Type} &
  \multicolumn{1}{l}{} &
  \multicolumn{1}{l}{} &
  \multicolumn{1}{l}{} &
  \multicolumn{1}{l}{} &
  \multicolumn{1}{l}{} &
  \multicolumn{1}{l}{} &
  \multicolumn{1}{l}{} &
  \multicolumn{1}{l}{} &
  \multicolumn{1}{l}{} \\
Computer Information &
  0.88 &
  0.94 &
  0.91 &
  30$|$483 &
  25/25 &
  24/25 &
  0.94 &
  0.91 &
  0.92 \\
Contact &
  0.94 &
  0.91 &
  0.92 &
  129$|$384 &
  24/25 &
  25/25 &
  0.91 &
  0.90 &
  0.91 \\
Cookies and Tracking Elements &
  0.99 &
  0.96 &
  0.98 &
  93$|$420 &
  23/25 &
  20/25 &
  0.95 &
  0.87 &
  0.91 \\
Demographic &
  0.94 &
  0.89 &
  0.92 &
  47$|$466 &
  24/25 &
  25/25 &
  0.90 &
  0.86 &
  0.88 \\
Financial &
  0.85 &
  0.77 &
  0.81 &
  39$|$474 &
  25/25 &
  25/25 &
  0.94 &
  0.90 &
  0.92 \\
Generic Personal Information &
  0.84 &
  0.83 &
  0.83 &
  196$|$317 &
  25/25 &
  25/25 &
  0.82 &
  0.81 &
  0.81 \\
Health &
  0.91 &
  0.95 &
  0.93 &
  10$|$503 &
  23/25 &
  25/25 &
  0.95 &
  0.66 &
  0.74 \\
IP Address and Device IDs &
  0.88 &
  0.91 &
  0.89 &
  36$|$477 &
  25/25 &
  24/25 &
  0.97 &
  0.89 &
  0.92 \\
Location &
  0.90 &
  0.75 &
  0.81 &
  14$|$460 &
  25/25 &
  24/25 &
  0.91 &
  0.85 &
  0.88 \\
Personal Identifier &
  0.67 &
  0.67 &
  0.67 &
  14$|$499 &
  25/25 &
  24/25&
  0.95 &
  0.77 &
  0.83 \\
Social Media Data &
  0.72 &
  0.80 &
  0.75 &
  5$|$511 &
  24/25 &
  25/25 &
  0.93 &
  0.82 &
  0.86 \\
% Survey Data &
%   0.54 &
%   0.59 &
%   0.56 &
%   5/508 &
%    &
%    \\
User Online Activities &
  0.86 &
  0.78 &
  0.81 &
  99$|$439 &
  24/25 &
  25/25 &
  0.88 &
  0.87 &
  0.88 \\
User Profile &
  0.63 &
  0.75 &
  0.67 &
  15$|$498 &
  23/25 &
  24/25 &
  0.90 &
  0.82 &
  0.86 \\
% Unspecified &
%   0.73 &
%   0.73 &
%   0.72 &
%   251/262 &
%    &
%    \\ 
   \hline
\multicolumn{1}{r}{Average} &
  0.82 &
  0.82 &
  0.81 &
  \multicolumn{1}{l}{} &
   &
   &
  0.86 &
  0.81 &
  0.83 \\ \hline
  & 
  \multicolumn{1}{l}{} &
  \multicolumn{1}{l}{} &
  \multicolumn{1}{l}{} &
  \multicolumn{1}{l}{} &
  \multicolumn{1}{l}{} &
  \multicolumn{1}{l}{} &
  \multicolumn{1}{l}{} &
  \multicolumn{1}{l}{} &
  \multicolumn{1}{l}{} \\
\textbf{Audience Type} &
  \multicolumn{1}{l}{} &
  \multicolumn{1}{l}{} &
  \multicolumn{1}{l}{} &
  \multicolumn{1}{l}{} &
  \multicolumn{1}{l}{} &
  \multicolumn{1}{l}{} &
  \multicolumn{1}{l}{} &
  \multicolumn{1}{l}{} &
  \multicolumn{1}{l}{} \\
Children &
  0.99 &
  0.99 &
  0.99 &
  35$|$33 &
  25/25 &
  25/25 
  0.99 &
  0.99 &
  0.99 \\ \hline
\multicolumn{1}{r}{Average} &
  0.99 &
  0.99 &
  0.99 &
  \multicolumn{1}{l}{} &
   &
   &
  0.99 &
  0.99 &
  0.99 \\ \hline
\multicolumn{1}{r}{} &
  \multicolumn{1}{l}{} &
  \multicolumn{1}{l}{} &
  \multicolumn{1}{l}{} &
  \multicolumn{1}{l}{} &
  \multicolumn{1}{l}{} &
  \multicolumn{1}{l}{} &
  \multicolumn{1}{l}{} &
  \multicolumn{1}{l}{} &
  \multicolumn{1}{l}{} \\ 
\textbf{Action First Party} &
  \multicolumn{1}{l}{} &
  \multicolumn{1}{l}{} &
  \multicolumn{1}{l}{} &
  \multicolumn{1}{l}{} &
  \multicolumn{1}{l}{} &
  \multicolumn{1}{l}{} &
  \multicolumn{1}{l}{} &
  \multicolumn{1}{l}{} &
  \multicolumn{1}{l}{} \\
Collect on website &
  0.90 &
  0.83 &
  0.86 &
  180$|$6 &
  23/25 &
  25/25 &
  0.77 &
  0.66 &
  0.67 \\
Collect in mobile app &
  0.97 &
  0.79 &
  0.85 &
  23$|$163 &
  25/25 &
  25/25 &
  0.82 &
  0.75 &
  0.78 \\ \hline
\multicolumn{1}{r}{Average} &
  0.93 &
  0.81 &
  0.85 &
  \multicolumn{1}{l}{} &
   &
   &
  0.80 &
  0.71 &
  0.73 \\ \hline
 &
  \multicolumn{1}{l}{} &
  \multicolumn{1}{l}{} &
  \multicolumn{1}{l}{} &
  \multicolumn{1}{l}{} &
  \multicolumn{1}{l}{} &
  \multicolumn{1}{l}{} &
  \multicolumn{1}{l}{} &
  \multicolumn{1}{l}{} &
  \multicolumn{1}{l}{} \\
\textbf{Action Third-Party} &
  \multicolumn{1}{l}{} &
  \multicolumn{1}{l}{} &
  \multicolumn{1}{l}{} &
  \multicolumn{1}{l}{} &
  \multicolumn{1}{l}{} &
  \multicolumn{1}{l}{} &
  \multicolumn{1}{l}{} &
  \multicolumn{1}{l}{} &
  \multicolumn{1}{l}{} \\
\cellcolor[HTML]{FFFFFF}Collect on first party website/app &
  0.90 &
  0.98 &
  0.94 &
  43$|$8 &
  25/25 &
  24/25 &
  0.84 &
  0.80 &
  0.82 \\
See &
  0.89 &
  0.87 &
  0.87 &
  14$|$47 &
  25/25 &
  25/25 &
  0.90 &
  0.73 &
  0.79 \\ \hline
\multicolumn{1}{r}{Average} &
  0.89 &
  0.92 &
  0.90 &
  \multicolumn{1}{l}{} &
   &
   &
  0.87 &
  0.77 &
  0.80 \\ \hline
 &
  \multicolumn{1}{l}{} &
  \multicolumn{1}{l}{} &
  \multicolumn{1}{l}{} &
  \multicolumn{1}{l}{} &
  \multicolumn{1}{l}{} &
  \multicolumn{1}{l}{} &
  \multicolumn{1}{l}{} &
  \multicolumn{1}{l}{} &
  \multicolumn{1}{l}{} 
\end{tabular}%
}
\end{table*}

\section{Case Studies of Privacy Policies}
\label{sec:case-studies-policies}

To further provide an understanding of the differences between policies and labels, we present a few interesting examples of popular apps and their privacy policies.

\paragraph{\textbf{Subsplash.}} A platform that develops and integrates multiple church services, including donations, memberships, and services, Subsplash~\cite{Subsplash} is used by 8,015 apps of local churches on the App Store  (examples,~\cite{Subsplash:FamilyLife, Subsplash:DappyTKeys}).

All of the hosted apps link to the same privacy policy~\cite{Subsplash:Policy} and share the same privacy label, i.e., a \textit{Data Not Linked to You} label, which states that the app collects \textit{Usage Data} for \textit{Analytics}, and \textit{Diagnostics} data for \textit{App Functionality}. Recall that the \textit{Data Not Linked to You} privacy type indicates that the data that is collected is aggregated or anonymized. Subsplash's policy states that they collect \textit{Contact Info}, \textit{Financial Info}, \textit{Purchases}, none of which are included in their privacy label. A snippet from their policy is provided below.
\begin{quote}
    \textit{When you interact with Subsplash, we may collect personal information relevant to the situation, such as your name, mailing address, phone number, email address, and contact preferences; your credit card information and information about the Subsplash products you own, such as their serial numbers and date of purchase; and information relating to a support or service issue.}
\end{quote}
The apps  additionally collects \textit{Location}, and \textit{Contacts} as stated in different segments but not included in the apps' privacy label. 

At the same time, there are some examples of the structure of the privacy policy that may lead Polisis classifiers to under- or over-represent some behaviors. One example is the treatment of anonymization of data. A single segment highlighting anonymization but does not specify which data types are anonymized. 
\begin{quote}
    \textit{Subsplash may use aggregated and anonymized forms of personal information for a variety of purposes, including, but not limited to, analyzing usage trends, fraud detection, and development of new Services.}
\end{quote}
As a result, Polisis is unable to match the data collection practice to anonymous linking and would classify most of the data collected by the app as {\em linked} rather than {\em not linked}. At the same time, since the policy is unclear on this point, it is difficult to fully know the data practices and if the labels are correct on this matter. 

Another example involves the format of Subsplash's privacy policy which includes some  data collection practices in varied visual formats, i.e., a table that includes different categories of data, examples of data types, and a column that states whether or not the stated data is collected. However, this table is implemented using \texttt{<div>} tags around each cell. The \texttt{readability} library interprets each of the cells as a separate paragraph, and makes it difficult to interpret the data presented here, potentially under-reporting some behavior as the segments are less complete. 

%Additionally, since the table is not available in a standardized format, it is difficult to develop countermeasures that would have accounted for such variations as we did with lists and headings in \S~\ref{sec:polisis}.

\paragraph{\textbf{ChowNow.}}
ChowNow~\cite{ChowNow} is an app platform used by 3,182 different apps of local restaurants to receive online orders for takeout and delivery (examples,~\cite{ChowNow:ElCharrito,ChowNow:HomeState,ChowNow:Bagelman}).

All apps using the ChowNow platform link to the same privacy policy~\cite{ChowNow:Policy} and apply the same privacy label. The label indicates that all data collection is not linked, indicating that the collected data is aggregated or anonymized. However, ChowNow's privacy policy states that they use contact information to manage user accounts and inform users about products through \textit{``electronic marketing communications''}. They also state that they use billing information, including card numbers, expiration date, security code, and billing address to process orders. Neither of these services can be provided in an anonymized manner, but the privacy labels lack a \textit{Data Linked to You} category.

ChowNow's privacy policy also states that they share information with advertisers, but their label does not include a \textit{Data Used to Track You} label. Additionally, the information that they share is mentioned as \textit{Other Information}, making it difficult for the Polisis framework to identify the data categories shared with third party services. The relevant snippet is provided below.
\begin{quote}
    \textit{We share Other Information about your activity in connection with your use of the Services with third-party advertisers and remarketers for the purpose of tailoring, analyzing, managing, reporting, and optimizing advertising you see on the Platforms, the Websites, the Apps, and elsewhere.}    
\end{quote}

ChowNow adds a content rating of 4+ to its apps on the App Store, making it accessible for children. Recall that developers choose a content rating according to Apple's guidelines~\cite{Apple:Guidelines}; this value is not assigned by Apple. However, ChowNow's privacy policy  absolve themselves of the responsibility of dealing with data collected from children, instead placing the burden of preventing such data collection on parents, guardians, and the children themselves. The relevant snippet is provided below.
\begin{quote}
    \textit{We do not knowingly collect personal information from children under the age of 13 through the Services. If you are under 13, please do not give us any personal information. We encourage parents and legal guardians to monitor their children’s Internet usage and to help enforce our Privacy Policy by instructing their children to never provide us personal information without their permission. If you have reason to believe that a child under the age of 13 has provided personal information to us, please contact us, and we will endeavor to delete that information from our databases.}
\end{quote}

\paragraph{\textbf{Walmart.}} A popular shopping and grocery delivery app with 6.6M user ratings, Walmart~\cite{Walmart} provides a large number of privacy labels on the App Store, which includes an extensive list of data categories across three privacy types, \textit{Data Used to Track You}, \textit{Data Linked to You}, and \textit{Data Not Collected}. 

Apple's description of sensitive information covers a list of example data types that are considered sensitive, providing  a general overview of possible values. 
Walmart's privacy label does \textit{not} state that it collects \textit{Sensitive Info}, which users may expect from a shopping and grocery delivery app. However, Walmart states in their privacy policy that they collect (i) demographic data, (ii) background \& criminal information, and (iii) audio, visual and other sensory information, all of which Apple may consider sensitive information.

\paragraph{\textbf{Credit Karma.}} A popular finance app with 5.4M user ratings on the App Store, Credit Karma~\cite{CreditKarma} does not use a \textit{Data Used to Track You} label on the App Store despite stating in their policy that they share personal information with 
% third parties, 
\textit{``other companies, lawyers, credit bureaus, agents, government agencies, and card associations in connection with issues related to fraud, credit, defaults, or debt collection''}. 

We also observed that multiple privacy policies, including others previously mentioned in this section, ask users to refer to the policies of third party providers that they use within their services. An example snippet from Credit Karma's policy is provided below.
\begin{quote}
    \textit{We may use third party API services, such as YouTube and Twilio, for certain product features. If you choose to use those features, you acknowledge and agree that you are also bound by the third party’s privacy policy, such as Google’s Privacy Policy for YouTube API services. You may manage your YouTube API data by visiting Google’s security settings page at \url{https://security.google.com/settings/security/permissions}. For more information about Twilio’s privacy practices, please visit \url{https://www.twilio.com/legal/privacy}.}
\end{quote}

This practice not only increases the burden of gathering additional information for users, but it also makes it difficult for Polisis to infer potentially missing information included in these additional external policies. As a result, the analysis of Credit Karma and similar apps may be a lower bound of the true privacy related behavior. 

\paragraph{\textbf{Aldi.}} A popular grocery store in the United States, Aldi, has an app available on the App Store, which is ranked \#59 in the Shopping category~\cite{Aldi}. The app offers a wide range of features, enabling users to conveniently order groceries, schedule deliveries or pickups, and make secure payments for their purchases. According to their privacy policy~\cite{Aldi:Policy}, Aldi collects (1)  payment information (such as credit or debit card or EBT number, security code, expiration date and billing address); (2) shopping list and purchase history information. It is worth noting, however, that their privacy label on the App Store does not include corresponding entries highlighting their collection of \textit{Financial Info} and \textit{Purchase History}.

\paragraph{\textbf{Axolochi.}} A popular application under the \textit{Games} category, Axolochi is ranked \#78 in the \textit{Trivia} sub-category~\cite{Axolochi}. The app's privacy policy~\cite{Axolochi:Policy} states the \textit{automatic} collection of various identifiers, such as a unique user ID, IP address, device IDs, hardware or operating system-based identifiers, and identifiers assigned to user accounts. Surprisingly, the app's privacy label on the App Store does not include the \textit{Identifiers} data category.

Furthermore, Axolochi offers in-app purchases for users. According to their privacy policy, when users make in-app purchases, the app collects ZIP or postal codes along with ``the amount of the transaction and records of purchases'' made by the user. However, it is worth noting that the privacy label on the App Store does not feature corresponding entries for \textit{Physical Address} or \textit{Purchase History}. This discrepancy may limit the visibility and transparency of the app's data practices, potentially leaving users with incomplete information regarding the collection and usage of their personal data within the app.

\paragraph{\textbf{WebMD.}} A widely known health-related service, WebMD hosts a flagship symptom checker app on the App Store~\cite{WebMD}. Their privacy policy~\cite{WebMD:Policy} explicitly mentions the collection of information from third-party vendors for targeted advertising purposes. 
\begin{quote}
\textit{Our ad network vendors use technologies to collect information about your activities on the WebMD Sites and \textbf{in our flagship WebMD App} to provide you cookie-based \textbf{targeted advertising} on our WebMD Sites and on third party websites based upon your \textbf{browsing activity and your interests}.}
\end{quote}

Surprisingly, the app does not include a specific privacy type entry for \textit{Data Used to Track You} in their privacy label. This absence in the privacy label highlights an instance of inconsistency in declaration of data collection practices across disclosures.

% \paragraph{\textbf{Chemistry \& Periodic Table.}} An app under the \textit{Education} category, this application assists users in solving chemical equations~\cite{Chemistry}. The app's privacy label on the App Store declares the collection of \textit{Identifiers}, \textit{Usage Data}, and \textit{Diagnostics}. However, their privacy policy~\cite{Chemistry:Policy} reveals that the app automatically collects information concerning the user's country and location. Unfortunately, the developers did not include corresponding entry for the \textit{Location} data category within the privacy label that they declared on App Store. 

\paragraph{\textbf{Pregnancy Tracker.}} The pregnancy tracking app developed by Fitness Labs has concerning discrepancies between its privacy label on the App Store~\cite{PregnancyTracker} and its privacy policy~\cite{PregnancyTracker:Policy}. The app's privacy label only includes a \textit{Data Not Linked to You} privacy type, mentioning the collection of \textit{Usage Data} and \textit{Diagnostics} data categories. However, the privacy policy reveals a much broader scope of data collection. The policy states:
they may collect personal information such as name, address, email address, phone numbers, payment information (credit or debit card), and other demographic information that can identify individuals or enable contact. 
\begin{quote}
    \textit{We may collect information about you such as: personal information including, for example, your name; home or business address; e-mail address; telephone, wireless or fax number; short message service or text message address or other wireless device address; instant messaging address; credit or debit card or other payment information; demographic information or other information that may \textbf{identify you as an individual} or allow online or offline contact with you as an individual.}
\end{quote}

Unfortunately, the app's privacy label fails to include the \textit{Data Linked to You} privacy type or indicate the collection of multiple data categories, including \textit{Identifiers}, \textit{Financial Information}, \textit{Contact Information}, and \textit{Sensitive Information}.

% \paragraph{Pandora.} Collect info about ``contact'', don't include in label. No Data Not Linked to You Label: Sharing of Deidentified, Aggregated, or Anonymized Information.
% \input{templates-appendix}

\trafficCollectionTable
\section{Network Traffic Collection}
\label{sec:traffic-collection}

We provide an overview of the analysis of 39 apps in~\autoref{tab:traffic-collection}.
% \onecolumn
\section{Additional Tables and Figures}
\label{sec:additional-figs}

We include additional tables and figures here.~\autoref{fig:ratings-counts-by-privacy} provides an overview of our findings based on apps' popularity.~\autoref{fig:app-genre-by-privacy} presents our findings based on app genres.~\autoref{tab:data-categories-overlap-browsing-history} details overlaps and discrepancies in disclosures across data categories in privacy labels and policies.

% \paragraph{\textbf{App Genre.}} We present an overview of our findings by app genre in \autoref{fig:app-genre-by-privacy}. We find that \textit{Games} apps are most likely to collect data used to track users (60\%) and linked to users (59\%) (see Fig.~\ref{fig:app-genre-by-privacy}; plots 1 \& 2; bar 8). Notably, while 83\% apps associated with the \textit{Stickers} genre stated on the App Store that they do not collect \textit{any} data, our analysis found that 66\% apps collected data linked to users (Fig.~\ref{fig:app-genre-by-privacy};plots 2 \& 4; bar 23). 

% \paragraph{\textbf{App Popularity.}} Since the App Store does not reveal the number of downloads for an app, we instead rely on the number of user ratings as a proxy for app popularity. To better represent their disclosures, we bin rating counts within the same order of magnitude in a single category, and present our findings in \autoref{fig:ratings-counts-by-privacy}. We find that with an increase in popularity, apps are more likely to declare data collection that is linked to users and used to track users. Our findings suggest that popular apps, which receive more scrutiny, are more likely to be more thorough in their declaration of data collection practices.  

% \privacyLabelPolicyOverlap
% \privacyTypesvsDataCategoriesBar
% \appCostsByPrivacyLabel

%\contentRatingByPrivacyLabel
% \ratingsCountByPrivacyLabel
% \appSizeByPrivacyLabel
% \releaseDateByPrivacyLabel
% \versionDateByPrivacyLabel
% \appStoreGenreTopFiveByPrivacyType

\begin{figure*}[t]
\centering
\input{figures/rating_counts_by_privacy_label_type.pgf}
\caption[Rating Counts By Privacy Label Type Ratios]{The ratios of the rating counts for each of the four \emph{Privacy Types}. The denominator is the number of apps with the designated rating counts that have a privacy label. Apps with a larger number of user ratings are more likely to collect data, including data used to track users. Ratings counts are not localized metadata and apps with low ratings counts in the US region may have higher counts elsewhere.\label{fig:ratings-counts-by-privacy}}
\end{figure*}

\begin{figure*}[ht]
\centering
\resizebox{!}{0.95\textheight}{\input{figures/app_genre_by_privacy_label_type.pgf}}
\caption[App Genre By Privacy Label Type Ratios]{The ratios of app store genres for each of the four \emph{Privacy Types}. The denominator is the number of apps with the designated app store genre that have a privacy label.}\label{fig:app-genre-by-privacy} 
\end{figure*}

\dataCategoriesOverlap

\clearpage
% \input{addlfigs}

%%%%%%%%%%%%%%%%%%%%%%%%%%%%%%%%%%%%%%%%%%%%%%%%%%%%%%%%%%%%%%%%%%%%%%%%%%%%%%%%
\end{document}